\documentclass[
rmp,aps,%
twocolumn,%
floatfix%
]{revtex4}

\usepackage{graphicx}
\setkeys{Gin}{width=3.375 in}

\usepackage{amsmath2000}

\def\prn#1{{\left(#1\right)}}
\def\cbrk#1{{\left\{#1\right\}}}
\def\sbrk#1{{\left[#1\right]}}
\def\abs#1{{\left|#1\right|}}
\def\abrk#1{{\left\langle#1\right\rangle}}
\def\cg#1#2{{\left\langle#1\vert#2\right\rangle}}
\def\bra#1{{\left\langle#1\right\vert}}
\def\ket#1{{\left\vert#1\right\rangle}}
\def\cg#1#2{{\left\langle#1\vert#2\right\rangle}}
\def\threej(#1,#2)(#3,#4)(#5,#6){\begin{pmatrix}#1&#3&#5\\#2&#4&#6\end{pmatrix}}
\def\sixj(#1,#2,#3)(#4,#5,#6){\begin{Bmatrix}#1&#2&#3\\#4&#5&#6\end{Bmatrix}}
\def\ninej(#1,#2,#3)(#4,#5,#6)(#7,#8,#9){\begin{Bmatrix}#1&#2&#3\\#4&#5&#6\\#7&#8&#9\end{Bmatrix}}

\DeclareMathOperator{\re}{Re}
\DeclareMathOperator{\im}{Im}
\DeclareMathOperator{\tr}{Tr}
\def\mr{\mathrm}
\def\mb{\mathbf}
\def\bs{\boldsymbol}
\def\mc{\mathcal}

\begin{document}
\title{Resonant nonlinear magneto-optical effects in atoms}
\thanks{This paper is dedicated to Professor Eugene D.
Commins on the occasion of his 70th birthday.}
\author{D. Budker}
\email{budker@socrates.berkeley.edu}
\affiliation{Department of
Physics, University of California, Berkeley, CA 94720-7300}
\affiliation{Nuclear Science Division, Lawrence Berkeley National
Laboratory, Berkeley CA 94720}
\author{W. Gawlik}
\email{gawlik@uj.edu.pl}
\affiliation{Instytut Fizyki im. M.
Smoluchowskiego, Uniwersytet Jagiello\'{n}ski, Reymonta 4, 30-059
Krakow, Poland}
\author{D. F. Kimball}
\email{dfk@uclink4.berkeley.edu}
\affiliation{Department of
Physics, University of California, Berkeley, CA 94720-7300}
\author{S. M. Rochester}
\email{simonkeys@yahoo.com}
\affiliation{Department of Physics,
University of California, Berkeley, CA 94720-7300}
\author{A. Weis}
\email{antoine.weis@unifr.ch} \affiliation{Depart\'{e}ment de
Physique, Universit\'{e} de Fribourg, Chemin du Mus\'{e}e 3,
CH-1700 Fribourg, Switzerland}
\author{V. V. Yashchuk}
\email{yashchuk@socrates.berkeley.edu} \affiliation{Department of
Physics, University of California, Berkeley, CA 94720-7300}
\date{\today}

\begin{abstract}
In this article, we review the history, current status, physical
mechanisms, experimental methods, and applications of nonlinear
magneto-optical effects in atomic vapors. We begin by describing
the pioneering work of Macaluso and Corbino over a century ago on
linear magneto-optical effects (in which the properties of the
medium do not depend on the light power) in the vicinity of atomic
resonances, and contrast these effects with various nonlinear
magneto-optical phenomena that have been studied both
theoretically and experimentally since the late 1960s. In recent
years, the field of nonlinear magneto-optics has experienced a
revival of interest that has led to a number of developments,
including the observation of ultra-narrow (1-Hz) magneto-optical
resonances, applications in sensitive magnetometry, nonlinear
magneto-optical tomography, and the possibility of a search for
parity- and time-reversal-invariance violation in atoms.
\end{abstract}
%\pacs{PACS 32.80.-t, 32.90.+a, 42.65.-k, 42.50.Gy}
%32.80.-t Photon interactions with atoms
%32.90.+a Other topics in atomic properties and interactions of atoms and ions with photons
%42.65.-k Nonlinear optics
%42.50.Gy Effects of atomic coherence on propagation, absorption, and amplification of light

\maketitle

\tableofcontents

\section{Introduction}
\label{section:Intro}

\begin{figure}
\includegraphics{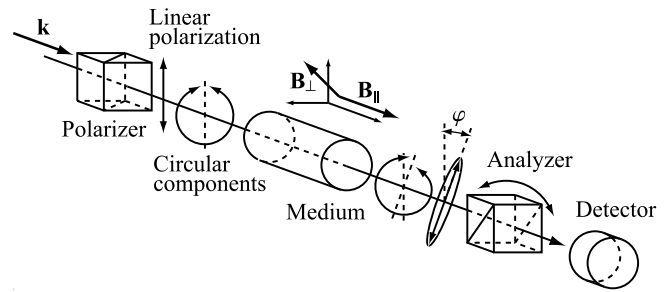}
\caption{The Faraday and Voigt effects. In the Faraday effect,
light, after passing through a linear polarizer, enters a medium
subjected to a longitudinal magnetic field $\mb{B}_\Vert\
(\mb{B}_\bot=0)$. Left- and right-circularly polarized components
of the light (equal in amplitude for linearly polarized light)
acquire different phase shifts, leading to optical rotation. A
difference in absorption between the two components induces
ellipticity in the output light. The intensity of the transmitted
light with a particular polarization, depending on the orientation
of the analyzer relative to the polarizer, is detected. Analyzer
orientation varies with the type of experiment being performed. In
forward-scattering experiments (Sec.\ \ref{subsection:FSandLC}),
the analyzer is crossed with the input polarizer, so that only
light of the orthogonal polarization is detected. In the
``balanced polarimeter'' arrangement (Sec.\
\ref{subsection:Polarim}), a polarizing beam splitter oriented at
$\pi/4$ to the input polarizer is used as an analyzer. In this
case, the normalized differential signal between the two channels
of the analyzer depends on the rotation of light polarization
while being insensitive to induced ellipticity. The Voigt effect
is similar except that instead of a longitudinal magnetic field, a
transverse field $\mb{B}_\bot\ (\mb{B}_\Vert=0)$ is applied. Here
optical rotation and induced ellipticity are due to differential
absorption and phase shifts of orthogonal linearly polarized
components of the input light (Sec.\
\ref{section:SymCons}).}\label{fig:FRE}
\end{figure}

Magneto-optical effects arise when light interacts with a medium
in the presence of a magnetic field. These effects have been
studied and used since the dawn of modern physics and have had a
profound impact on its development.\footnote{Magneto-optics were
listed among the most important topics in Physics at the World
Congress of Physics in Paris in 1900 \citep{wcp1900}.} Most
prominent among the magneto-optical effects are the
\citet{Far1846a,Far1846b,Far1855} and the \citet{Voi01} effects,
i.e., rotation of light's polarization plane as it propagates
through a medium placed in a longitudinal or transverse magnetic
field, respectively (Fig.\ \ref{fig:FRE}).
%----------------------------------------------------------------
%----------------------------------------------------------------
The linear (Secs.\ \ref{section:LinearMO},
\ref{section:linvsnonlin}) near-resonance Faraday effect is also
known as the Macaluso-Corbino
(\citeyear{Mac1898a,Mac1898b,Mac1899}) effect. The
\citeauthor{Voi01} effect is sometimes called the Cotton-Mouton
(\citeyear{Cot07,Cot11}) effect, particularly in condensed-matter
physics.

The remarkable properties of resonant (and, particularly,
nonlinear) magneto-optical systems---as compared to the well-known
transparent condensed-matter magneto-optical materials such as
glasses and liquids---can be illustrated with the Faraday effect.
The magnitude of optical rotation per unit magnetic field and unit
length is characterized by the Verdet constant $V$. For typical
dense flint glasses that are used in commercial Faraday
polarization rotators and optical isolators,
$V\simeq3\times10^{-5}\ \mr{rad\,G^{-1}\,cm^{-1}}$. In subsequent
sections, we will describe experiments in which nonlinear
magneto-optical rotation corresponding to $V\simeq10^{4}\
\mr{rad\,G^{-1}\,cm^{-1}}$ is observed in resonant rubidium vapor
(whose density, $\sim$$3\times10^9\ \mr{cm^{-3}}$, satisfies the
definition of very high vacuum). Taking into account the
difference in density between glass and the rarified atomic vapor,
the latter can be thought of as a magneto-optical material with
some $10^{20}$ greater rotation ``per atom'' than heavy flint.

In this paper, we briefly review the physics and applications of
resonant linear magneto-optical effects, and then turn to our main
focus---the nonlinear effects (i.e., effects in which optical
properties of the medium are modified by interaction with light).
We will also discuss various applications of nonlinear
magneto-optics in atomic vapors, and the relation between
nonlinear magneto-optics and a variety of other phenomena and
techniques, such as coherent population trapping,
electromagnetically induced transparency, nonlinear electro-optics
effects, and self-rotation.

\section{Linear magneto-optics}
\label{section:LinearMO}

In order to provide essential background for understanding
nonlinear magneto-optical effects (NMOE), we first review
\emph{linear} near-resonant magneto-optics of atoms and molecules.
(See Sec.\ \ref{section:linvsnonlin} for a discussion of the
difference between the linear and nonlinear magneto-optical
effects.)

\subsection{Mechanisms of the linear magneto-optical effects}
\label{subsection:LinMOEmechs}

At the conclusion of the 19th century,
\citet{Mac1898a,Mac1898b,Mac1899}, studying absorption spectra of
the alkali atoms in the presence of magnetic fields, discovered
that the Faraday effect (magneto-optical activity) in the vicinity
of resonance absorption lines has a distinct resonant character
\citep[see also work by][]{For64}.

The principal mechanism of the linear Macaluso-Corbino effect can
be illustrated by the case of an $F=1 \rightarrow F'=0$ transition
(Fig.\ \ref{fig:F1toF0}), where $F,F'$ are the total angular
momenta.{\footnote{Throughout this article, we designate as $F,F'$
total angular momenta of the lower and the upper states of the
transition, respectively. For atoms with zero nuclear spin, $F,F'$
coincide with the total electronic angular momenta $J,J'$.}
%--------------------------------------------------------------
\begin{figure}
\includegraphics[width=2in]{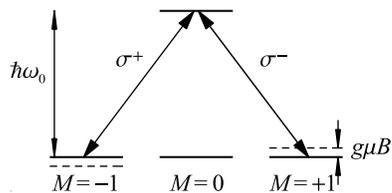}
\caption{An $F=1\rightarrow F'=0$ atomic transition. In the
presence of a longitudinal magnetic field, the Zeeman sublevels of
the ground state are shifted in energy by $g\mu BM$. This leads to
a difference in resonance frequencies for left- ($\sigma^+$) and
right- ($\sigma^-$) circularly polarized light.}\label{fig:F1toF0}
\end{figure}
%--------------------------------------------------------------
Linearly polarized light incident on the sample can be decomposed
into two counter-rotating circular components $\sigma^{\pm}$. In
the absence of a magnetic field, the $M=\pm1$ sublevels are
degenerate and the optical resonance frequencies for $\sigma^+$
and $\sigma^-$ coincide. The real part of the refractive index $n$
associated with the atomic medium is shown in Fig.\
\ref{fig:MCspectr} as a function of the light frequency detuning
$\Delta$ (the solid dispersion curve).
%--------------------------------------------------------------------
\begin{figure}
\includegraphics[width=2.75in]{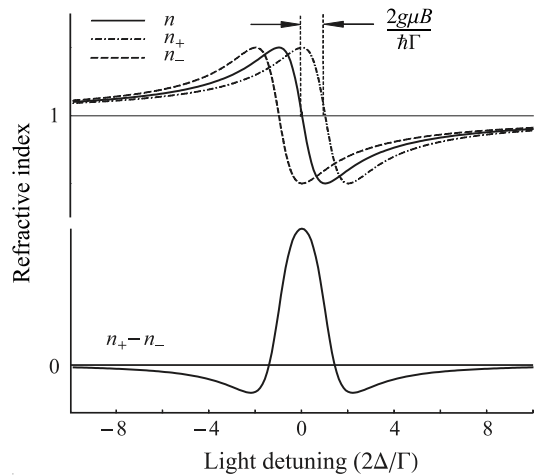}
\caption{The dependence of the refractive index on light frequency
detuning $\Delta$ in the absence $(n)$ and in the presence
$(n_\pm)$ of a magnetic field. Shown is the case of
$2g\mu{B}=\hbar\Gamma$ and a Lorentzian model for line broadening.
The lower curve shows the difference of refractive indices for the
two circular polarization components. The spectral dependence of
this difference gives the characteristic spectral shape of the
linear magneto-optical rotation (the Macaluso-Corbino
effect).}\label{fig:MCspectr}
\end{figure}
%--------------------------------------------------------------------
The refractive index is the same for the two circular components.

When a magnetic field is applied, however, the Zeeman
shifts\footnote{The connection between the Faraday and the Zeeman
effects was first established by \citet{Voi1898a}, who also
explained the observations of Macaluso and Corbino
\citep{Voi1898b}.} lead to a difference between the resonance
frequencies for the two circular polarizations. This displaces the
dispersion curves for the two polarizations as shown in Fig.\
\ref{fig:MCspectr}. A characteristic width of these dispersion
curves, $\Gamma$, corresponds to the spectral width (FWHM) of an
absorption line. Under typical experimental conditions in a vapor
cell this width is dominated by the Doppler width and is on the
order of 1 GHz for optical transitions. The difference between
$n_+$ and $n_-$ (Fig.\ \ref{fig:MCspectr}) signifies a difference
in phase velocities of the two circular components of light and,
as a result, the plane of polarization rotates through an angle
%--------------------------------------------------------------------
\begin{equation}
\label{l1} \varphi=\pi(n_+-n_-)\frac{l}{\lambda}.
\end{equation}
%--------------------------------------------------------------------
Here $l$ is the length of the sample, and $\lambda$ is the
wavelength of light. In addition to the difference in refraction
for the two circular polarizations (circular birefringence), there
also arises a difference in absorption (circular dichroism). Thus
linear light polarization before the sample generally evolves into
elliptical polarization after the sample. For nearly monochromatic
light (i.e., light with spectral width much smaller than the
transition width), and for zero frequency detuning from the
resonance, the optical rotation in the sample as a function of
magnetic field $B$ can be estimated from Eq.\ (\ref{l1})
as\footnote{Explicit formulae for $n_{\pm}$ are given, for
example, in \citet[Appendix VII]{Mit71}; analogous expressions can
also be obtained for induced ellipticity.}
%--------------------------------------------------------------------
\begin{equation}
\varphi\simeq\frac{2g\mu{B}/\hbar\Gamma}{1+\prn{2g\mu{B}/\hbar\Gamma}^2}\frac{l}{l_0}.
\label{Eqn:phi_vs_B}
\end{equation}
%--------------------------------------------------------------------
Here $g$ is the Land\'{e} factor, $\mu$ is the Bohr magneton, and
$l_0$ is the absorption length. This estimate uses for the
amplitude of each dispersion curve (Fig.\ \ref{fig:MCspectr}) the
resonance value of the imaginary part of the refractive index
(responsible for absorption). The Lorentzian model for line
broadening is assumed. The Voigt model \citep[discussed by, for
example,][]{Dem96}, which most accurately describes a Doppler- and
pressure-broadened line, and the Gaussian model both lead to
qualitatively similar results. The dependence of the optical
rotation on the magnitude of the magnetic field [Eq.\
(\ref{Eqn:phi_vs_B})] has a characteristic dispersion-like shape:
$\varphi$ is linear with $B$ at small values of the field, peaks
at $2g\mu B\simeq\hbar\Gamma$, and falls off in the limit of large
fields.

For atoms with nonzero nuclear spin, mixing of different hyperfine
components (states of the same $M$ but different $F$) by a
magnetic field also leads to linear magneto-optical effects
[\citet{Nov77}; \citet{Rob80}; \citet{Khr91}; \citet{Pap94}]. The
contribution of this mechanism is comparable to that of the
level-shift effect discussed above in many practical situations,
e.g., linear magneto-optical rotation in the vicinity of the
alkali $D$-lines \citep{Che87}. For the Faraday geometry and when
$g\mu B\ll\hbar\Gamma\ll\Delta_\text{hfs}$, the amplitude of the
rotation can be estimated as
%--------------------------------------------------------------------
\begin{equation}
\varphi\simeq\frac{g\mu B}{\Delta_\text{hfs}}\frac{l}{l_0},
\label{l3}
\end{equation}
%--------------------------------------------------------------------
where $\Delta_\text{hfs}$ is the separation between hyperfine
levels. Since hyperfine mixing leads to a difference in the
magnitude of $n_+$ and $n_-$ (and not the difference in resonance
frequencies as in the level-shift effect), the spectral profile of
the rotation for the hyperfine-mixing effect corresponds to
dispersion-shaped curves centered on the hyperfine components of
the transition.

There exists yet another mechanism in linear magneto-optics called
the \emph{paramagnetic} effect. The populations of the
ground-state Zeeman sublevels that are split by a magnetic field
are generally different according to the Boltzmann distribution.
This leads to a difference in refractive indices for the
corresponding light polarization components. For gaseous media,
this effect is usually relatively small compared to the other
mechanisms. However, it can be dramatically enhanced by creating a
nonequilibrium population distribution between Zeeman sublevels.
This can be accomplished by \emph{optical pumping}, a nonlinear
effect that will be discussed in detail in Sec.\
\ref{subsection:OpticalPumping}.

\subsection{Forward scattering and line-crossing}
\label{subsection:FSandLC}

Magneto-optics plays an important role in the study of resonant
light scattering in the direction of the incident light
(\emph{forward scattering}, FS) whose detection is normally
hindered experimentally by the presence of a strong incident light
beam that has identical properties (frequency, polarization,
direction of propagation) as the forward-scattered light. In order
to investigate this effect, \citet*{Cor66} used two crossed
polarizers (Fig.\ \ref{fig:FRE}). In this arrangement, both the
direct unscattered beam and the scattered light of unchanged
polarization are blocked by the analyzer. Only the light that
undergoes some polarization change during scattering is detected.
If the medium is isotropic and homogeneous and there is no
additional external perturbation or magnetic field, the
forward-scattered light has the same polarization as the primary
light and cannot be detected. The situation changes, however, when
an external magnetic field $\mb{B}$ is applied. Such a field
breaks the symmetry of the $\sigma^+$ and $\sigma^-$ components of
the propagating light in the case of $\mb{B}\parallel\mb{k}$ (and
$\pi$ and $\sigma$ components when $\mb{B}\perp\mb{k}$) and
results in a nonzero component of light with opposite polarization
that is transmitted by the analyzer.

In the late 1950s, it was determined \citep{Col59,Fran61} that
coherence between atomic sublevels (represented by off-diagonal
elements of the density matrix, see Sec.\ \ref{subsection:DMCalc})
affects lateral light scattering. For example, in the
\citet{Han24} or level-crossing effects \citep{Col59,Fran61}, the
polarization or spatial distribution of fluorescent light changes
as a function of relative energies of coherently excited atomic
states. Another example is that of \emph{resonance narrowing} in
lateral scattering [\citet{Gui57}; \citet{Bar59}]. In this
striking phenomenon the width of resonance features (observed in
the dependence of the intensity of scattered light on an applied
dc magnetic field or the frequency of an rf field) was seen to
decrease with the increase of the density of the sample; this
appeared as an effective increase in the upper state lifetime
despite collisional broadening that usually results from elevated
density. This effect is due to multiple light scattering
(radiation trapping) which transfers excitation from atom to atom,
each of the atoms experiencing identical evolution in the magnetic
field \citep[a more detailed discussion is given by, for
example,][]{Cor88}.

In lateral scattering, the resonance features of interest are
usually signatures of interference between various sublevels in
each individual atom. In forward scattering, the amplitudes of
individual scatterers add in the scattered light
\citep{Cor66,Dur72}. Thus forward scattering is coherent, and
interference can be observed between sublevels belonging to
\emph{different} atoms. The forward-scattered light has the same
frequency as the incident light. However, the phase of the
scattered light depends on the relative detuning between the
incident light and the atom's effective resonance frequency. This
leads to inhomogeneous broadening of the FS resonance features; at
low optical density, their width (in the magnetic field domain) is
determined by the Doppler width of the spectral
line.\footnote{Obviously, homogeneous broadening (e.g., pressure
broadening) also affects these widths.} For this reason, the
corresponding FS signals associated with the linear
magneto-optical effects can be regarded as Doppler-broadened,
multi-atom Hanle resonances. Here the Hanle effect is regarded
either as a manifestation of quantum-mechanical interference or
atomic coherence, depending on whether it appears in emissive or
dispersive properties of the medium (see Sec.\
\ref{subsubsection:Hanle_LC}). If the optical density of the
sample increases to the extent that multiple scattering becomes
important, substantial narrowing of the observed signals results,
interpreted by \citet{Cor66} as coherence narrowing.
Forward-scattering signal narrowing was observed in Hg by
\citet{Cor66} and in Na by \citet{Kro72}. Further exploring the
relation between manifestations of single- and multi-atom
coherence, \citeauthor{Cor66} analyzed the phenomenon of
\emph{double resonance} in the context of FS. This is a two-step
process in which first optical excitation by appropriately
polarized resonant light populates atomic states. Subsequently,
transitions are induced among the excited states by a resonant
radio-frequency (rf) field. \citeauthor{Cor66} also studied double
resonance in the limiting case in which the upper states have the
same energy, so that the second resonance occurs at zero
frequency. Such a zero-frequency resonant field is simply a
constant transverse magnetic field. Thus the double-resonance
approach is applied to the interpretation of the Voigt effect in
an unorthodox manner.

The fact that in forward scattering light scattered by different
atoms is coherent makes it possible to study the phenomenon of
\emph{line crossing}. Whereas in level crossing
\citep{Col59,Fran61}, signals in lateral light scattering are
observed when different sublevels of single atoms cross (for
example, in an external magnetic field), in a line-crossing
experiment interference occurs due to crossing of sublevels of
different atoms. This effect was first demonstrated by
\citet{Hac70}. These authors observed interference in the FS
signals due to crossing of Zeeman sublevels of different Hg
isotopes contained in one cell. \citet{Chu73} showed that line
crossing occurs even when atoms of different kinds are contained
in separate cells. \citet{Hac70}, \citet{Chu73}, \citet{Sta74},
and \citet{Sie76} investigated the possibility of applying the
line-crossing effect to precise measurements of isotope shifts.
This idea is based on the fact that line crossings occur when the
applied magnetic field is such that the Zeeman shifts compensate
for the initial field-free isotope shifts. It was hoped that
strong coherence narrowing of the line-crossing resonance would
significantly increase precision of such measurements. However,
these authors found that various complications (arising, for
example, from pressure broadening of the signals), render this
method impractical for isotope shift measurements.

Forward-scattering signals for weak-intensity light can be written
as
%--------------------------------------------------------------------
\begin{multline}
I_{FS}=\frac{1}{4}\int\xi(\omega)\sbrk{e^{-\frac{A_+\omega l}{c}}-e^{-\frac{A_-\omega l}{c}}}^2d\omega\\
+\int\xi(\omega)\sin^2\cbrk{(n_+-n_-)\frac{\omega{l}}{2c}}e^{-\frac{(A_++A_-)\omega{l}}{c}}d\omega,\label{l4}
\end{multline}
%--------------------------------------------------------------------
where $\xi(\omega)$ is the spectral density of the incident light,
$A_{\pm}$ and $n_{\pm}$ denote amplitude absorption coefficients
and refractive indices for the $\sigma^{\pm}$ components of the
incident light beam, respectively, and $l$ is the sample length.
We assume ideal polarizers here.

In general, the two terms in Eq.\ (\ref{l4}) give comparable
contributions to the FS signal. The first term is due to
differential absorption of the $\sigma^+$ and $\sigma^-$
components of the incident light (circular dichroism), and the
second term is due to differential dispersion (circular
birefringence). The two contributions have different frequency
dependence. This can be illustrated with the simple case of the
$F=0 \rightarrow F'=1$ transition, for which the $\sigma^{\pm}$
resonance frequencies are split by a longitudinal magnetic field
(Fig.\ \ref{fig:FSspectr}).
%-----------------------------------------------------------------
\begin{figure}
\includegraphics{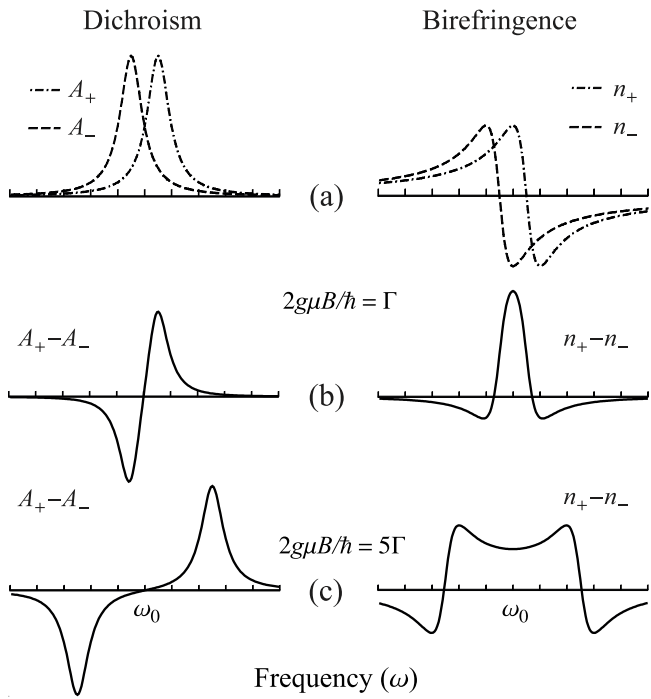}
\caption{Spectral dependences of the circular dichroic ($A_+-A_-$)
and birefringent ($n_+-n_-$) anisotropies that determine the
forward-scattering signal for a $F=0\rightarrow{F'=1}$ transition.
The $\sigma^+$ and $\sigma^-$ resonance frequencies are split by a
longitudinal magnetic field $B$. For curves (a) and (b) the
magnitude of the splitting is equal to the resonance width
$\Gamma$. For the curves (c), it is five times
larger.}\label{fig:FSspectr}
\end{figure}
%-----------------------------------------------------------------
While the function in the first integral goes through zero at
$\omega=\omega_0$ (since the function in the square brackets is
anti-symmetric with respect to detuning), the second
(birefringence) term is maximal at zero detuning (for small
magnetic fields). For a narrow-band light source, it is possible
to eliminate the dichroic contribution by tuning to the center of
the resonance. One is then left with only the second integral,
representing Malus's law, where $\varphi=(n_+-n_-)\omega
l/\prn{2c}$ is the Faraday rotation angle [Eq.\ (\ref{l1})].

When the density-length product for the medium is sufficiently
high, the range of variation of $\varphi$ can easily exceed $\pi$,
and intensity transmitted through the apparatus shown in Fig.\
\ref{fig:FRE} oscillates as a function of the magnetic field
(Fig.\ \ref{fig:FSsignals}).
%--------------------------------------------------------------------
\begin{figure}
\includegraphics{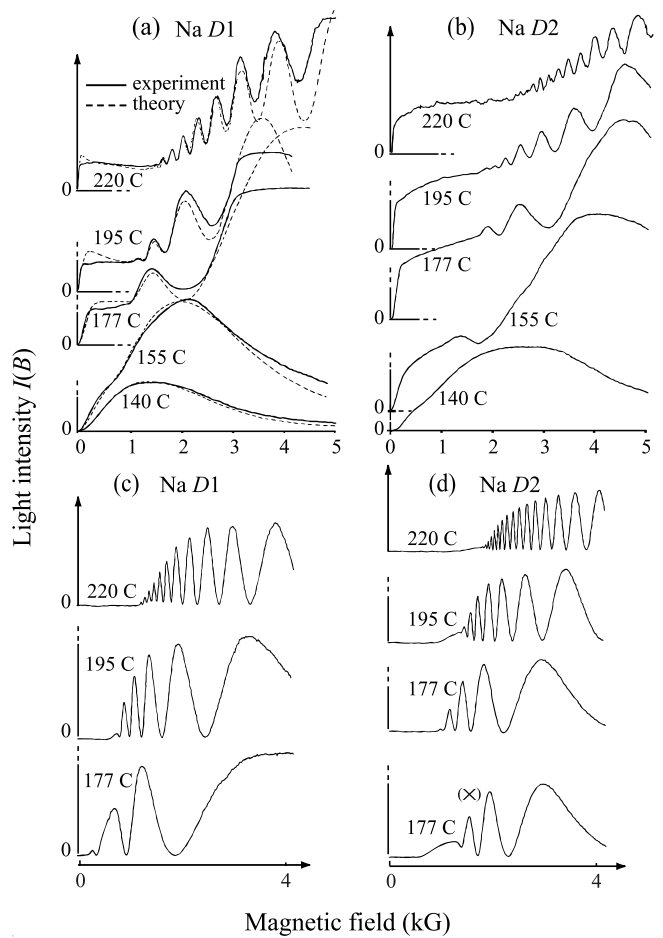}
\caption{Forward-scattering signals $I(B)$ observed on the sodium
$D1$ [(a),(c)] and $D2$ [(b),(d)] lines by \citet{Gaw75b}. Curves
(a) and (b) are signals obtained with a conventional spectral lamp
(single $D$ line selected by a Lyot filter). Curves (c) and (d)
are signals obtained with single-mode cw dye-laser excitation. The
laser is tuned to the atomic transition in all cases except the
one marked ($\times$). Curve ($\times$) was recorded with the
laser detuned by about 600 MHz to demonstrate the influence of
residual dichroism. In plot (a), flattening of some curves at high
fields is due to detector saturation---the dashed lines represent
calculated signals.}\label{fig:FSsignals}
\end{figure}
%--------------------------------------------------------------------
These oscillations are clearly seen despite the proximity of the
absorption line because the refractive indices drop with detuning
slower than the absorption coefficients. When $\xi(\omega)$
represents a narrow spectral profile, the modulation contrast can
be high, particularly when the magnetic field is strong enough to
split the $A_{\pm}$ profiles completely. Such case is shown
schematically in Fig.\ \ref{fig:FSspectr}(c).

At the center of a Zeeman-split resonance, absorption drops off
with the magnitude of the splitting much faster than optical
rotation does. Thus, using a dense atomic vapor in a magnetic
field and a pair of polarizers, it is possible to construct a
transmission filter for resonant radiation, which can be turned
into an intensity modulator by modulating the magnetic field
[\citet{For64}; \citet{Ale65}].

\subsection{Applications in spectroscopy}

Magneto-optical effects can provide useful spectroscopic
information. For example, given a known atomic density and sample
length, a measurement of Faraday rotation for a given transition
can be used to determine the oscillator strength $f$ for that
transition. Conversely, when $f$ is known, Faraday rotation may be
used to determine vapor density. \citet{Vli2001} found that
Faraday rotation measurements with high alkali metal densities
($\sim$$10^{15}$--$\,10^{16}\ \mr{cm}^{-3}$) are free from certain
systematic effects associated with measurement of absorption. An
earlier review of applications of magneto-optical rotation was
given by \citet{Ste89}.

\subsubsection{Analytical spectroscopy and trace analysis, investigation of weak transitions}

An example of an application of magneto-optical rotation to
molecular spectroscopy is the work of \citet{Aub66} who, using a
multi-pass cell, demonstrated that the magneto-optical rotation
spectrum of NO is easier to interpret than the absorption
spectrum. Molecular magneto-optical spectra are in general much
simpler than the absorption spectra \citep[see discussion by][Ch.\
V,5]{Her89}. This is because significant magneto-optical effects
are only present for transitions between molecular states of which
at least one has nonzero electronic angular momentum. In addition,
since molecular $g$-values decrease rapidly with the increase of
the rotational quantum number $J$ \citep[see discussion by][Ch.\
7.2]{Khr91}, only a small part of the rotational band produces
magneto-optical effects. Magneto-optical rotation has been used to
identify atomic resonance lines in Bi against a complex background
of molecular transitions \citep{Bar80,Rob80}.

Magneto-optical rotation, in particular, the concept of forward
scattering, was also applied in analytical spectroscopy for trace
element detection. This was first done by \citet{Chu74}, who,
using FS signals, showed sensitivity an order of magnitude higher
than could be obtained with absorption measurements. The improved
sensitivity of magneto-optical rotation compared to absorption is
due to almost complete elimination of background light and a
corresponding reduction of its influence on the signal noise. Such
noise is the main limitation of the absorption techniques.
Following this work, \citet{Ito77} also employed both Faraday and
Voigt effects for trace analysis of various elements.

The detection of weak transitions by magneto-optical rotation
(with applications to molecular and analytical spectroscopy) was
spectacularly advanced by employment of lasers. Using tunable
color-center lasers, \citet{Lit80} demonstrated 50 times better
sensitivity in detection of NO transitions in the vicinity of
$2.7\ \mr{\mu m}$ with magneto-optical rotation compared to
absorption spectroscopy. Similar results were obtained by
\citet{Yam86} who obtained 200-fold enhancement in sensitivity
over absorption spectroscopy in their work with a pulsed dye laser
and the sodium $D2$ line. Another interesting result was reported
by \citet{Hin82} who worked in the mid-infrared range with NO
molecules and a CO laser. These authors also demonstrated that
magneto-optical rotation improves sensitivity in either of the
basic configurations, i.e., in the Faraday as well as the Voigt
geometry.

The advent of high-sensitivity laser spectropolarimeters, allowing
measurement of optical rotation at the level of $10^{-8}$ rad and
smaller (Sec.\ \ref{subsection:Polarim}), made possible the
sensitive detection of species with low concentration. Detection
of on the order of hundreds of particles per cubic centimeter was
reported by \citet{Vas78}. It is important to note that, while it
is beneficial from the point of view of the photon shot noise of
the polarimeter to operate at a high light power, great care
should be taken to make sure that the atoms of interest are not
bleached by nonlinear saturation effects (Sec.\
\ref{subsection:satpar}). As a practical way to optimize a trace
analysis setup, we suggest the use of a buffer gas to pressure
broaden the homogenous width of the transition up to the point
when this width becomes comparable to the Doppler width. This way,
the linear absorption and Faraday rotation are not compromised,
but the light intensity constraints due to nonlinearities are
relaxed by many orders of magnitude.

Due to its high sensitivity, the magneto-optical rotation method
can also be applied to the study of weak transitions, such as
magnetic dipole transitions with small transition magnetic moments
\citep{Bar89a}.

\subsubsection{Measurement of oscillator strengths}
\label{subsection:f}

\citet{For64} performed some of the earliest work in which
resonant magneto-optical rotation was used to measure oscillator
strengths.\footnote{Early measurements of oscillator strength are
described in a monograph by \citet{Mit71}.} They measured light
dispersion in Hg vapor at about five Doppler widths from the
center of the 253.7-nm line. They used an electrodeless discharge
$^{198}$Hg lamp placed in a solenoid as a light source tunable
over eight Doppler widths and measured the dispersion using a
Mach-Zehnder interferometer. In addition to the observation of
Faraday rotation in excess of 7 rad, they also demonstrated the
inversion of the sign of the dispersion in a vapor with inverted
population, and the feasibility of using Faraday rotation for
narrow-band modulatable optical filters.

The advent of tunable lasers has enabled significant improvement
of magneto-optical-rotation methods for measuring absolute and
relative oscillator strengths. This is illustrated by the results,
shown in Fig.\ \ref{fig:FSsignals}, of an experiment performed by
\citet{Gaw75b}. Here curves (a) and (b) refer to a FS experiment
performed with a sodium spectral lamp and a Lyot filter that
selected one of the Na $D$ lines. The light from the lamp had an
asymmetric spectral profile with some \emph{self-reversal} (a
line-profile perturbation due to re-absorption of resonance light
at the line center) and FWHM of about 8 GHz. From the results for
the $D1$ and $D2$ lines, one can see how the modulation period
depends on the product of atomic density and oscillator strength
of the investigated transitions and how the oscillatory
birefringent contribution (see also Fig.\ \ref{fig:FS-FR}) becomes
overwhelmed by the structureless dichroic one, particularly at
small magnetic fields.
%--------------------------------------------------------------------
\begin{figure}
\includegraphics{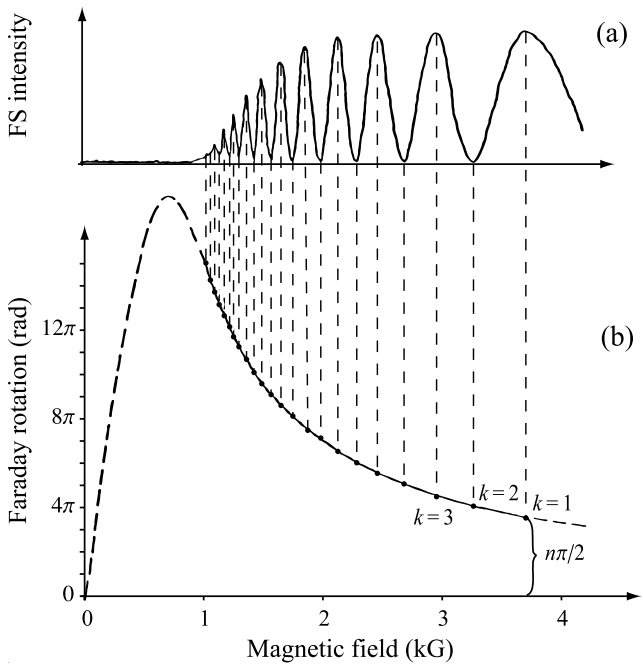}
\caption{(a) Signals obtained by \citet{Gaw79} from forward
scattering of single-mode cw laser light tuned to the Na $D1$
line. (b) Magnetic-field dependence of the Faraday rotation angle
$\varphi$ obtained from the extrema of the $I(B)$ curve of figure
(a).}\label{fig:FS-FR}
\end{figure}
%--------------------------------------------------------------------
Figures \ref{fig:FSsignals} (c) and (d) represent results from the
same experiment but with, in place of a lamp, a narrow-band cw dye
laser tuned to the centers of gravity of the Na $D1$ and $D2$
lines, respectively [see also \citet{Gaw77,Gaw79}]. Several
important changes are seen: first, the modulation depth is now
100\%---this allows accurate determination of
$\varphi(nl,\omega,B)$, where $\omega$ is the light frequency
(Fig.\ \ref{fig:FS-FR}); second, the dichroic contribution can be
fully eliminated by appropriately tuning the laser (this is
because the dichroic effect has a nearly dispersive spectral
dependence which goes through zero when the Faraday rotation
contribution is nearly maximal); third, as seen by the envelope of
the oscillatory pattern, total absorption
$\exp\sbrk{-\prn{A_++A_-}\omega{l}/c}$ plays a role only for small
$B$, in agreement with the above considerations [see Figs.
\ref{fig:FSsignals}(c) and (d)]. Fast oscillation in the region of
low total absorption makes it possible to determine the absorption
profile (envelope) simultaneously with the measurement of
$\varphi$. This technique allows one to simultaneously determine
the real and imaginary parts of the complex index of refraction,
which is useful for measuring collisional parameters, for example.

A number of experiments have measured oscillator strengths by
scanning the light frequency (at a fixed magnetic field). For
example, employing synchrotron radiation and superconducting
magnets, oscillator strengths of ultraviolet lines of Rydberg
series of several elements were determined by \citet{Gar83},
\citet{Con83}, and \citet{Ahm96}. An interesting experimental
method, employing pulsed magnets and UV lasers, was developed by
Connerade and coworkers for studies of Faraday rotation in
autoionizing resonances and transitions to Rydberg states
\citep{Con88,Con92} and of collisional broadening \citep{War94}.

It should be remembered that most methods based on light
dispersion measure not just the oscillator strength $f$ but the
product of $f$ and the optical density.\footnote{One interesting
exception is the early work of \citet{Wein31} who devised a method
of measuring the Lorentzian line width independently of the
optical density. The method relies on using a white-light source
and recording the intensity of light transmitted through an
optically dense medium placed between two crossed polarizers in a
longitudinal magnetic field $B$. When $B\neq0$, the spectrum of
the transmitted light has the form of a bright line on a dark
background (the Macaluso-Corbino effect), but when the polarizers
are uncrossed by appropriate angles, the line merges with the
background. For a given atomic transition and magnetic field, the
values of the uncrossing angles allow determination of the
Lorentzian width without prior knowledge of density.} Knowledge of
the optical density is hence a central issue in absolute
measurements. On the other hand, these methods can be very useful
for high-precision relative measurements.

\subsubsection{Investigation of interatomic collisions}
\label{subsection:collisions}

The effect of pressure broadening on the magneto-optical rotation
spectra of resonant gaseous media was theoretically analyzed
within the impact approximation by \citet{Gir75}. Later,
collisional effects in magneto-optical rotation in a
$F=0\rightarrow F=1$ atomic transition were treated with the
density-matrix formalism by \citet{Sch87}. This work considered
atomic motion and light fields of arbitrary intensity (and thus
included nonlinear effects), but was limited to optically thin
samples. The case of high optical density appears to be
particularly difficult. This is because for high atomic densities,
the impact approximation is not appropriate---the effect of
quasi-molecular behavior of the perturbed atoms becomes
significant. Consequently, optical transitions contributing to the
rotation spectra can no longer be regarded as isolated. Faraday
rotation produced under such circumstances was studied in an
experiment of \citet{Kri94} with the $D2$ line of Rb atoms that
were highly diluted in Xe buffer gas at densities of
$\sim$$10^{20}\ \text{cm}^{-3}$. It was found necessary to include
the effect of Born-Oppenheimer coupling of Rb-Xe
long-range-collision pairs in order to explain the magnitude of
the observed rotation. Born-Oppenheimer coupling is the locking of
the total electronic angular momentum to the collision axis at
densities at which the mean free path falls below about 1 nm. Such
angular-momentum locking hinders the Larmor precession, and hence
reduces Faraday rotation.

Faraday rotation spectroscopy is now widely used for measuring
collisional broadening and shift of spectral lines. For example,
\citet{Bog86}, \citet{Bog87}, and \citet{Bog88} used this method
to study broadening and shift of the 648-nm magnetic-dipole
transition in Bi and self-broadening, foreign-gas-broadening and
shift, and electron-impact broadening of transitions between
excited states in Cs. \citet{Bar88a,Bar89a} studied buffer-gas
broadening of transitions in atomic Sm. One of the important
advantages of this method is that it provides the ability to
modulate the signal by changing the magnetic field, thus
distinguishing the signal from the background.

For measuring line broadening, e.g., collisional broadening, it is
important that the characteristic spectral profile of the
Macaluso-Corbino effect for an isolated line has two zero
crossings (Fig.\ \ref{fig:MCspectr}). The separation between zeros
is linearly dependent on the Lorentzian width of the transition
even when this width is much smaller than the Doppler width of the
transition.

An experimental study of collisional broadening of Rydberg spectra
with the use of Faraday rotation was performed by \citet{War94}.
This work demonstrated that the resonant Faraday effect and the FS
technique can be used for studies of collisional perturbations of
Rydberg states with much greater precision than that of earlier
photoabsorption experiments.

\subsubsection{Gas lasers} Magneto-optical rotation of a weak probe
light passing through a gas discharge is a useful tool for
studying amplifying media in gas lasers. Early examples of this
are the papers by \citet{Ale72} who observed large rotations in a
$^{136}$Xe discharge and by \citet{Men73} who determined
population and polarization relaxation rates of the Ne levels
making up the 3.39-$\mu$m laser transition. The technique is
applicable to transitions which exhibit light absorption as well
as amplification depending on particular population ratios. Since
the Faraday angle is proportional to the population difference for
the two levels of a transition and changes sign when population
inversion is achieved, such studies allow precise determination of
population differences and their dependences on parameters such as
gas pressure and discharge current. An example is the work of
\citet{Win77} who studied the 632.8-nm Ne line. The method's
sensitivity makes it useful for optimizing the efficiency of gas
lasers, particularly those utilizing weak transitions.

\subsubsection{Line identification in complex spectra and search for ``new'' energy levels}
\label{subsubsection:LineIdent}

We have already mentioned the utility of magneto-optical rotation
for identification of molecular spectra and for distinguishing
atomic resonances from molecular ones. Another, more unusual,
application of Faraday-rotation spectroscopy was implemented by
Barkov \emph{et al.} (\citeyear{Bar87}, \citeyear{Bar88a},
\citeyear{Bar89a}). They were looking for optical magnetic-dipole
transitions from the ground term of atomic samarium to the levels
of the first excited term of the same configuration whose energies
were not known. A study of the spectrum revealed hundreds of
transitions that could not be identified with the known energy
levels (most of these were due to transitions originating from
thermally populated high-lying levels). Drawing an analogy with
infrared transitions between the ground-term levels for which
unusually small pressure broadening had been
observed\footnote{This was attributed to shielding of the valence
$4f$-electrons by other atomic shells. Note that the same
mechanism is responsible for appearance of ultra-narrow lines in
the optical spectra of rare-earth-doped crystals \citep[for
example,][]{Thi2001}.} by \citet{Ved86,Ved87}, \citeauthor{Bar87}\
predicted small pressure broadening for the sought-after
transitions as well. They further noticed that in the limit of
$\gamma\gg\Gamma_D$ (where $\gamma$ and $\Gamma_D$ are the
Lorentzian and Doppler widths, respectively), the peak Faraday
rotation is $\propto\gamma^{-2}$ (in contrast to absorption, for
which the amplitude is $\propto\gamma^{-1})$. In fact, at high
buffer-gas pressures (up to 20 atm), the magneto-optical rotation
spectra showed that almost all transitions in the spectrum were so
broadened and reduced in amplitude that they were practically
unobservable. The spectrum consisted of only the sought-after
transitions, which had unusually small pressure broadening.

\subsubsection{Applications in parity violation experiments}
\label{subsubsection:PNC}

Magneto-optical effects have played a crucial role in experiments
measuring weak-interaction induced parity-violating optical
activity of atomic vapors.\footnote{See, for example, a review of
early experiments by \citet{Khr91}, and more recent reviews by
\citet{Bou97} and \citet{BudPNCRev}.} Parity-violating optical
activity is usually observed in the absence of external static
fields near magnetic-dipole (M1) transitions with transition
amplitudes on the order of a Bohr magneton. The magnitude of the
rotation is typically $10^{-7}$ rad per absorption length for the
heavy atoms (Bi, Tl, Pb) in which the effect has been observed.

The Macaluso-Corbino effect allows calibration of the optical
rotation apparatus by applying a magnetic field to the vapor. On
the other hand, because the parity-violating optical rotation
angles are so small, great care must be taken to eliminate
spurious magnetic fields, since sub-milligauss magnetic fields
produce Macaluso-Corbino rotation comparable in magnitude to the
parity-violating rotation, albeit with a different lineshape. The
lack of proper control over spurious magnetic fields apparently
was a problem with some of the early measurements of
parity-violating optical activity in the late 1970s whose results
were inconsistent with subsequent more accurate
measurements.\footnote{The most accurate optical rotation
measurements to date \citep{Vet95,Mee95} have reached the level of
uncertainty of $\sim$1\% of the magnitude of the effect.}

The origin of the parity-violating optical rotation lies in the
mixing, due to the weak interaction, of atomic states of opposite
parity,\footnote{Parity-violating interactions mix states of the
same total angular momentum and projection. As a consequence of
time-reversal invariance, the mixing coefficient is purely
imaginary, which is the origin of the chirality that leads to
optical rotation.} which leads to an admixture of an
electric-dipole ($E1$) component to the dominant $M1$ amplitude of
the transition. The observed effect is the result of the
interference of the two components of the amplitude. In general,
there is also an electric-quadrupole ($E2$) component of the
transition amplitude. However, there is no net contribution from
$E1$--$E2$ interference to the parity-violating optical rotation,
since the effect averages to zero for an unpolarized atomic
sample. Nevertheless, knowledge of the $E2$ contribution is
crucial for accurate determination of the parity-violating
amplitude, as it affects the sample absorption and the
Macaluso-Corbino rotation used for calibration. In addition, this
contribution is important for interpretation of the
parity-violation experiments, since the $E2/M1$ ratio can serve as
an independent check of atomic theory. Fortunately, investigation
of Macaluso-Corbino lineshapes allows direct determination of the
$E2/M1$ ratio \citep{Rob80,Tre86,Bog86,Maj99}. These measurements
exploit the difference in the selection rules for these two types
of transitions, and more generally, the difference between line
strengths for different hyperfine components of a transition.

\subsubsection{Investigations with synchrotron radiation sources}
\label{subsubsection:SynchrRad}

Synchrotron radiation has been used for a number of years as a
source of ultraviolet radiation to study magneto-optical effects
in transitions to high-lying Rydberg and autoionization states [as
in work by, for example, \citet{Gar83}; \citet{Con83,Ahm96}].

In recent years, with the advent of synchrotron sources of intense
radiation in the extreme ultra-violet ($\hbar\omega=30-250$ eV),
and soft x-ray ($\hbar\omega=250$ eV to several keV) ranges
\citep{Attwood}, the studies and applications of resonant
magneto-optical effects have been extended to atomic transitions
involving excitation of core electrons.

Linear polarizers in this photon energy range are based on
polarization-selective reflection at near-Brewster angles from
multi-layer coatings \citep[see discussion by][Ch.\
4.5.5]{Attwood}, while additional polarization control (for
example, conversion of linear polarization to circular) can be
achieved using magneto-optical effects themselves
\citep{Kor97,Kor99}.

Resonant magneto-optical effects, due to their element-specific
character \citep{Kor95} are proving to be a useful tool for the
study of heterogeneous magnetic materials \citep{Kor00,Hel00}.

\subsection{Related phenomena}
\label{section:linRelPhenom}

\subsubsection{Magnetic depolarization of fluorescence: Hanle-effect and level-crossing}
\label{subsubsection:Hanle_LC}

Apart from magneto-optical rotation, much attention has been
devoted to studies of polarization properties of resonance
fluorescence and its magnetic-field dependence. In particular,
\citet{Woo22} and \citet{Woo23,Woo24} noted strong magnetic-field
dependence of polarization of the 253.7-nm resonance fluorescence
of Hg. These studies were continued and interpreted by
\citet{Han24}.\footnote{An English translation of the Hanle's
paper as well as interesting historical remarks can be found in a
book edited by \citet{Mor91}. \citet{Han26} also observed the
influence of an electric field on the polarization of resonance
fluorescence.} Hanle interpreted his observations in terms of the
induced dipole moments precessing in an external magnetic field. A
weak field causes slow Larmor precession, so the dipoles have no
time to change their orientation before they spontaneously decay.
Consequently, re-emitted fluorescence preserves the polarization
of the incident excitation light. On the other hand, in a high
field, fast precession causes rapid averaging of the dipoles'
orientation, i.e., there is efficient depolarization of the
re-emitted light. The degree of polarization depends on the rates
of spontaneous decay of the induced dipoles and Larmor precession.
Hence, the product of the excited-state lifetime $\tau$ and the
Land\'{e} factor $g$ can be determined by measuring the
magnetic-field dependence of the degree of polarization.

Figure \ref{fig:HanleA}(a) shows the energy-level scheme of the Hg
$253.7$-nm transition between the ground state with $F=0$ and the
excited state with $F'=1$ (and Zeeman components $M'=0,\pm1$).
%------------------------------------------------------------------
\begin{figure}
\includegraphics[width=3in]{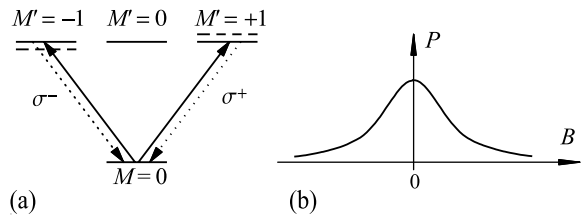}
\caption{(a) An ${F=0}\rightarrow{F'=1}$ atomic transition. In the
presence of a longitudinal magnetic field, the Zeeman sublevels of
the exited state become nondegenerate. This leads to a
magnetic-field-dependent phase difference for $\sigma^+$ and
$\sigma^-$ circular-polarization components of the re-emitted
light and to the alteration of the net re-emitted light
polarization. (b) Dependence of the re-emitted light degree of
polarization $P$ on the applied longitudinal magnetic field
$B$.}\label{fig:HanleA}
\end{figure}
%------------------------------------------------------------------
Light linearly polarized along a direction perpendicular to the
quantization axis (solid arrows) excites the $M'=\pm1$ sublevels;
spontaneous emission of $\sigma^+$ and $\sigma^-$ light (dashed
arrows) follows. When $B=0$ and the excited state is degenerate,
the relative phases of the $\sigma^+$ and $\sigma^-$ components
are the same and the re-emitted light has a high degree of
polarization $P$ [Fig.\ \ref{fig:HanleA}(b)]. When $B\ne0$, the
$M'=\pm1$ sublevels are split and the $\sigma^{\pm}$ components
acquire a $B$-dependent phase difference which alters the net
polarization of the re-emitted light and reduces $P$. The width of
the $P(B)$ dependence is inversely proportional to the product
$\tau g$.

The interpretation of the Hanle effect in terms of a classical
Lorentz oscillator was far from satisfactory as it did not
explain, for example, why $D2$ lines in atoms with one electron
above closed shells exhibited the Hanle effect, while $D1$ lines
did not. (The modern explanation is that the upper state of $D1$
lines has $J=1/2$ and thus, neglecting hyperfine interactions,
polarization moments higher than orientation do not occur, while
$D2$ lines have $J=3/2$ in the upper states and alignment is
possible; see Appendix \ref{Appendix:Multipoles} for a discussion
of atomic polarization moments.) This and other difficulties
inspired many eminent theoreticians to analyze the Hanle
experiment. \citet{Boh24} discussed the role of quantum state
degeneracy and \citet{Hei25} formulated the principle of
spectroscopic stability. Other important contributions were made
by \citet{Opp27}, \citet{Wei31}, \citet{Kor32}, and
\citet{Bre32,Bre33}.

The Hanle effect can be viewed as a generic example of quantum
interference. In the basis in which the quantization axis is along
the magnetic field, the two circular components of the incident
light coherently excite atoms to different Zeeman sublevels of the
upper level. Due to the Zeeman splitting, light spontaneously
emitted from different sublevels acquires different phase shifts,
leading to magnetic-field-dependent interference effects in the
polarization of the emitted light. The time-dependence of the
fluorescence polarization is referred to as quantum beats
\citep{Har76,Dod78,Ale93}.

The Hanle method was also extended to degeneracies occurring at
$B\ne0$, which is the basis of \emph{level-crossing spectroscopy}
\citep{Col59}. Level-crossing spectroscopy has been widely used
for determination of fine- and hyperfine-structure intervals,
lifetimes, $g$-factors, and electric polarizabilities of atomic
levels \citep{Mor91,Ale93}. An advantage of the level-crossing
method is that its spectral resolution is not limited by Doppler
broadening because it measures the frequency difference between
two optical transitions for each atom.

\subsubsection{Magnetic deflection of light}
\label{subsubsection:MagDefl}

A uniaxial crystal generally splits a light beam into two distinct
light beams with orthogonal linear polarizations. This effect
(linear birefringence) vanishes when the $\mb{k}$-vector of the
incident beam is either parallel or perpendicular to the crystal
axis.

If, instead of a preferred axis, an axial vector $\mb{a}$ (also
called a gyrotropic axis) determines the symmetry of the medium,
another type of birefringence occurs. Here, a light beam is split
into linearly polarized beams propagating in the
$(\hat{k},\hat{a})$ and $(\hat{k},\hat{k}\times\hat{a})$ planes,
respectively. The birefringent splitting in this case is maximal
when $\mb{k}$ is perpendicular to $\mb{a}$. Examples of media with
such a gyrotropic axial symmetry are isotropic media in an
external magnetic field or samples with net spin orientation
(magnetization).

The dielectric tensor of a medium in a magnetic field $\mb{B}$ is
given by \citep{LLElCont}
%--------------------------------------------------------------------
\begin{equation}
\varepsilon_{ij}\prn{\mb{B}}=\tilde{n}^2\delta_{ij}+i\tilde{\gamma}\epsilon_{ijk}B_k,
\end{equation}
%--------------------------------------------------------------------
where $\tilde{n}$ is the complex index of refraction in the
absence of the magnetic field. The real and imaginary parts of
$\tilde{\gamma}$ are responsible for the circular birefringence
and circular dichroism of the medium, respectively. A
straightforward calculation yields the direction of the Poynting
vector $\mb{P}$ of a linearly polarized plane wave in the case of
a weakly absorbing medium
($\im\tilde{n}^2,\abs{\im\tilde{\gamma}B}\ll\re\tilde{n}^2\simeq1$):
%--------------------------------------------------------------------
\begin{equation}
\mb{P}\propto\hat{k}+\im\tilde{\gamma}\sin\beta\prn{\mb{B}\cos\beta+\hat{k}\times\mb{B}\sin\beta},
\end{equation}
%--------------------------------------------------------------------
where $\beta$ is the angle between the polarization and the field.
A transversely polarized beam ($\beta=\pi/2$) will thus be
deflected by an angle $\alpha\simeq\im\tilde{\gamma}B$. This
results in a parallel beam displacement upon traversal of a sample
with parallel boundary surfaces.\footnote{The order of magnitude
of the displacement for sufficiently weak magnetic field can be
estimated as $\frac{g\mu B}{\Gamma}\frac{l}{l_0}\lambda$ (with the
usual notations).} The effect was first observed in resonantly
excited Cs vapor in fields up to 40 G \citep{Sch92}. The small
size of the effect (maximal displacements were a few tens of
nanometers) impeded the use of low light intensities, and the
experimental recordings show a strong nonlinear component,
presumably due to optical pumping. These nonlinearities manifest
themselves as dispersively shaped resonances in the magnetic-field
dependence of the displacement, which are similar to those
observed in the nonlinear Faraday effect (see Sec.\ \ref
{section:PhysMechNMO}). Unpublished results obtained by A. Weis at
high laser intensities have revealed additional resonances---not
observed in magneto-optical rotation experiments---whose origin
may be coherences of order higher than $\Delta M=2$. A theoretical
analysis of the results of \citeauthor{Sch92}\ based on the
Poynting-vector picture and the density-matrix method described in
Sec.\ \ref{subsection:DMCalc} was attempted by \citet{Roc2001MD}.
It was found that, while the theory reproduces the magnitude and
spectral dependence of deflection at low light power reasonably
well, it completely fails to reproduce the saturation behavior.
This suggests existence of some additional nonlinear mechanism for
deflection, perhaps at the interface between the vapor and the
glass.

As mentioned above, a spin-polarized medium has the same symmetry
properties as a medium subjected to an external magnetic field;
beam deflection was observed in a spin-polarized alkali vapor by
\citet{Bla92}.

The magnetic-field-induced displacement of light bears a certain
resemblance to the Hall effect---the deflection of a current
$\mb{j}$ of electrons (holes) by a field $\mb{B}$ in the direction
$\mb{j\times B}$---and might thus be called the ``photonic Hall
effect.'' However, this term was actually used to designate a
related observation of a magnetic-field-induced asymmetry in the
lateral diffusion of light from condensed-matter samples in
Tesla-sized fields \citep{Rik96}. Recently, the description of the
propagation of light in anisotropic media in terms of a
Poynting-vector model was questioned following a null result in a
magneto-deflection experiment on an aqueous solution of Nd$^{3+}$
\citep{Rik97}. More experimental and theoretical work is needed to
clarify these issues.

\subsubsection{The mechanical Faraday effect}
\label{subsubsection:The mechanical Faraday effect}

Since the effect of applying a magnetic field to a sample can be
interpreted, by Larmor's theorem, as a rotation of the atomic wave
function, a question can be raised: can optical rotation be
induced by macroscopic rotation of the whole sample? If the
electrons bound in the sample ``feel'' the macroscopic rotation, a
\emph{mechanical} Faraday effect results. This rotation can only
be felt if strong coupling exists between atomic electrons and
their environment. The mechanical Faraday effect has been
demonstrated in a rotating solid-state sample by \citet{Jon76}
(see also \citet{BarZel79} for a theoretical discussion), but
there is still the question of whether one can rotate atoms in the
gas phase. \citet{Woe92}; \citet{Nie92}; \citet{Nie94} found that
an isolated atom cannot be rotated but if diatomic complexes are
formed during its interaction with the environment (collisions),
mechanical Faraday rotation may be possible. Such binary complexes
in high-density atomic vapor were observed in the experiment of
\citet{Kri94} described in Sec.\ \ref{subsection:collisions}.
These authors observed rotational locking of the electronic wave
function to the instantaneous interatomic collision axes that
could possibly be used for a demonstration of the mechanical
Faraday effect. Estimates by \citet{Nie94} show that the ratio of
the rotation due to the mechanical and magnetic effects, when the
angular velocity equals the Larmor precession frequency, is of the
order of the collision frequency multiplied by the duration of the
collision. As far as the authors of this review are aware, such
effect has not yet been observed in the gas phase.

\section{Linear vs. nonlinear light-atom interactions}
\label{section:linvsnonlin}

In its broadest definition, a nonlinear optical process is one in
which the optical properties of the medium depend on the light
field itself. The light field can consist either of a single beam
that both modifies the medium and probes its properties, or of
multiple beams (e.g, a pump-probe arrangement). In certain
situations, light-induced modifications of the optical properties
of the medium can persist for a long time after the perturbing
light is turned off (these are sometimes referred to as \emph{slow
nonlinearities}). In such cases, the pump and probe interactions
can be separated in time, with the effect of the pump interaction
accumulating over a time limited by the relaxation rate of the
slow nonlinearity. This is the situation in optical pumping (Sec.\
\ref{subsection:OpticalPumping}), and in much of the NMOE work
discussed in this Review. In this Section, we provide a
classification of optical properties based on the perturbative
approach, and introduce saturation parameters that are ubiquitous
in the description of nonlinear optical processes near resonance.

\subsection{Perturbative approach}
\label{subsection:linvsnonlin:perturb}

The classification of atom-light interactions in terms of linear
and nonlinear processes can be done using a description of the
atomic ensemble by its density matrix $\rho$. The knowledge of
$\rho$ allows one to calculate the induced polarization $\mb{P}$
(the macroscopic electric-dipole moment) using the relation
%--------------------------------------------------------------------
\begin{equation}
\mb{P}=\tr\rho\prn{\bs{\mc{E}}}\mb{d}, \label{PvsE}
\end{equation}
%--------------------------------------------------------------------
which is the response of the medium to the optical field
$\bs{\mc{E}}$, where $\mb{d}$ is the electric-dipole operator. The
evolution of $\rho$ is described by the Liouville equation
\citep[see, for example,][]{Cor88}, which is linear in the
Hamiltonian describing the interaction of the ensemble with the
external magnetic field and the optical field. As discussed, for
example, by \citet{Ste84} and \citet{Boy92}, the solutions $\rho$
of the Liouville equation can be expanded in powers of the optical
field amplitude
%--------------------------------------------------------------------
\begin{equation}
\rho=\sum_{n=0}^{\infty}\rho^{(n)}\mc{E}^n. \label{rhoexpand}
\end{equation}
%--------------------------------------------------------------------

Figure {\ref{fig:nonlinrho} is a diagramatic representation of
this perturbative expansion for light interacting with a two-level
atom, in which each level consists of a number of degenerate
Zeeman sublevels.
%--------------------------------------------------------------------
\begin{figure}
\includegraphics[width=3in]{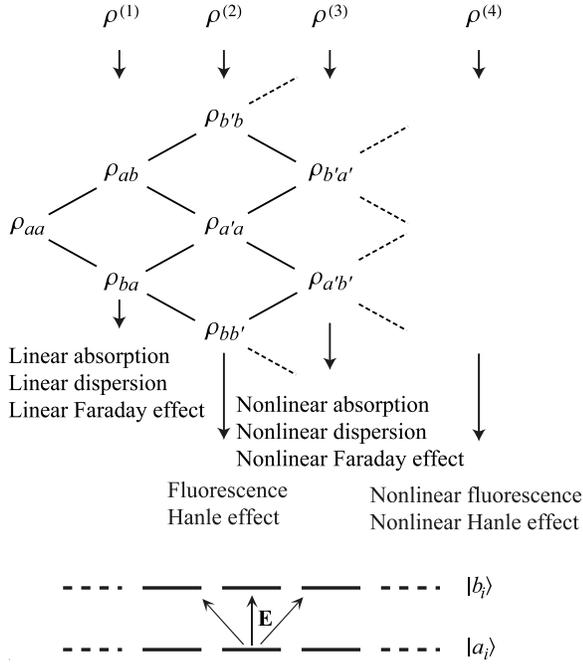}
\caption{Perturbative evolution of the density matrix of two-level
atoms interacting with an optical field. Note that for any given
order, the primed quantities label any of the magnetic sublevels
or coherent superpositions thereof which may be reached from the
previous order by a single interaction respecting the angular
momentum selection rules. The set of all $\rho_{aa'}$ also
contains $\rho_{aa}$.}\label{fig:nonlinrho}
\end{figure}
%--------------------------------------------------------------------
In the absence of light and in thermodynamic equilibrium, the
density matrix contains only lower-level populations $\rho_{aa}$.
The lowest-order interaction of the light field with the atomic
ensemble creates a linear polarization characterized by the
off-diagonal matrix element $\rho_{ab}^{(1)}$. This polarization
is responsible for the linear absorption and dispersion effects,
such as the linear Faraday effect. In the next order of
perturbation, excited-state populations $\rho_{bb}^{(2)}$ and
coherences $\rho_{aa'}^{(2)}$ and $\rho_{bb'}^{(2)}$ in the ground
and excited states appear. They are needed to describe
fluorescence and quantum interference effects (e.g, quantum
beats), excited-state level crossings and the Hanle effect (Sec.\
\ref{subsubsection:Hanle_LC}). Generally, even-order terms of
$\rho$ are stationary or slowly oscillating, while odd-order terms
oscillate with the frequencies in the optical range. In
particular, the next order $\rho_{ab}^{(3)}$ is the lowest order
in which a nonlinear polarization at the optical frequency
appears. It is responsible for the onset of the nonlinear
(saturated) absorption and dispersion effects, such as the
nonlinear Faraday and Voigt effects. The appearance of
$\rho^{(4)}$ marks the onset of nonlinear (saturated) fluorescence
and the nonlinear Hanle effect. In terms of quantum interference
(Sec.\ \ref{subsubsection:Hanle_LC}), these effects can be
associated with opening of additional interference channels (in
analogy to a multi-slit Young's experiment).

The Liouville equation shows how subsequent orders of $\rho$ are
coupled \citep[see discussion by][]{Boy92}:
%--------------------------------------------------------------------
\begin{equation}
\dot{\rho}_{ab}^{(n)}=-\prn{i\omega_{ab}+\gamma_{ab}}\rho_{ab}^{(n)}-\frac{i}{\hbar}\sbrk{-\mb{d}\cdot\bs{\mc{E}},\rho^{(n-1)}}_{ab},
\label{Liouville}
\end{equation}
%--------------------------------------------------------------------
where $\hbar\omega_{ab}$ is the energy difference between levels
$a$ and $b$. The density matrix $\rho$ to a given order
$\rho^{(n)}$ can thus be expressed in terms of the elements
$\rho^{(n-1)}$ of the next lowest order. Odd orders of $\rho$
describe optical coherences and even orders describe level
populations and Zeeman coherences [according to Eq.\
(\ref{Liouville}) the latter are the source terms for the former
and vice-versa]. The magnitude of the linear absorption
coefficient, governed by $\rho_{ab}^{(1)}$, can be expressed in
terms of the populations $\rho_{aa}^{(0)}$. Similarly,
$\rho_{ab}^{(3)}$, which determines the lowest-order saturated
absorption and the lowest-order nonlinear Faraday effect, is
driven by $\rho_{aa}^{(2)}$ and $\rho_{bb}^{(2)}$ (excited- and
ground-state populations), $\rho _{bb'}^{(2)}$ (excited-state
coherences) and $\rho_{aa'}^{(2)}$ (ground-state coherences).}

In the language of nonlinear optics, the induced polarization can
be expanded into powers of $\bs{\mc{E}}$:
%--------------------------------------------------------------------
\begin{equation}
\mb{P}=\sum_{n=1}^{\infty}\mb{P}^{(n)}=\sum_{n=1}^{\infty}\bs{\chi}^{(n)}\!\cdot\bs{\mc{E}}^n,
\label{Pexpand}
\end{equation}
%--------------------------------------------------------------------
where the $\bs{\chi}^{(n)}$ are the $n$-th order electric
susceptibilities. Expression (\ref{Pexpand}) is a particular case
of the general expansion used in nonlinear optics in which $n$
optical fields $\bs{\mc{E}}_i$ oscillating at frequencies
$\omega_i$ generate $n$-th order polarizations oscillating at all
possible combinations of $\omega_i$. In the present article, we
focus on the case of a single optical field oscillating at
frequency $\omega$. Here the only polarizations oscillating at the
same frequency are of odd orders in $P$. Coherent forward
scattering is governed by $P$ and is thus directly sensitive only
to odd orders of $\rho$ or $\chi$. Even orders can be directly
probed by observing (incoherently emitted) fluorescence light.
Linear dispersion and absorption effects such as the linear
Faraday effect are $\chi^{(1)}$ processes, while the nonlinear
Faraday effect arises first as a $\chi^{(3)}$ process.

In forward-scattering experiments, the absorbed power is
proportional to $\langle\mb{\dot{P}}\cdot\bs{\mc{E}}\rangle$,
which is given in (odd) order $l$ by
$(\bs{\chi}^{(l)}\!\cdot\bs{\mc{E}}^l)\!\cdot\bs{\mc{E}}\propto\chi^{(l)}I^{(l+1)/2}$,
where $I$ is the light intensity.\footnote{This follows directly
from the Maxwell's equations in a medium \citep[see discussion by,
for example,][]{She84}.} Forward-scattering in lowest order
($l=1$) involves $\chi ^{(1)}$ and is a linear process in the
sense that the experimental signal is proportional to $I$. In
light-induced fluorescence experiments, the scattered power is
proportional to the excited-state population, which, to lowest
order [$\rho_{bb}^{(2)}$] is also proportional to $I$. We adopt
the following classification (which is consistent with the
commonly used notions): Processes which involve $\rho^{(1)}$ and
$\rho^{(2)}$ will be called \emph{linear processes}, while all
higher-order processes will be referred to as \emph{nonlinear
processes}.\footnote{The formalism of \emph{nonlinear wave mixing}
\citep[see discussion by][]{She84,Boy92} allows one to describe
\emph{linear} electro- and magneto-optical effects as three- or
four-wave-mixing processes, in which at least one of the mixing
waves is at zero frequency. For example, linear Faraday rotation
is seen in this picture as mixing of the incident linearly
polarized optical field with a zero-frequency magnetic field to
produce an optical field of orthogonal polarization. While this is
a perfectly consistent view, we prefer to adhere to a more common
practice of calling these processes linear. In fact, for the
purpose of the present review, the term \emph{nonlinear} will only
refer to the dependence of the process on \emph{optical} fields.
Thus, we also consider as linear a broad class of optical-rf and
optical-microwave double-resonance phenomena.}

This perturbative approach is useful for understanding the onset
of a given effect with increasing light power. It works well in
most cases involving nonresonant light fields. In the case of
resonant laser excitation, however, caution must be used, since
there is no way to make meaningful distinctions between different
perturbative orders once the transition is saturated [the series
(\ref{Liouville}) and (\ref{Pexpand}) are non-converging and all
orders of these expansions are coupled.]

This discussion is generalized in a straightforward manner to the
case of pump-probe experiments (i.e., in which separate light
beams are used for pumping and probing the medium). Here the
signals that depend on both beams are, to lowest order,
proportional to the product of the intensities of these beams and
thus correspond to a $\chi^{(3)}$ nonlinear process.

\subsection{Saturation parameters}
\label{subsection:satpar}

Nonlinear optical processes are usually associated with high light
intensities. However, nonlinear effects can show up even with weak
illumination---conventional spectral lamps were used in early
optical pumping experiments with resonant vapors to demonstrate
nonlinear effects (Sec.\ \ref{subsection:OpticalPumping}). It is
important to realize that the degree of nonlinearity strongly
depends on the specific mechanism of atomic saturation by the
light field as well as the relaxation rate of atomic polarization.

A standard discussion of optical nonlinearity and saturation for a
two-level atom is given by \citet{AllenEberly}. The degree of
saturation of a transition is frequently characterized by a
\emph{saturation parameter} of general form
%--------------------------------------------------------------------
\begin{equation}
\kappa=\frac{\text{excitation rate}}{\text{relaxation rate}}.
\label{SPGEN}
\end{equation}
%--------------------------------------------------------------------
The saturation parameter is the ratio of the rates of coherent
light-atom interactions (responsible for Rabi oscillations) and
incoherent relaxation processes (e.g., spontaneous decay).

Consider a two-level system in which the upper state decays back
to the lower state and light is tuned to resonance. The saturation
parameter is given by
%--------------------------------------------------------------------
\begin{equation}
\kappa_1=\frac{d^2\mc{E}^2_0}{\hbar^2\gamma_0^2}. \label{kappa1}
\end{equation}
%--------------------------------------------------------------------
Here $d$ is the transition dipole moment, $\mc{E}_0$ is the
amplitude of the electric field of the light, and $\gamma_0$ is
the homogeneous width of the transition. (For simplicity, we
assume that $\gamma_0$ is determined by the upper state decay.)
The excitation rate is $d^2\mc{E}_0^2/(\hbar^2\gamma_0)$, and the
relaxation rate is $\gamma_0$.

When $\kappa_1\ll1$, the spontaneous rate dominates. Atoms mostly
reside in the lower state. Occasionally, at random time intervals,
an atom undergoes an excitation/decay cycle; the average fraction
of atoms in the upper state is $\sim$$\kappa_1$. On the other
hand, when $\kappa_1\gg1$, atoms undergo regular Rabi oscillations
(at a rate much greater than that of spontaneous emission), and
the average populations of the upper and the lower states
equalize.

If the upper state predominantly decays to states other than the
lower state of the transition, the character of saturation is
quite different. If there is no relaxation of the lower level of
the transition, light of any intensity will eventually pump all
atoms out of the lower and upper states into the other states.
When there is lower-state relaxation at rate $\gamma$ in the
absence of light, the saturation parameter is
%--------------------------------------------------------------------
\begin{equation}
\kappa_2=\frac{d^2\mc{E}^2_0}{\hbar^2\gamma_0\gamma}.
\label{kappa2}
\end{equation}
%--------------------------------------------------------------------
Here the excitation rate is the same as for $\kappa_1$, but the
relaxation rate is now $\gamma$. In many experiments, $\gamma$ is
much smaller than $\gamma_0$, so that nonlinear effects can appear
at lower light intensities than for systems characterized by
$\kappa_1$. Note that the parameter $\kappa_1$ is also significant
for open systems as it gives the ratio of the populations of the
upper and lower states.

When the upper and lower levels of the transition are composed of
Zeeman sublevels, the relevant saturation parameter for nonlinear
effects depends on the specifics of the transition structure and
the external (optical and other) fields. For example, if a
``dark'' state is present in the lower level, the system is
effectively open (atoms are pumped into the dark state in a
process known as \emph{coherent population trapping}; see Sec.\
\ref{subsection:lambdares}), and the optical pumping saturation
parameter $\kappa_2$ applies. For an $F=1/2\rightarrow{F'=1/2}$
transition pumped with linearly polarized light, on the other
hand, there is no dark state and the saturation parameter
$\kappa_1$ applies. If circularly polarized light is used for
optical pumping, however, or if a magnetic field is used to break
the degeneracy between transitions for right- and left-circularly
polarized light, the transition becomes effectively open and the
parameter $\kappa_2$ applies.

The parameters $\kappa_1$ and $\kappa_2$ are also of use in the
characterization of the magnitude of power broadening and of light
shifts of the transition.

\section{Early studies of nonlinear magneto-optical effects}
\label{section:EarlyStudies}

In this Section, we trace the development of the study of NMOE
from the early optical pumping work of the 1950s to the
experiments with tunable lasers carried out in the 1980s and early
1990s that led to the formulation of quantitative theories of
magneto-optics in simple systems and in the alkalis \citep{Kan93}.

\subsection{Optical pumping}
\label{subsection:OpticalPumping}

Research on polarized light-atom interactions (Sec.\
\ref{subsubsection:Hanle_LC}) initiated by
\citet{Woo22,Woo23,Woo24,Han24} was not pursued much further at
the time. Twenty-odd years later, \citet{Kas50} had the idea to
make use of angular momentum conservation in light-matter
interaction \citep{Rub18} to create spatially oriented atomic
angular momenta by absorption of circularly polarized light
(optical pumping). Anisotropy created by optical pumping can be
detected by monitoring of the intensity and/or polarization of the
transmitted light, monitoring of the fluorescence intensity and/or
polarization, or state selection in atomic beams. For transmission
monitoring, either transmission of the pump beam or of a probe
beam can be recorded. The latter method has been most often
applied with resonant probe beams for absorption monitoring
\citep{Deh57}. In particular, \citet{Bel57,Bel61a} employed
Dehmelt's method for detection of the ground-state coherence
induced by an rf field and by an intensity-modulated pump beam.

It was soon realized that probe transition monitoring could be
used with off-resonant light for dispersion detection.
\citet{Goz62} suggested observing the rotation of the polarization
plane of a weak probe beam to detect the population imbalance in
atomic sublevels created by optical pumping. This rotation (which
occurs in zero magnetic field) is caused by amplitude asymmetry
between the $\sigma^+$ and $\sigma^-$ refractive indices,
analogous to that occurring in paramagnetic samples (Sec.
\ref{subsection:LinMOEmechs}); hence it is sometimes called the
paramagnetic Faraday effect.\footnote{Strictly speaking, the
paramagnetic Faraday effect is not a magneto-optical effect, since
it occurs even when there is no magnetic field present. Still, it
affects atomic dispersion, and so can play a role in
magneto-optical effects.} Many experiments were performed using
this idea \citep[reivewed by][]{Hap71}. For optical pumping
studies, off-resonance detection of dispersive properties has the
advantage over absorption monitoring that the probe beam interacts
only weakly with the pumped atoms. Thus high probe-beam
intensities can be used, enhancing the signal-to-noise ratio of
the detection, without loss of sensitivity due to power
broadening. On the other hand, a probe beam detuned from exact
resonance could induce light shifts\footnote{The quantum theory of
light shifts in the context of optical pumping was developed by
\citet{Bar61} \citep[also see discussion by][]{Coh68}. A simple
classical derivation by \citet{Pan66} indicated a relation to
atomic dispersion by showing that the light shift is given by a
convolution of the spectral profile of the light field with the
real part of the atomic susceptibility (refractive index). Since
the real part of the refractive index is an anti-symmetric
function of the light detuning from atomic resonance, the light
shift vanishes for resonant light beams.} of the atomic energy
levels, which may not be acceptable in high-precision work.

Because the optical properties of the atomic vapor are altered by
the pump beam, and these changes are detected by optical means,
optical pumping is classified as a nonlinear process. However, it
should be realized that, in most of the pre-laser cases, the
effect was due to the cumulative action of several optical-pumping
cycles, i.e., sequences of absorption and spontaneous emission,
occurring on a time scale shorter than the lower state relaxation
time, but longer than the upper-state lifetime. Thus no
appreciable upper-state population was produced.

Nonlinear effects related to upper-state saturation can arise even
with lamps as light sources, however, provided that the light
intensity is strong enough. An example of this can be found in the
work of \citet{Sch70} who investigated level-crossing signals in
the fluorescence of Rb and Cs vapors. At high light intensities,
narrow features appeared in the magnetic-field dependence near
$B=0$. They were ascribed to the appearance of the ground-state
Hanle signal in the excited-state fluorescence due to nonlinear
coupling of the ground- and excited-state coherences. As with the
Hanle experiment (Sec.\ \ref{subsubsection:Hanle_LC}), the width
of the resonance depended on the relaxation time of the atomic
state; however, in contrast to the original Hanle effect, here it
was the ground-state relaxation, rather than the upper-state
relaxation, that was relevant. Very narrow widths (on the order of
1 $\mu$G) were seen by \citet{Dup69}, who observed the
ground-state Hanle effect in transmission. In this work, rubidium
atoms contained in a paraffin-coated vapor cell (see Sec.\
\ref{subsect:ARCoatedCells}) were optically pumped by a circularly
polarized beam from a Rb lamp, and thus acquired macroscopic
orientation (magnetization) in their ground state. In the
transverse-magnetic-field dependence of the pump-beam
transmission, there was a Lorentzian resonance due to the
ground-state Hanle effect. In order to get a dispersive resonance
whose steep slope near the zero crossing could be used to detect
small magnetic field changes, \citeauthor{Dup69}\ applied
modulation to the magnetic field and performed lock-in detection
of the transmitted light intensity at the modulation frequency.
The ground-state Hanle effect studied was applied to
ultra-sensitive magnetometry by \citet{Coh69b,Dup70}, providing
sensitivity $\sim$$10^{-9}\ \mr{G\,Hz^{-1/2}}$.

\citet{BouGro66} performed experiments that were an important
precursor to later NMOE work with lasers (particularly, that which
utilized separated pump and probe light fields). They illuminated
an alkali-vapor cell subject to a magnetic field with pump and
probe lamp-light sources in order to carry out detailed studies of
spin-relaxation and spin-exchange processes. An interesting
feature of this setup is that the pump and probe light fields
could be resonant with different transitions of the same atom, or
even with transitions in different isotopes or species. In the
latter cases, atomic polarization induced by the pump light in one
isotope or species is transferred to the other via spin-exchange
collisions.

\subsection{Nonlinear magneto-optical effects in gas lasers}
\label{subsection:GasLasers}

When a magnetic field is applied to a gas laser medium, several
nonlinear magneto-optical effects can be observed by monitoring
laser intensity or the fluorescence light laterally emitted from
the laser tube.\footnote{Such effects were extensively studied in
the early days of laser physics, both experimentally
\citep{Cul64a,Cul64b,Cul64c,ForHarPol64,Dum64,Sch66,Kru66,Tom67}
and theoretically \citep{Dya66,Hae67,Sar67}.} These effects
include nonlinear Hanle and level-crossing effects (Sec.\
\ref{subsection:FSandLC}) and also a specific coherence effect,
known as \emph{mode crossing}. Mode crossing occurs when the
frequency separation between two laser modes matches the Zeeman
splitting of a pair of magnetic sublevels. The two sublevels are
then coherently driven by the two laser modes, resulting in a
resonance feature in the magnetic dependence of the output power
[\citet{Sch66}; \citet{Her72}; \citet{Dum72}; \citet{Her77}]. Mode
crossing is an example of a three-level coherence resonance,
related to various effects described in Sec.\
\ref{subsection:lambdares}.

Most gas lasers have tubes with Brewster windows and so possess
high polarization anisotropy. Such lasers are sensitive to
magneto-optical effects since the influence of light polarization
on the laser output intensity is strongly enhanced by the cavity.
Thus even minute changes in polarization are transformed into
easily detected intensity variations. Early experimental
demonstrations of such effects were described by
\citet{Her67,Her68,Her72,LeF71,LeF72}.\footnote{Additional
information on magneto-optical effects coupled to intra-cavity
anisotropy can be found in a book by \citet{Voy84} and in a
topical issue of Journal of Optics B (Semiclassical and Quantum
Optics), \textbf{3} (2001).}

It is possible to use a gas laser as a magnetometer by minimizing
anisotropy in the cavity and applying a longitudinal magnetic
field. The eigenmodes generated by the laser are two oppositely
circularly polarized components with different frequencies
(dependent on the magnetic field). If the two components have the
same intensity, the resultant output of the laser is linearly
polarized, with the direction of polarization rotating at half the
difference frequency between the modes. The time dependence of the
intensity transmitted through a linear analyzer gives a
measurement of the magnetic field \citep{Bre92,Bre95}.
\citeauthor{Bre92}\ demonstrated a magnetometric sensitivity of a
few $\mu$G$\,\text{Hz}^{-1/2}$ for fields on the order of one
Gauss.

\subsection{Nonlinear effects in forward scattering}
\label{subsect:nonlin_FS}

Forward scattering in the linear regime was discussed in Sec.\
\ref{subsection:FSandLC}. At high light intensity, nonlinear
effects appear in FS [the quantities $A$ and $n$ in Eq.\
(\ref{l4}) become intensity dependent]. There are two principal
mechanisms for the nonlinearity. One is velocity-selective
modifications of the atomic population distributions by
narrow-band light (see Sec.\ \ref{subsection:BenStr}); the other
is related to light-induced coherences between Zeeman sublevels.
Often, both of these mechanisms operate simultaneously.

In the early 1970s came the advent of the tunable laser, an ideal
tool for optical pumping and generation of atomic coherences. The
creation of atomic coherences by a strong light field was
extensively analyzed, and is reviewed by several authors
\citep[e.g.,][]{Coh74,Dec76,Gaw94}.

The first observations of laser-induced coherences in FS (as well
as the first observations of the nonlinear Faraday effect with
lasers) were made by \citet{Gaw74b,Gaw74a}, who used both cw and
pulsed lasers. Very strong dependence of the FS signal intensity
on the magnetic-field dependence was observed \citep{Gaw74b}, in
contrast to the prediction of the linear theory and earlier
observations with spectral lamps \citep{Cor66,Dur71,Kro72}. The
observed curves, reproduced in Fig.\ \ref{fig:FirstFS}, are
measures of $\varphi^2(B)$ if the optical density is not too high.
%-------------------------------------------------------------------
\begin{figure}
\includegraphics{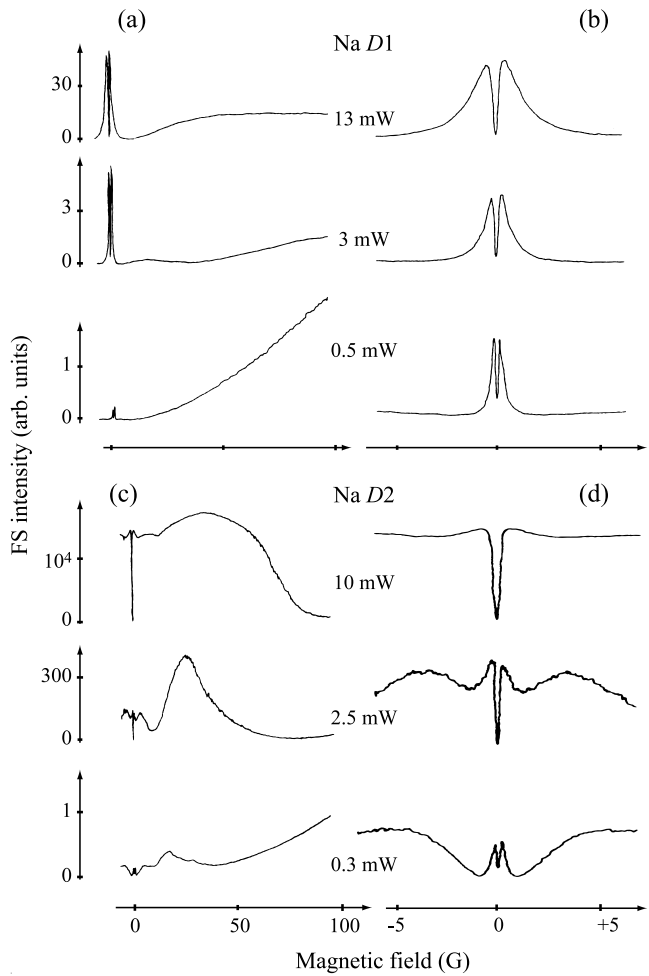}
\caption{Forward-scattering signals (FS intensity versus magnetic
field) from the sodium (a) $D1$ and (c) $D2$ lines (note the
additional broad structure arising from excited-state coherence)
obtained for various laser intensities by \citet{Gaw74b}. Plots
(b) and (d) show signals near zero magnetic
field.}\label{fig:FirstFS}
\end{figure}
%-------------------------------------------------------------------

The structures reported in \citep{Gaw74b} were \emph{subnatural},
i.e., they were narrower than would be expected if they were due
to the natural width of the optical transition (Sec.\
\ref{subsubsection:Hanle_LC}). They were also not subject to
Doppler broadening. This indicates that these structures were
associated with ground-state coherences whose effective relaxation
rate was determined by the finite atomic transit time across the
light beam.

The amplitude of the signals depended nonlinearly on the light
intensity and the signals were easily power-broadened. They were
also quite sensitive to small buffer-gas admixtures, which
indicated that optical coherences played a role in their creation
[collisions destroy optical coherences faster than they do
populations and ground-state coherences \citep{Gaw74b}]. The
observed FS signals exhibited a complex structure that was similar
to that seen by \citet{Duc73} in fluorescence. \citeauthor{Duc73}\
attributed this structure to hexadecapole moments (associated with
the coherence between the $M=\pm2$ sublevels of the $F=2$
ground-state component; see Appendix \ref{Appendix:Multipoles} for
a discussion of atomic multipoles). The experiment of
\citet{Gaw74b} was initially interpreted in terms of such
multipoles. This interpretation raised some controversy since, as
pointed out by \citet{Gir82b,Gir85b},\footnote{This was also
discussed by \citet{Jun89}; \citet{Sta90}; \citet{Hol95}.} signals
like those observed in \citet{Gaw74b} could be explained by
third-order perturbation theory (see Sec.\
\ref{subsection:linvsnonlin:perturb}) in which quadrupoles are the
highest-rank multipoles, i.e., without invoking the hexadecapole
moment. On the other hand, under the conditions of the experiment
of \citet{Gaw74b}, perturbation theory was clearly invalid
\citep{Gaw82}. Moreover, hexadecapole moments were unambiguously
observed in several other experiments with fluorescence detection
\citep{Duc73,Fis82,War86} performed under similar conditions as
those of \citet{Gaw74b}. An intriguing question arose: do the
higher-order multipoles affect the FS signals, and if so, why are
simple perturbative calculations so successful in reproducing the
experimental results? Answering this question was not
straightforward because higher-order multipoles exist in systems
with high angular momenta, and thus a large number of coupled
sublevels. Hence, it took quite some time until the above
controversy was resolved. \citet{Lob96} performed detailed
nonperturbative calculations for the Na $D1$ line, taking all
possible Zeeman coherences into account. They found that
hexadecapoles can indeed affect FS signals but only when a single
hyperfine structure (hfs) component is selected ($F=2\rightarrow
F'=2$ in the case of the Na $D1$ transition). Other hfs components
of the Na $D1$ line are largely insensitive to the hexadecapole
coherence. In later work by \citet{Lob97}, this result was
interpreted in terms of two competing trap states (Sec.\
\ref{subsection:lambdares}) contributing to the forward-scattered
signal.

Apart from the above-mentioned nonlinear magneto-optical effects
at near-zero magnetic fields, some attention was also devoted to
high magnetic fields. \citet{Gib74} investigated Faraday rotation
in the vicinity of the Na $D1$ transition under the conditions of
saturation with cw light and, in a separate set of measurements,
under the conditions of \emph{self-induced transparency}
\citep{McC69}. These experiments were done at magnetic fields in
the range 1--10 kG. \citeauthor{Gib74}\ found that Faraday
rotation in these nonlinear regimes is actually similar to the
rotation in the linear regime. The utility of the nonlinear
regimes is that they afford a significant reduction in absorption
of the light, making it possible to achieve large rotation angles
[in excess of $3\pi/2$ rad in the work of \citet{Gib74}].
\citet{Aga97} studied the influence of propagation effects (high
optical density) and high light intensity on the FS spectra taken
at high magnetic fields. This analysis is important for precision
measurements of oscillator strengths (Sec.\ \ref{subsection:f}).

\subsection{``Rediscoveries'' of the nonlinear magneto-optical
effects}

Some ten years after the first investigations of nonlinear
magneto-optical effects in forward scattering
\citep{Gaw74b,Gaw74a}, similar effects were independently
``rediscovered'' by several groups, in certain instances in a
somewhat dramatic manner.\footnote{These ``rediscoveries''
prompted systematic investigations of the nonlinear
magneto-optical effects, both experimental
\citep{Bad84,Sta85,Dra86,Lan86,Dra88,Davies87,Bue87,Bai89,Bar89a,Che90a,Sta90,Wei93}
and theoretical \citep{Gir85a,Gir85b,Fom87,Sch87,Jun89,Kan93}.}
For example, \citet{Davies87} passed a linearly polarized laser
beam tuned near an atomic resonance through a samarium vapor cell
placed in a longitudinal magnetic field. The direction of light
polarization after the cell was analyzed. Normally, a small amount
of buffer gas was present in the cell; the resulting magnitude and
frequency dependence of the Faraday rotation were well understood.
However, rather accidentally, it was discovered that, if the
buffer gas was removed from the cell, the magnitude of the peak
rotation changed sign and its size increased by some three orders
of magnitude! The frequency and magnetic-field dependences of the
rotation also changed radically. Similar observations were made by
other groups, but a consistent explanation of the phenomenon was
lacking. In order to clarify the situation, \citet{Bark89b}
performed an experiment on several low-angular-momentum
transitions of atomic Sm and identified two mechanisms (Fig.
\ref{fig:Rediscovery}) responsible for the enhanced rotation:
Bennett-structure formation in the atomic velocity distribution
(Sec.\ \ref{subsection:BenStr}) and Zeeman coherences (Sec.\
\ref{subsection:CohEf}). The latter effect produces a larger
enhancement of rotation in small magnetic fields; \citet{Bark89b}
observed rotation $\sim$$10^4$ times larger than that due to the
linear Macaluso-Corbino effect. These experimental results were
subsequently compared to theoretical predictions, showing good
agreement \citep{Koz89,Zet92}. At about the same time, precision
measurements and detailed calculations of the nonlinear Faraday
effect on the cesium $D2$ line were performed
\citep{Che90a,Wei93,Kan93}. Because Cs has hyperfine structure and
high angular momentum, the situation was somewhat more complicated
than for transitions between low-angular-momentum states of the
nuclear-spin-less isotopes of Sm (see Sec.\ \ref{section:TheorMod}
for a description of the theoretical approaches to NMOE).
Nevertheless, excellent quantitative agreement between theory and
experiment was obtained.
%-------------------------------------------------------------------
\begin{figure}
\includegraphics{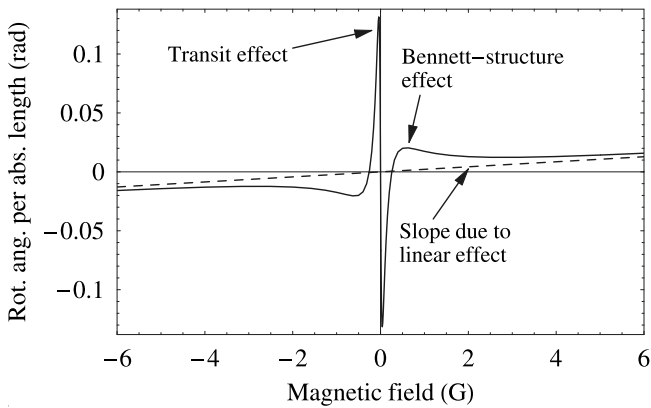}
\caption{Density-matrix calculation (\citet{Bud2002Bennett}; Sec.
\ref{subsection:DMCalc}) of light polarization rotation as a
function of magnetic field for the case of the samarium 571 nm
($J=1\rightarrow J'=0$) line, closely reproducing experimental
results described by \citet{Bark89b}; see also
\citet{Zet92}.}\label{fig:Rediscovery}
\end{figure}
%-------------------------------------------------------------------

\section{Physical mechanisms of nonlinear magneto-optical effects}
\label{section:PhysMechNMO}

When a nonlinear magneto-optical effect occurs, the properties of
both the medium and the light are affected. Our description of the
physical mechanisms that cause NMOE deals with the effects on the
medium and the light separately, first detailing how the
populations of and coherences between atomic states are changed
(optical pumping) and then how the light polarization is
subsequently modified (optical probing). If, as is the case in
many of the experiments discussed in Secs.\
\ref{section:EarlyStudies}, \ref{section:NMOEinSpecSit},
\ref{Sect:NMORApplications}, a single laser beam is used for both
pumping and probing, these processes occur simultaneously and
continuously. However, comparison with full density-matrix
calculations and experimental results shows that the essential
features of NMOE can be understood by considering these processes
separately.

As discussed above, we distinguish between two broad classes of
mechanisms of NMOE: first, Bennett-structure effects, which
involve the perturbation of populations of atomic states during
optical pumping, and second, ``coherence'' effects, which involve
the creation and evolution of atomic polarization (although in
some cases, with a proper choice of basis, it is possible to
describe these effects without explicit use of coherences, see
Sec. \ref{subsection:KanWeiTheory}).

In the case of nonlinear Faraday rotation, the primary empirical
distinction between these two effects is the magnitude of the
magnetic field $B_\text{max}$ at which optical rotation reaches a
maximum. This magnetic-field magnitude is related to a line width
by the formula [see Eq.\ (\ref{Eqn:phi_vs_B})]:
%--------------------------------------------------------------------
\begin{equation}
    B_\text{max}=\frac{\hbar\Gamma}{2g\mu}. \label{linewidth}
\end{equation}
%--------------------------------------------------------------------
The smallest achievable line width for Bennett-structure-related
NMOR corresponds to the natural width of the atomic transition
(typically $2\pi\times$1--10 MHz for allowed optical
electric-dipole transitions). In the case of the coherence
effects, the line width $\Gamma$ is determined by the rate of
atomic depolarization. The smallest NMOR line widths
$\sim$$2\pi\times1$ Hz have been observed by \citet{Bud98} for
atoms in paraffin-coated cells.

\subsection{Bennett-structure effects}
\label{subsection:BenStr}

We first consider NMOR caused by Bennett-structures---``holes''
and ``peaks'' in the atomic velocity distributions of the
ground-state sublevels resulting from optical pumping
\citep{Ben62}. As an example, we discuss a simple case of
Bennett-structure-related (BSR) NMOR in which optical pumping and
probing are performed in separate regions (\citet{Bud2002Bennett};
Fig.\ \ref{fig:TwoRegions}).
%------------------------------------------------------------------
\begin{figure}
\includegraphics{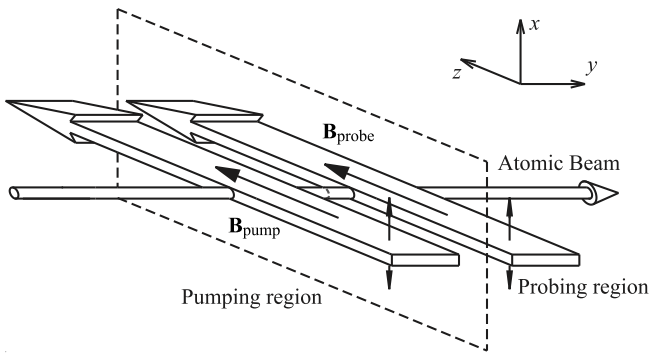}
\caption{Conceptual two-region experimental arrangement with
separated optical pumping and probing regions used to treat
Bennett-structure effects in NMOR. Pump and probe light beams are
initially linearly polarized along $x$, the atomic beam propagates
in the $\mb{\hat{y}}$ direction, and the magnetic fields are
oriented along the direction of light propagation
($\mb{\hat{z}}$).}\label{fig:TwoRegions}
\end{figure}
%------------------------------------------------------------------
In this example there is $\mb{\hat{z}}$-directed magnetic field
$\mb{B}_\text{probe}$ in the probe region ($\mb{B}_\text{pump}$ is
set to zero). In the pumping region, atoms interact with a
near-resonant $x$-polarized narrow-band laser beam of saturating
intensity [the optical pumping saturation parameter (Sec.\
\ref{subsection:satpar})
$\kappa=d^2\mc{E}_0^2/(\hbar^2\gamma_0\gamma_t)\gg1$, where
$\gamma_0$ is the natural width of the transition, and $\gamma_t$
is the rate of atoms' transit through the pump laser beam]. In the
probing region, a weak ($\kappa\ll1$) light beam, initially of the
same polarization as the pump beam, propagates through the medium.
The resultant polarization of the probe beam is subsequently
analyzed.

Consider an $F=1/2\rightarrow{F'=1/2}$ transition. This system is
especially straightforward, because it exhibits no coherence
magneto-optical effects with linearly polarized light. Thus NMOR
in this system is entirely due to Bennett structures. Suppose that
the upper state decays to levels other than the ground state. For
atoms in the resonant velocity group, the $\ket{M=\pm1/2}$
lower-state Zeeman sublevels are depopulated by optical pumping,
creating ``holes'' in the atomic velocity distributions of the two
states [Fig.\ \ref{fig:HoleBurn}(a)].
%------------------------------------------------------------------
\begin{figure}
\includegraphics{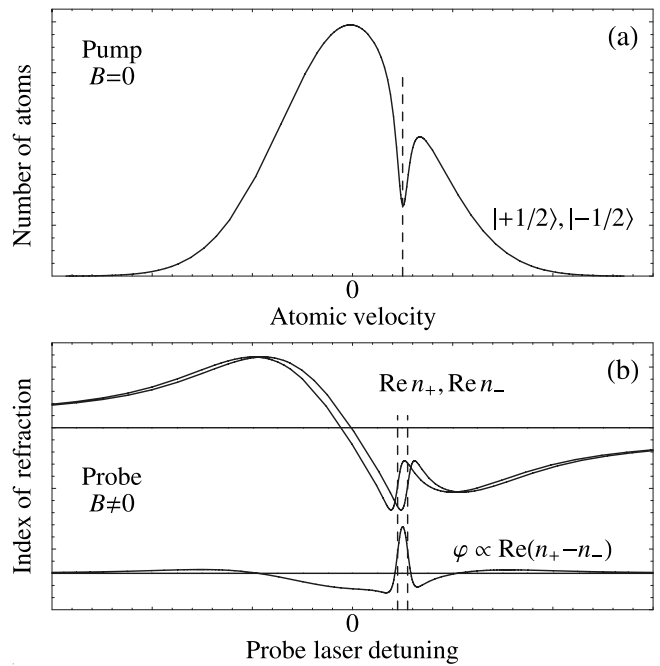}
\caption{The Bennett-structure effect on a $1/2\rightarrow1/2$
transition in which the upper state decays to levels other than
the lower state; $\mb{B}_\text{pump}=0$,
$\mb{B}_\text{probe}\neq0$. (a) In the pump region, monochromatic
laser light produces Bennett holes in the velocity distributions
of atoms in the lower state $\ket{+1/2}$, $\ket{-1/2}$ sublevels.
Since there is no magnetic field, the holes occur in the same
velocity group (indicated by the dashed line) for each sublevel.
(b) In the probe region, a magnetic field is applied, shifting
$n_+$ and $n_-$ relative to each other (upper trace). The real
part of the indices of refraction are shown; so that the features
in plot (a) correspond to dispersive shapes in this plot. The
shifted central detunings of the BSR features are indicated by the
dashed lines. Polarization rotation of the probe laser light is
proportional to the difference $\re\prn{n_+-n_-}$ (lower trace).
Features due to the Doppler distribution and the Bennett holes can
be seen. Since the Bennett related feature is caused by the
removal of atoms from the Doppler distribution, the sign of
rotation due to this effect is opposite to that of the linear
rotation. From \citet{Bud2002Bennett}.}\label{fig:HoleBurn}
\end{figure}
%------------------------------------------------------------------
Consequently, there are sub-Doppler features (with minimum width
$\gamma_0$) in the indices of refraction $n_-$ and $n_+$ for
right- and left-circularly polarized ($\sigma^-$ and $\sigma^+$,
respectively) light. In the probe region [Fig.\
\ref{fig:HoleBurn}(b)], a small magnetic field
$\mb{B}_\text{probe}<\hbar\gamma_0/(2g\mu)$ is applied. (For
simplicity, the Land\'{e} factor for the upper state is assumed to
be negligible.) The indices of refraction $n_-$ and $n_+$ are
displaced relative to each other due to the Zeeman shift, leading
to optical rotation of the probe beam. In the absence of Bennett
structures, this gives the linear Faraday effect. When Bennett
holes are produced in the pump region, the resultant Faraday
rotation can be thought of as rotation produced by the
Doppler-distributed atoms without the hole (linear Faraday
rotation) minus the rotation that would have been produced by the
pumped out atoms. Thus the rotation due to Bennett holes has the
opposite sign as that due to the linear effect.

As described in detail by \citet{Bud2002Bennett}, the mechanism of
BSR NMOR depends critically on the details of the experimental
situation. Depending on whether a magnetic field is present in the
pump region, and whether the excited state decays to the ground
state or other levels, BSR NMOR can be due to holes, (and have one
sign), be due to ``peaks'' generated by spontaneous decay to the
ground state (and have the opposite sign), or not be present at
all.

Note also that mechanical action of light on atoms could, under
certain conditions, lead to redistribution of atoms among velocity
groups, thus deforming the Bennett structures and modifying the
nonlinear optical properties of the medium \citep{Kaz86}.

\subsection{Coherence effects}
\label{subsection:CohEf}

The coherence effects can produce even narrower widths than the
Bennett-structure-related NMOE, thus leading to significantly
higher small-field Faraday rotation [\citet{Gaw94};
\citet{Ari96}]. At low light power, the effect of the pump light
can be conceptually separated from that of the magnetic field, so
that the coherence effect can be thought of as occurring in three
stages. First, atoms are optically pumped into an aligned state,
causing the atomic vapor to acquire linear dichroism; second, the
atomic alignment precesses in the magnetic field, rotating the
axis of dichroism; third, the light polarization is rotated by
interaction with the dichroic atomic medium, since the alignment
is no longer along the initial light polarization. The third,
``probing,'' step does not require high light intensity, and can
be performed either with a weak probe beam or with the same pump
light as used in the first step.

Consider atoms with total angular momentum $F=1$ that are not
aligned initially subject to linearly polarized laser light with
frequency corresponding to a transition to a $F'=0$ state (Fig.
\ref{fig:F1toF0}). One can view the atoms as being in an
incoherent mixture of the following states: $\ket{M=0}$,
$\prn{\ket{M=1}\pm\ket{M=-1}}/\sqrt{2}$. The first of these states
can be excited to the $F'=0$ state only by $z$-polarized
radiation; it is decoupled from $x$- and $y$-polarized light.
Similarly, the other two states (which are coherent superpositions
of the Zeeman sublevels) are $y$- and $x$-absorbing states,
respectively. Suppose the laser light is polarized along the
$x$-axis. Optical pumping by this light causes depletion of the
$x$-absorbing state, leaving atoms in the ``dark'' $y$- and
$z$-absorbing states.\footnote{This process is known as coherent
population trapping \citep{Ari96} because, as a result of optical
pumping by linearly polarized light, atoms in the $\ket{M=\pm1}$
substates are not completely pumped out as would seem to be the
case at first glance (Fig.\ \ref{fig:F1toF0}), but largely remain
in a dark coherent superposition.} The medium becomes transparent
for the $x$-polarized radiation; however, it can still absorb and
refract light of an orthogonal polarization.  The medium is
\emph{aligned}\footnote{In general, alignment can exist for an
$F\ge1$ state even when there is no dark state [i.e., for
$F\rightarrow F+1$ transitions \citep{Kaz84}].} (Appendix
\ref{Appendix:Multipoles}) and possesses linear dichroism and
birefringence.\footnote{The linear birefringence turns out not to
be important for the coherence effect because the refractive index
is unity on resonance. Note however, that at high light powers,
\emph{circular} birefringence emerges as a dominant effect
responsible for nonlinear Faraday rotation (see Sec.\
\ref{subsection:AOC}).}

In the presence of a magnetic field, the atomic alignment axis
precesses around the direction of the field at the Larmor
frequency [precession of atomic alignment in a magnetic field is
explained in detail in, for example, \citet[Ch.\ 15]{Cor88}]. One
can understand the effect of this precession on the light
polarization by thinking of the atomic medium as a layer of
polarizing material like a polaroid film [\citet{Kan93};
\citet{Bud99}]. It is easy to show that the rotation of the
``polarizer'' around the $z$-axis causes the linear output
polarization to be rotated by an angle proportional to the optical
density of the sample and to $\sin2\theta$, where $\theta$ is the
angle between the transmission axis of the rotated ``polarizer''
and the direction of initial light polarization (Fig.\
\ref{fig:RotPol}).
%-----------------------------------------------------------------
\begin{figure}
\includegraphics[width=2.5in]{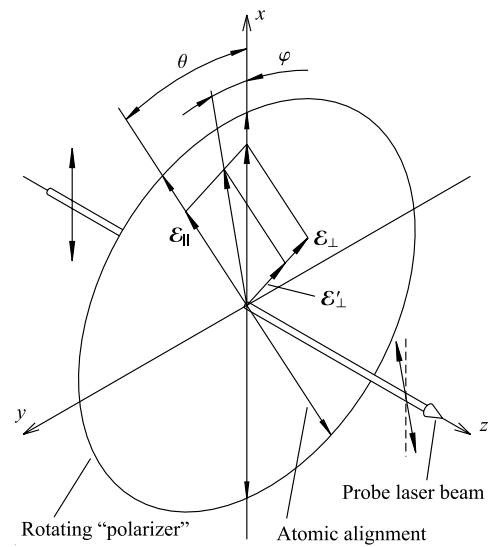}
\caption{ An optically thin sample of aligned atoms precessing in
a magnetic field can be thought of as a thin rotating polaroid
film that is transparent to light polarized along its axis
($\mc{E}_\parallel$), and slightly absorbent for the orthogonal
polarization ($\mc{E}_\perp$). $\mc{E}_\parallel$ and
$\mc{E}_\perp$ are the light electric field components. The effect
of such a ``polarizer'' is to rotate light polarization by an
angle $\varphi\propto\sin2\theta)$. The figure is drawn assuming
that a magnetic field is directed along $\mb{\hat{z}}$. Adapted
from \citet{Bud99}.}\label{fig:RotPol}
\end{figure}
%-----------------------------------------------------------------
In order to describe optical rotation of cw laser light, one has
to sum the effect of atomic ``polarizers'' continually produced by
the light, each rotating for a finite time and then
relaxing.\footnote{In the ``transit'' effect, at low light power,
this time is the atoms' time of flight between pumping and
probing, either within one laser beam (Sec.\
\ref{subsect:buff_gas_free_cells}) or between two (Sec.\
\ref{subsection:FRS}). In the ``wall-induced Ramsey effect,''
atoms leave the laser beam after optical pumping and travel about
the cell, returning to the beam after colliding with the
antirelaxation-coated cell walls (Sec.\
\ref{subsect:ARCoatedCells}). The relaxation time is thus
ultimately determined by collisional relaxation and spin
exchange.} Using this model, one can show \citep{Wei93,Kan93} that
for sufficiently low light power and magnetic field, the rotation
due to the coherence effect is once again described by Eq.\
(\ref{Eqn:phi_vs_B}), but now the relevant relaxation rate is that
of the ground-state alignment.

Calculations based on the rotating polarizer model reproduce the
magnitude and the characteristic details of the low-power
line-shape of NMOE quite well \citep{Kan93}, even when complicated
by the presence of transverse magnetic fields \citep{Bud98}. To
account for transverse fields, the model must include two
independent atomic subsamples with alignments corresponding to
polarizers with transmission axes directed mutually
perpendicularly. These subsamples arise due to optical pumping
through different hyperfine transitions. For zero transverse
field, the NMOE dependences on the longitudinal magnetic field for
each of these subsamples have symmetrical dispersive shapes of the
same widths, but of different signs and magnitudes. Thus the
resulting sum curve has a symmetrical shape. If a transverse field
is applied, the features for the two alignments acquire different
asymmetrical shapes. This leads to the asymmetrical sum curve seen
in the experiments of \citet{Bud98}.

\subsection{Alignment-to-orientation conversion}
\label{subsection:AOC}

In Sec.\ \ref{subsection:CohEf}, the coherence NMOE were described
as arising due to Larmor precession of optically induced atomic
alignment [Fig.\ \ref{fig:AOCpeanut}(a)].
%------------------------------------------------------------------
\begin{figure}
\includegraphics{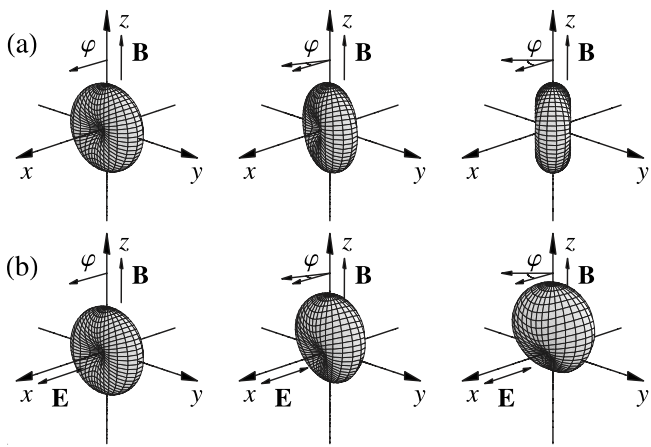}
\caption{(a) Sequence showing the evolution of optically pumped
ground-state atomic alignment in a longitudinal magnetic field for
an $F=1\rightarrow{F'=1}$ transition at low light powers (time
proceeds from left to right). The distance from the surface to the
origin represents the probability of finding the projection $M=F$
along the radial direction (Appendix
\ref{AppendixSubsect:AtPolVis}). In the first plot, the atoms have
been optically pumped into an aligned state by $x$-polarized
light. The magnetic field along $\mb{\hat{z}}$ creates a torque on
the polarized atoms, causing the alignment to precess (second and
third plots). This rotates the medium's axis of linear dichroism,
which is observed as a rotation of the polarization of transmitted
light by an angle $\varphi$ with respect to the initial light
polarization. (b) Sequence showing the evolution of the same
ground-state atomic alignment in a longitudinal magnetic field for
high light power. The light frequency is slightly detuned from
resonance to allow for a nonzero light shift. The combined action
of the magnetic and light fields produces orientation along the
$\mb{\hat{z}}$ axis.}\label{fig:AOCpeanut}
\end{figure}
%------------------------------------------------------------------
However, for sufficiently strong light intensities---when the
light shifts become comparable to or exceed the ground-state
relaxation rate---a more complicated evolution of the atomic
polarization state occurs \citep{Bud2000AOC}. Under these
conditions, the combined action of the magnetic and optical
electric fields causes atoms to acquire orientation along the
direction of the magnetic field, in a process known as
alignment-to-orientation conversion (AOC). As discussed in Sec.\
\ref{subsection:LinMOEmechs}, an atomic sample oriented along the
direction of light propagation causes optical rotation via
circular birefringence, since the refractive indices for
$\sigma^+$ and $\sigma^-$ light are different.

The evolution of atomic polarization leading to AOC-related NMOR
is illustrated for an $F=1\rightarrow{F'=1}$ transition in Fig.\
\ref{fig:AOCpeanut}(b). In the first plot, the atoms have been
optically pumped into an aligned state by $x$-polarized light
(they have been pumped out of the ``bright'' $x$-absorbing state
into the dark states). If the atomic alignment is parallel to the
optical electric field, the light shifts have no effect on the
atomic polarization---they merely shift the energies of the bright
and dark states relative to each other. However, when the magnetic
field along $\mb{\hat{z}}$ causes the alignment to precess, the
atoms evolve into a superposition of the bright and dark states
(which are split by the light shifts), so
optical-electric-field-induced quantum beats occur. These quantum
beats produce atomic orientation along $\mb{\hat{z}}$ (appearing
in the second plot and growing in the third plot) causing optical
rotation due to circular birefringence.

The atomic orientation produced by AOC is proportional to the
torque $(\overline{\mb{P}\!\!\times\!\bs{\mc{E}}})$, where
$\mb{P}$ is the macroscopic induced electric-dipole moment (Eq.\
\ref{PvsE}), and the bar indicates averaging over a light
cycle.\footnote{$\mb{P}$ is not collinear with $\bs{\mc{E}}$ due
to the presence of the magnetic field. Note also that since
$(\overline{\mb{P}\!\!\times\!\bs{\mc{E}}})$ is linear in $\mb{B}$
(at sufficiently small magnetic fields), it is a T-odd
pseudovector, as orientation should be.} The quantity
$(\overline{\mb{P}\!\!\times\!\bs{\mc{E}}})$ is proportional to
the light shift, which has an antisymmetric dependence on detuning
of the light from the atomic resonance. Thus (in the Doppler-free
case), net orientation can only be produced when light is detuned
from resonance.  This is in contrast to NMOR at low light powers,
which is maximum when light is tuned to the center of a
Doppler-free resonance.

It turns out that in applications to magnetometry (Sec.\
\ref{subSect:Magnetom}), the light power for which optimum
magnetometric sensitivity is obtained is sufficient to produce
significant AOC, so this effect is important for understanding the
properties of an NMOR-based magnetometer. For example, if two
atomic species (e.g., Rb and Cs) are employed in an NMOR-based
magnetometer, AOC generates a longitudinal spin polarization.
Spin-exchange collisions can then couple the polarizations of the
two atomic species. [Related AOC-induced coupling of polarization
of different ground-state hfs components in $^{85}$Rb was studied
by \citet{Yas99Sep}].

Although the phenomenon of AOC has been studied in a variety of
different contexts [\citet{Lom69}; \citet{Pin82}; \citet{Hil94};
\citet{Dov2001}; \citet{Kun2002}], its role in NMOE was only
recently recognized by \citet{Oku2000} and \citet{Bud2000AOC}.
However, as is usual with ``new'' phenomena, a closely related
discussion can be found in the classic literature
\citep{Coh69AOC}.

\section{Symmetry considerations in linear and nonlinear
magneto-optical effects} \label{section:SymCons}

The interaction of light with a dielectric medium is characterized
by the electric polarizability of the medium, which may be
equivalently described by various frequency-dependent complex
parameters: the electric susceptibility ($\chi$), the dielectric
permittivity ($\varepsilon$) and the index of refraction
($\tilde{n}$). The imaginary parts of these quantities determine
light absorption by the medium, while their real parts describe
the dispersion, i.e., the phase shifts that a light wave
traversing the medium experiences. For isotropic media, the above
parameters are scalar quantities and the interaction of the medium
with the light is independent of the light polarization. In
anisotropic media, the interaction parameters are tensors, and the
incident optical field and the induced electric polarization are
no longer parallel to each other. The induced polarization acts as
a source of a new optical field with a polarization (and
amplitude) that differs from the incident field and which adds
coherently to this field as light propagates through the medium.
This leads to macroscopic phenomena such as dichroism and
birefringence. Thus linear and nonlinear magneto-optical effects
in vapors can be regarded as the result of the symmetry breaking
of an initially isotropic medium due to its interaction with an
external magnetic field, and for the nonlinear effects, intense
polarized light.

An incident light field propagating along $\mb{k}$ can be written
as a superposition of optical eigenmodes (i.e., waves that
traverse the medium without changing their state of polarization,
experiencing only attenuation and phase shifts) determined by the
symmetry properties of the medium. Suppose that the medium is
symmetric about $\mb{k}$, and that the light is weak enough so
that it does not affect the optical properties of the medium. Then
there is no preferred axis orthogonal to $\mb{k}$, so the optical
eigenmodes must be left- and right-circularly polarized waves. If,
in addition, the medium has the symmetry of an axial vector
directed along $\mb{k}$ (generated, for example, by a magnetic
field in the Faraday geometry), the symmetry between the two
eigenmodes is broken and the medium can cause differential
dispersion of the eigenmodes (circular birefringence) leading to
optical rotation, and differential absorption (circular dichroism)
causing the light to acquire elliptical polarization (Fig.\
\ref{fig:FRE}).

If the medium does possess a preferred axis orthogonal to $\mb{k}$
(generated, for example, by a magnetic field in the Voigt
geometry), the eigenmodes must be fields linearly polarized along
and perpendicular to the preferred axis. There is clearly
asymmetry between the two eigenmodes; the medium possesses linear
birefringence and dichroism.\footnote{The effect of deflection of
a linearly polarized light beam by a medium with a transverse
axial symmetry was discussed in detail in Sec.
\ref{subsubsection:MagDefl}.} Since changing the sign of the
magnetic field does not reverse the asymmetry between the
eigenmodes, there should be no magneto-optical effects that are
linear in the applied field. Indeed, the lowest-order Voigt effect
is proportional to $B^2$.

If the light field is strong enough to alter the medium
susceptibility, it imposes the symmetry of the optical field onto
the medium, i.e., if the light is linearly polarized, the
eigenmodes are linearly polarized waves. In the absence of a
magnetic field, the input light field is an eigenmode, and no
rotation or ellipticity is induced. However, if a weak magnetic
field is present (and the light field is not too strong), it
effectively rotates the axis of symmetry of the medium, changing
the polarization of the eigenmodes. The input light field is no
longer an eigenmode, and rotation can occur. At high light powers,
the combined optical and magnetic fields cause a more complicated
evolution to occur (Sec.\ \ref{subsection:AOC}).

Consideration of the symmetry of the medium can be used to choose
the quantization axis that is most convenient for the theoretical
description of a particular effect. The choice of the orientation
of the quantization axis is in principle arbitrary, but in many
situations calculations can be greatly simplified when the
quantization axis is chosen along one of the symmetry axes of the
system. Thus for the linear Faraday and Voigt effects, a natural
choice is to orient $\mb{\hat{z}}$ parallel to the magnetic field.
For the nonlinear effects with low-power linearly polarized light,
on the other hand, it may be advantageous to choose the
quantization axis along the light polarization direction.

\section{Theoretical models}
\label{section:TheorMod}

The first theoretical treatment of NMOE was performed by
\citet{Gir80,Gir82b,Gir82a,Gir85a,Gir85b} who obtained expressions
to lowest order in the Rabi frequency (atom-light coupling) for
the nonlinear Faraday effect. In this perturbative approach NMOR
is viewed as a $\chi^{(3)}$ process (Sec.\
\ref{subsection:linvsnonlin:perturb}), and the NMOR angle is hence
proportional to the light intensity. Although the model presented
by \citeauthor{Gir85a}\ is valid for arbitrary angular momenta, it
can not be directly applied to alkali atoms, because it considers
only open systems, i.e., two-level atoms with Zeeman substructure
in which the excited states decay to external levels. For alkali
atoms under monochromatic excitation, repopulation and
hyperfine-pumping effects have to be taken into account (only the
depopulation effects were considered by \citeauthor{Gir85a}).
\citet{Che89,Che90a} applied the perturbation method developed by
\citet{Dra86,Dra88} for $F=1\rightarrow F'=0$ transitions to the
multilevel case of Cs. They extended the results obtained by
\citeauthor{Gir85a}\ by including repopulation effects, i.e., the
transfer of coherences from the excited state into the ground
state by fluorescence, allowing the first quantitative comparison
of theoretical predictions with experimental data. The agreement
was only partially satisfactory, since hyperfine pumping processes
were not properly treated---the important role they played was not
realized at the time. These processes were later properly included
in a calculation by \citet{Kan93} (see Sec.\
\ref{subsection:KanWeiTheory}), which turned out to be extremely
successful for the description of the coherence NMOE at low light
powers.

The steady-state density-matrix equations for a transition subject
to light and a static magnetic field \citep{Coh74} can be solved
nonperturbatively to obtain a theoretical description of the
transit, Bennett-structure, and linear effects for arbitrary light
power. This was done for systems consisting of two states with low
angular momentum ($F,F'=0,1/2,1$) by \citet{Sch87};
\citet{Davies87}; \citet{Izm88}; \citet{Bai89}; \citet{Sch89};
\citet{Koz89}; \citet{Fom91}; \citet{Sch95}; \citet{Hol95};
\citet{Sch99}. Good qualitative agreement with experiments in
samarium was demonstrated by \citet{Davies87,Bai89,Koz89,Zet92}.
The nonperturbative approach was generalized by \citet{Roc2001SR}
to systems with hyperfine structure for calculation of
self-rotation (Sec. \ref{subsection:self-rotation}) in Rb. This
method (discussed in Sec.\ \ref{subsection:DMCalc}) was applied to
Faraday rotation calculations, and quantitative agreement with
experiment was demonstrated \citep{Bud2000AOC,Bud2002Bennett}.

\subsection{Kanorsky-Weis approach to low-power nonlinear magneto-optical rotation}
\label{subsection:KanWeiTheory}

Quantitative agreement between theory and experiment for NMOE was
achieved by \citet{Kan93}, who calculated Faraday rotation
assuming both a weak magnetic field and low light power. Under
these conditions, the Bennett-structure (Sec.\
\ref{subsection:BenStr}) and linear (Sec.\
\ref{subsection:LinMOEmechs}) effects, as well as saturation
effects and alignment-to-orientation conversion (Sec.\
\ref{subsection:AOC}) are negligible. With these approximations,
and with an appropriate choice of quantization axis (Sec.\
\ref{section:SymCons}), it is possible to perform the calculation
making no explicit use of sublevel coherences. Optical-pumping
rate equations for the sublevel populations are solved, neglecting
the effect of the magnetic field, and then precession due to the
magnetic field is taken into account. This procedure is a
quantitative version of the description of the coherence NMOE as a
three-stage process (Sec.\ \ref{subsection:CohEf}).

As described by \citet{Kan93}, the complex index of refraction for
light of frequency $\omega$ and polarization $\mb{\hat{e}}$ can be
written in terms of level populations:
%------------------------------------------------------------------
\begin{equation}
\tilde{n}-1=\frac{2\pi}{\hbar}\sum_{M,M'}{\rho_{FM}\frac{\abs{\bra{F'M'}\mb{\hat{e}}\cdot\mb{d}\ket{FM}}^2}{\Delta-i\gamma_0}},
\end{equation}
%------------------------------------------------------------------
where $\mb{d}$ is the electric-dipole operator, $\Delta$ is the
light frequency detuning from resonance, and $\rho_{FM}$ are the
ground-state Zeeman-sublevel populations. The rotation angle is
given by
%------------------------------------------------------------------
\begin{equation}
    \varphi=\im\prn{\tilde{n}_\parallel-\tilde{n}_\perp}\frac{\omega{l}}{2c}\sin2\theta,
\end{equation}
%------------------------------------------------------------------
where $l$ is the sample length and $\theta$ is the average
orientation of the axis of dichroism with respect to the linear
polarization. The angle $\theta$ is calculated by time-averaging
the contributions of individual atoms, each of which precesses in
the magnetic field before relaxing at rate $\gamma_t$, with the
result [cf. Eq.\ (\ref{Eqn:phi_vs_B})]
%------------------------------------------------------------------
\begin{equation}
    \varphi=\im\prn{\tilde{n}_\parallel-\tilde{n}_\perp}\frac{\omega{l}}{2c}\frac{x}{x^2+1}, \label{Eqn:KanWei_phi_vs_B}
\end{equation}
where the dimensionless parameter
$x=2g\mu{B}/\prn{\hbar\gamma_t}$.
%------------------------------------------------------------------

\subsection{Density-matrix calculations}
\label{subsection:DMCalc}

This section describes a nonperturbative density-matrix
calculation \citep{Roc2001SR,Bud2000AOC,Bud2002Bennett} of NMOE
for a transition $\xi J\rightarrow\xi' J'$ in the presence of
nuclear spin $I$. Here $J$ is total electronic angular momentum,
and $\xi$ represents additional quantum numbers.

The calculation is carried out in the collision-free approximation
and assumes that atoms outside the volume of the laser beam are
unpolarized. Thus the calculation is valid for an atomic beam or a
cell without antirelaxation wall coating containing a low-density
vapor. It is also assumed that light is of a uniform intensity
over an effective area, and that the passage of atoms through the
laser beam can be described by one effective relaxation rate
$\gamma_t$.

The time evolution of the atomic density matrix $\rho$ is given by
the Liouville equation \citep[see discussion by, for
example,][]{Ste84}:
%---------------------------------------------------------------
\begin{equation}
\frac{d\rho}{dt}=\frac{1}{i\hbar}\sbrk{H,\rho}-\frac{1}{2}\cbrk{\Gamma_R,\rho}+\Lambda,
  \label{Louiville}
\end{equation}
%--------------------------------------------------------------------
where the square brackets denote the commutator and the curly
brackets the anticommutator. The total Hamiltonian $H$ is the sum
of the light-atom interaction Hamiltonian
$H_L=-\mb{d}\cdot\bs{\mc{E}}$ (where $\bs{\mc{E}}$ is the electric
field vector, and $\mb{d}$ is the electric dipole operator), the
magnetic-field--atom interaction Hamiltonian
$H_B=-\bs{\mb{\mu}}\cdot\mb{B}$ (where $\mb{B}$ is the magnetic
field and $\bs{\mb{\mu}}$ is the magnetic moment), and the
unperturbed Hamiltonian $H_0$. The Hamiltonian due to interaction
with a static electric field can also be included if necessary.
$\Gamma_R$ is the relaxation matrix (diagonal in the
collision-free approximation)
%--------------------------------------------------------------------
\begin{equation}
\begin{gathered}
  \bra{{\xi}JFM}\Gamma_R\ket{{\xi}JFM}=\gamma_t,\\
  \bra{\xi'J'F'M'}\Gamma_R\ket{\xi'J'F'M'}=\gamma_t+\gamma_0.
\end{gathered}
\end{equation}
%--------------------------------------------------------------------
$\Lambda=\Lambda^0+\Lambda^{repop}$ is the pumping term, where the
diagonal matrix
%--------------------------------------------------------------------
\begin{equation}
\bra{{\xi}JFM}\Lambda^t\ket{{\xi}JFM}=\frac{\gamma_t\rho_0}{\prn{2I+1}\prn{2J+1}}
\end{equation}
%--------------------------------------------------------------------
describes incoherent ground-state pumping ($\rho_0$ is the atomic
density), and
%--------------------------------------------------------------------
\begin{widetext}
\begin{equation}
\begin{split}
\bra{{\xi}JFM_1}\Lambda^{repop}\ket{{\xi}JFM_2}=\gamma_0&\sum_{F'}\prn{2J'+1}\prn{2F+1}\sixj(J,F,I)(F',J',1)^2\\
&\times\sum_{M'_1,M'_2,q}\cg{J,M_1,1,q}{J',M'_1}\cg{J,M_2,1,q}{J',M'_2}\rho_{\xi'J'F'M'_1,\xi'J'F'M'_2}
\end{split}
\end{equation}
\end{widetext}
%--------------------------------------------------------------------
describes repopulation due to spontaneous relaxation from the
upper level \citep[see, for example, discussion by][]{Rautian}.
Here $\cg{\ldots}{\ldots}$ are the Clebsch-Gordan coefficients and
the curly brackets represent the six-J symbol. A solution for the
steady-state density matrix can be found by applying the
rotating-wave approximation and setting $d\rho/dt=0$. For the
alkali-atom $D$ lines, one can take advantage of the well-resolved
ground-state hyperfine structure by treating transitions arising
from different ground-state hyperfine sublevels separately and
then summing the results.

The electric field vector is written \citep[see discussion by, for
example,][]{Huard}
%--------------------------------------------------------------------
\begin{align}\label{lightfield}
\bs{\mc{E}}=&\frac{1}{2}\sbrk{\mc{E}_0e^{i\phi}\prn{\cos\varphi\cos\epsilon-i\sin\varphi\sin\epsilon}e^{i\prn{\omega{t}-kz}}+c.c.}\mb{\hat{x}}\nonumber\\
&{}+\frac{1}{2}\sbrk{\mc{E}_0e^{i\phi}\prn{\sin\varphi\cos\epsilon+i\cos\varphi\sin\epsilon}e^{i\prn{\omega{t}-kz}}+c.c.}\mb{\hat{y}},\nonumber\\
\end{align}
%--------------------------------------------------------------------
where $\omega$ is the light frequency, $k=\omega/c$ is the vacuum
wave number, $\mc{E}_0$ is the electric field amplitude, $\varphi$
is the polarization angle, $\epsilon$ is the ellipticity
(arctangent of the ratio of the major and minor axes of the
polarization ellipse), and $\phi$ is the overall phase (the field
can also be written in terms of the Stokes parameters; see
Appendix \ref{ApStokes}). By substituting (\ref{lightfield}) into
the wave equation
%--------------------------------------------------------------------
\begin{equation}
\prn{\frac{\omega^2}{c^2}+\frac{d^2}{dz^2}}\bs{\mc{E}}=-\frac{4\pi}{c^2}\frac{d^2}{dt^2}\mb{P},
\end{equation}
%--------------------------------------------------------------------
where $\mb{P}=\tr\rho\mspace{1mu}\mb{d}$ is the polarization of
the medium, the absorption, rotation, phase shift, and change of
ellipticity per unit distance for an optically thin medium can be
found in terms of the density-matrix elements:
%--------------------------------------------------------------------
\begin{widetext}
\begin{subequations}
\begin{align}
\frac{d\varphi}{dz}&=-\frac{2\pi\omega}{\mc{E}_0c}\sec2\epsilon\sbrk{\cos\varphi\prn{P_1\sin\epsilon+P_4\cos\epsilon}+\sin\varphi\prn{-P_2\cos\epsilon+P_3\sin\epsilon}},\\
\frac{d\mc{E}_0}{dz}&=-\frac{2\pi\omega}{c}\sbrk{\sin\varphi\prn{-P_1\sin\epsilon+P_4\cos\epsilon}+\cos\varphi\prn{P_2\cos\epsilon+P_3\sin\epsilon}},\\
\frac{d\phi}{dz}&=-\frac{2\pi\omega}{\mc{E}_0c}\sec2\epsilon\sbrk{\cos\varphi\prn{P_1\cos\epsilon+P_4\sin\epsilon}+\sin\varphi\prn{-P_2\sin\epsilon+P_3\cos\epsilon}},\\
\frac{d\epsilon}{dz}&=\frac{2\pi\omega}{\mc{E}_0c}\sbrk{\sin\varphi\prn{P_1\cos\epsilon+P_4\sin\epsilon}+\cos\varphi\prn{P_2\sin\epsilon-P_3\cos\epsilon}},
\end{align}
\end{subequations}
\end{widetext}
%--------------------------------------------------------------------
where the components of polarization are defined by
%--------------------------------------------------------------------
\begin{equation}
\begin{split}
 \mb{P}=&\frac{1}{2}\sbrk{\prn{P_1-iP_2}e^{i\prn{\omega{t}-kz}}+c.c.}\mb{\hat{x}}\\
 &{}+\frac{1}{2}\sbrk{\prn{P_3-iP_4}e^{i\prn{\omega{t}-kz}}+c.c.}\mb{\hat{y}}.
\end{split}
\end{equation}
%--------------------------------------------------------------------
Since we neglect collisions between atoms, the solutions for a
Doppler-free medium can be generalized to the case of Doppler
broadening by simply convolving the calculated spectra with a
Gaussian function representing the velocity distribution. In
addition, an integration along the light path can be performed to
generalize the thin-medium result to media of arbitrary optical
thickness.

Results of such calculations have been instrumental for
understanding of various phenomena discussed throughout this
Review (see, for example, Figs.\ \ref{fig:Rediscovery} and
\ref{fig:AOCdata_theor}).

\section{Nonlinear magneto-optical effects in specific situations}
\label{section:NMOEinSpecSit}

In preceding sections, we have outlined the basic NMOE
phenomenology and the methods used for theoretical description of
these effects. In this section, we discuss the specific features
of NMOE in a wide variety of physical situations.

\subsection{Buffer-gas-free uncoated vapor cells}
\label{subsect:buff_gas_free_cells}

\subsubsection{Basic features}

There have been numerous studies of NMOR in buffer-gas-free vapor
cells without antirelaxation coating [by, for example,
\citet{Bark89b}; \citet{Kan93}; \citet{Bud2000AOC} and references
therein]. In alkali-atom cells at room temperature, the atomic
density is low enough so that the mean free path of the atoms is
orders of magnitude larger than the cell dimensions. In addition,
because collisions with the walls of the cell destroy atomic
polarization, atoms entering the light beam are unpolarized. In
such cells, the coherence NMOR (Sec.\ \ref{subsection:CohEf}) is
due to the ``transit'' effect---the pumping, precession, and
subsequent probing of atoms during a single pass through the laser
beam [Bennett-structure-related NMOR (Sec.\
\ref{subsection:BenStr}) can also be observed in these cells].
Only the region illuminated by the light beam need be considered
in a calculation, and the relaxation rate of atomic coherence is
determined by the transit time of atoms through the beam. Since
velocity-changing collisions rarely occur, the atoms have a
constant velocity as they make a single pass through the light
beam. Thus when calculating NMOR spectra (Sec.\
\ref{section:TheorMod}), averaging over the atomic velocity
distribution is equivalent to averaging over the Doppler-free
spectral profiles. At sufficiently low light power, the coherence
transit effect is well described by the Kanorsky-Weis model (Sec.\
\ref{subsection:KanWeiTheory}). At higher light powers, where
saturation effects and the alignment-to-orientation conversion
effect (Sec.\ \ref{subsection:AOC}) are important, a
density-matrix approach (Sec.\ \ref{subsection:DMCalc}) can be
used to describe both the coherence and Bennett-structure effects
in uncoated cells.

In addition to being many orders of magnitude greater for small
fields, NMOR can have a considerably different spectrum than
linear magneto-optical rotation (Sec.\ \ref{section:LinearMO}).
One of the signature features of the low-light-power NMOR transit
effect is the dependence of the sign of optical rotation on the
nature of the transition (Fig.\ \ref{fig:AOCdata_theor}).
%-----------------------------------------------------------------
\begin{figure}
\includegraphics{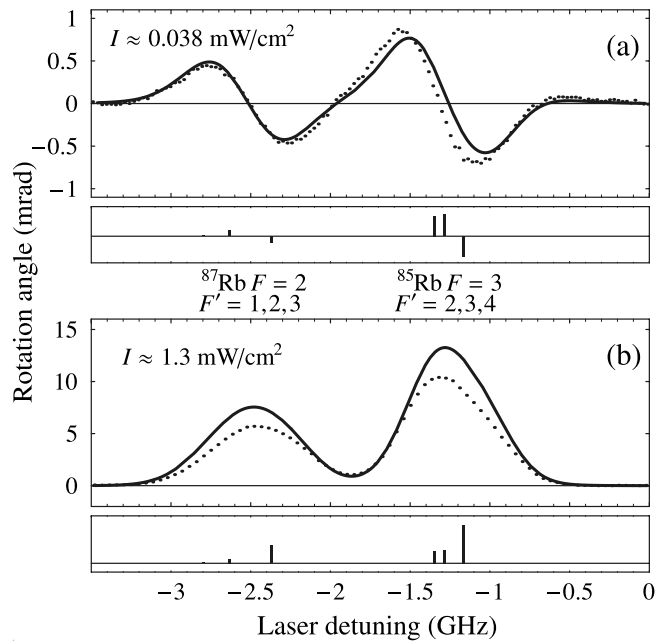}
\caption{Comparison of experimental NMOR spectra obtained by
\citet{Bud2000AOC} to density-matrix calculations performed in the
same work for a natural mixture of Rb isotopes contained in an
uncoated buffer-gas-free vapor cell. Dots -- data points, solid
curves -- theory (without free parameters). Offset vertical bars
indicate the central frequencies and calculated relative
contributions of different hyperfine components
($F\rightarrow{F'}$) to the overall rotation. Laser detuning is
relative to the $D2$ line center. Magnetic field is $\sim$0.1 G
(at which NMOR is relatively large and coherence effects
dominate), laser beam diameter is $\sim$3.5 mm, and Rb density is
$\sim$$10^{10}\ \mr{cm^{-3}}$. Residual discrepancies between data
and theory may be due to nonuniform spatial distribution of light
intensity. Note that at low light power (a), the sign of rotation
for $F\rightarrow{F+1}$ transitions is opposite to that for
$F\rightarrow{F-1,F}$ transitions, whereas at high light power (b)
the sign of rotation is the same for all hyperfine
transitions.}\label{fig:AOCdata_theor}
\end{figure}
%-----------------------------------------------------------------
The sign of optical rotation for $F\rightarrow{F-1,F}$ transitions
is opposite to that for closed $F\rightarrow{F+1}$ transitions
(for ground states with the same Land\'{e} factors). This is
because linearly polarized light resonant with
$F\rightarrow{F-1,F}$ transitions pumps atoms into an aligned dark
state (see Sec.\ \ref{subsection:CohEf}), while atoms are pumped
on $F\rightarrow{F+1}$ transitions into a bright state which
interacts with the light field more strongly \citep{Kaz84}. Thus
when the magnetic field causes the ground-state alignment to
precess, the transmission of light polarized along the initial
axis of light polarization decreases for $F\rightarrow{F-1,F}$
transitions, and increases for $F\rightarrow{F+1}$ transitions. By
considering the ``rotating polarizer'' model (Fig.\
\ref{fig:RotPol}), one can see that the sign of rotation is
opposite in the two cases (for an $F\rightarrow{F+1}$ transition,
the initial ``polarizer'' is crossed with the light polarization).
At higher light powers at which alignment-to-orientation becomes
the dominant mechanism for NMOR (Sec.\ \ref{subsection:AOC}), the
sign of optical rotation is the same for all types of transitions.

\subsubsection{Peculiarities in the magnetic-field dependence}

All theories described in the literature predicted the shape of
the $\varphi(B)$ dependence of the NMOR transit effect signal to
be a dispersive Lorentzian [Eq.\ (\ref{Eqn:KanWei_phi_vs_B})],
linear near the zero crossing at $B=0$. However, experimental
measurement of $\varphi(B)$ for small $B$ by \citet{Che89}
revealed a slight deviation from this anticipated behavior: the
signal had higher than expected derivative at $B=0$.

This observation was eventually explained as follows. The shape of
$\varphi(B)$ was expected to be Lorentzian as a consequence of an
assumed exponential relaxation of atomic coherences in time. In
room-temperature buffer-gas-free vapor cells, however, the width
of the NMOR resonance was determined by the \emph{effective}
relaxation due to the finite interaction time of the pumped atoms
with the light beam. The nonexponential character of this
effective relaxation led to the observed discrepancy. The
simplicity of the theoretical model of the NMOR transit effect
developed by \citeauthor{Kan93}\ (Sec.
\ref{subsection:KanWeiTheory}) allowed them to treat these
time-of-flight effects quantitatively by averaging over the
Gaussian-distributed light beam intensity and the Maxwellian
velocity distribution. Their approach produced quasianalytical
expressions for $\varphi(B)$ which were beautifully confirmed in
an experiment by \citet{Pfl93}.

\subsubsection{NMOE in optically thick vapors}
\label{subsubsection:OptThick}

For a given medium used in studies of near-resonant NMOE, the
optimum optical thickness depends on the light power employed; if
most of the light is absorbed upon traversal of the medium, there
is a loss in statistical sensitivity. At low light power, when the
saturation parameter $\kappa$ (Sec.\ \ref{subsection:satpar}) is
not much greater than unity, it is generally useful to use samples
of no more than a few absorption lengths. However, at high light
powers, the medium may bleach due to optical pumping, and greater
optical thicknesses may be used to enhance the NMOE without unduly
compromising statistical sensitivity. Nonlinear Faraday rotation
produced by a optically thick vapor of rubidium was studied by
\citet{Sau2000,NovLargePol2001,MatRadTrap2001,Nov2002Mag}.
\citet{NovLargePol2001} used a 5-cm-long vapor cell containing
$^{87}$Rb and a 2.5-mW, 2-mm-diameter laser beam tuned to the
maximum of the NMOR $D1$-line spectrum. They measured maximum
polarization rotation $\varphi_\text{max}$ (with respect to laser
frequency and magnetic field) as a function of atomic density. It
was found that $\varphi_\text{max}$ increases essentially linearly
up to $n\simeq3.5\times10^{12}\ \mr{cm^{-3}}$. At higher
densities, $d\varphi_\text{max}/dn$ decreases and eventually
becomes negative. The maximum observed rotation, $\sim$10 rad, was
obtained with magnetic field $\sim$0.6 G. (This shows the effect
of power broadening on the magnetic-field dependence; at low light
power, the maximum rotation would occur at a magnetic field about
an order of magnitude smaller.)

A model theory of NMOR in optically thick media, accounting for
the change in light power and polarization along the light path
and for the high-light-power effects (see Sec.\
\ref{subsection:AOC}), was developed by several authors
\citep[e.g,][and references therein]{Mat2002Rad,Roc2002}. However,
a complete theoretical description of NMOR in dense media must
also consider radiation trapping of photons spontaneously emitted
by atoms interacting with the laser light.
\citet{MatRadTrap2001,Mat2002Rad} concluded that for densities
such that the photon absorption length is comparable to the
smallest dimension of the cell, radiation trapping leads to an
increase in the effective ground-state relaxation rate.

\subsection{Time-domain experiments}
\label{subsection:TimeDomain}

In many NMOE experiments, the same laser beam serves as both pump
and probe light. More complex experiments use separate pump and
probe beams that can have different frequencies, spatial
positions, directions of propagation, polarizations, and/or
temporal structures. A wealth of pump-probe experimental
techniques in the time and frequency domains have been developed,
including spin nutation, free induction decay, spin echoes,
coherent Raman beats or Raman heterodyne spectroscopy, to name
just a few.\footnote{For a detailed discussion we refer the reader
to a comprehensive overview of the underlying mechanisms and
applications in a book by \citet{Sut97}.} In this section, we give
an example of an experiment in the time domain; in Sec.\
\ref{subsection:FRS} we discuss frequency domain experiments with
spatially separated light fields.

\citet{Zibrams2001,ZibRAms2002} applied two light pulses to a
$^{87}$Rb vapor in sequence, the first strong and linearly
polarized and the second weak and circularly polarized. A
particular elliptical polarization of the transmitted light from
the second pulse was detected. When a longitudinal magnetic field
was applied to the cell, the resulting intensity displayed
oscillations (within the duration of the probe pulse) at twice the
Larmor frequency. The authors interpreted these experiments in
terms of Raman-scattering of the probe light from atoms with
coherent Zeeman-split lower-state sublevels.

Here we point out that the ``rotating polarizer'' picture of
coherence NMOR discussed in Sec.\ \ref{subsection:CohEf} can be
used to provide an equivalent description of these experiments.
The linearly polarized pump pulse induces alignment and
corresponding linear dichroism of the atomic medium, i.e.,
prepares a polaroid out of the medium. The polaroid proceeds to
rotate at the Larmor frequency in the presence of the magnetic
field. Circularly polarized probe light incident on such a
rotating polaroid produces linearly polarized light at the output
whose polarization plane rotates at the same frequency. This is
the same as saying there are two circularly polarized components
with a frequency offset.

\subsection{Atomic beams and separated light fields, Faraday-Ramsey Spectroscopy}
\label{subsection:FRS}

In the coherence effects (Sec.\ \ref{subsection:CohEf}), the
resonance width is determined by the relaxation rate of
ground-state polarization. In the transit effect, in particular,
this relaxation rate is given by the time of flight of atoms
between optical pumping and probing. When one laser beam is used
for both pumping and probing, this time can be increased,
narrowing the resonance, by increasing the diameter of the beam.
Another method of increasing the transit time, however, is to use
two beams, spatially separating the pump and probe regions. For
this technique, the model (Sec.\ \ref{subsection:CohEf}) of the
coherence effects in terms of three sequential steps (pumping,
precession, probing) becomes an exact description.

As this experimental setup (Fig.\ \ref{fig:SepBeams}) bears some
similarities to the conventional Ramsey arrangement (see Sec.\
\ref{subsection:NMOR_Ramsey_Connection}) and as the coherence
detection in the probe region involves polarimetric detection of
polarization rotation, this technique is called
\emph{Faraday-Ramsey spectroscopy}.
%-------------------------------------------------------------
\begin{figure}
\includegraphics{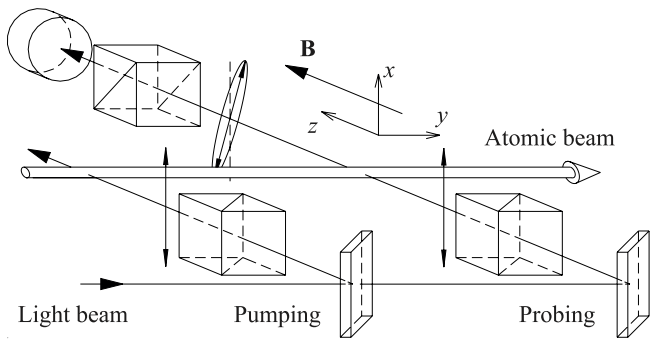}
\caption{Faraday rotation technique utilizing separated light
beams and an atomic beam. The resonance line width is determined
by the time-of-flight between the pump and the probe beams.
Variations of this technique include the use of circularly
polarized light both for pumping and probing, and the detection of
absorption or induced ellipticity instead of polarization
rotation.}\label{fig:SepBeams}
\end{figure}
%-------------------------------------------------------------

\subsubsection{Overview of experiments}

Several experiments studying NMOE using spatially separated light
fields have been performed, using various detection techniques.
The first atomic beam experiment of this sort was performed by
\citet{Sch74} with a Na beam using fluorescence detection.
\citet{Mly88} observed Ramsey fringes in a Sm beam on the
570.68-nm $F=1\rightarrow{F=0}$ transition using \emph{Raman
heterodyne spectroscopy}, a technique involving application of
both light and rf magnetic fields to the atoms \citep[see, for
example, a book by][]{Sut97}, with light beam separations of up to
$L=2.2$ cm. The fringe widths were found to scale as $L^{-1}$ with
the beam separation, as anticipated, but signal quality was
strongly degraded as $L$ increased \citep{Mly87}.\footnote{Other
examples of early work in the field are the work of \citet{Ber86},
who used fluorescence detection to observe nonlinear
level-crossing Ramsey fringes in a metastable Ca beam, the work of
\citet{Nak80}, who observed Ramsey fringes on the $D1$ line in a
Na vapor cell using laser beams separated by up to 10 mm, and an
experiment performed by \citet{Bor84} in the time domain using a
monokinetic Ba$^+$ beam, which showed a large number of
higher-order fringes, normally damped in thermal beams due to the
large velocity spread. This last experiment determined the ratio
of $g$-factors of the $5d\;^2\!D_{3/2}$ and $5d\;^2\!D_{5/2}$
states of Ba$^+$.}

The first beam experiment with a long (34.5-cm) precession region
was done by \citet{The91} using $D2$-line excitation on a Cs beam
with fluorescence detection. In similar work, using detection of
forward-scattered light rather than fluorescence monitoring,
\citet{Sch93,Wei93EDM} applied pump-probe spectroscopy to a Rb
beam using $D2$-line excitation. The experiments were performed
with interaction lengths of $L=5$ cm \citep{Sch93} and $L=30$ cm
\citep{Wei93EDM}, respectively. Improving the sensitivity by going
to longer interaction lengths is difficult because of signal loss
due to the finite atomic-beam divergence. This can be counteracted
by laser collimation of the beam (this technique is being used in
work towards a 2-m apparatus at the University of Fribourg), or by
using atoms that are free-falling from a magneto-optical trap or
an atomic fountain (Sec.\ \ref{subsect:AtomTraps}). Hypersonic and
laser-slowed beams can also be used to increase the atomic flux
and transit time, respectively.

\subsubsection{Line shape, applications}

In Faraday-Ramsey spectroscopy, one measures the magneto-optical
rotation angle $\varphi$ of the linearly polarized probe beam. For
a beam with Maxwellian velocity distribution, the magnetic-field
dependence of $\varphi$ is obtained from the three-step model
discussed in Sec.\ \ref{subsection:CohEf} in a manner similar to
that described in Sec.\ \ref{subsection:KanWeiTheory}:
%-------------------------------------------------------------
\begin{equation}
\varphi\prn{B}\propto\int_0^\infty\sin\prn{\frac{B}{\mc{B}_0^R}\frac{1}{u}+2\theta_\text{pp}}u^{2}e^{-u^2}du.
\label{eq:FRSshape}
\end{equation}
%-------------------------------------------------------------
where $\theta_\text{pp}$ is the relative orientation of the pump
and probe polarizations, $u=v/v_0$ is the dimensionless velocity,
and $v_0$ is the most probable velocity in the atomic beam oven,
and
%-------------------------------------------------------------
\begin{equation}
\mc{B}_0^R=\frac{\hbar{v_0}}{2g\mu{L}} \label{eq:B0RAM}
\end{equation}
%-------------------------------------------------------------
is a scaling field ($L$ is the distance between pump and probe
beams). The line shape is a damped oscillatory Ramsey fringe
pattern centered at $B=0$ (the scaling field $\mc{B}_0^R$ is a
measure of the fringe width). Figure \ref{fig:FarRamSignal} shows
an experimental Faraday-Ramsey spectrum recorded with a thermal Cs
beam ($T=400$ K) with a precession region of $L=30$ cm, for which
$\mc{B}_0^R=170\ \mr{\mu G}$.
%-------------------------------------------------------------
\begin{figure}
\includegraphics[width=3in]{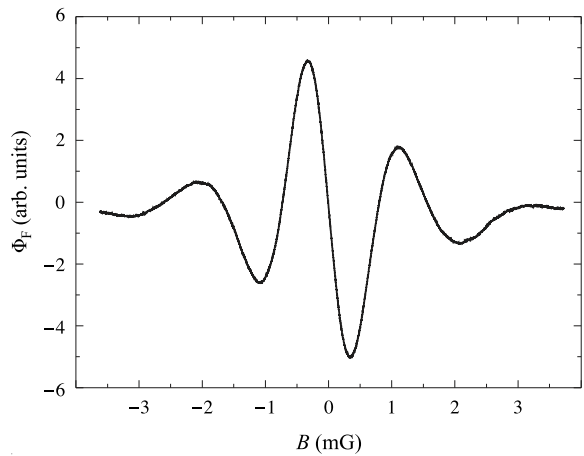}
\caption{Experimental Faraday-Ramsey signal from a thermal Cs beam
($T=400$ K) and a precession length of $L=30$ cm. Polarimetric
detection of the rotation angle $\varphi$ of the probe beam is
used.}\label{fig:FarRamSignal}
\end{figure}
%-------------------------------------------------------------

Faraday-Ramsey spectroscopy has proven to be an extremely
sensitive tool for investigating spin-coherence perturbations by
magnetic and/or electric\footnote{Electric interactions do not
couple to spin directly; however, they affect the orbital angular
momentum. Thus spin coherences are affected by an electric field
via spin-orbit and hyperfine interactions.} fields under
collision-free conditions. Any additional external perturbation
which affects spin coherence will lead to an additional phase
shift in the fringe pattern given by Eq.\ (\ref{eq:FRSshape}).
Faraday-Ramsey spectroscopy has been applied in a variety of
situations, including an investigation of the Aharonov-Casher
effect in atoms (Sec.\ \ref{ExV}), a precision electric field
calibration \citep{Ras2001}, and a measurement of the forbidden
electric tensor polarizabilities (Sec.\ \ref{subsect:TensPol}) of
the ground state of Cs \citep{Ras2001,Wei2001}. Applications of
the technique of Faraday-Ramsey spectroscopy with atomic beams to
the search for parity- and time-reversal-invariance violation, in
particular, for a permanent electric dipole moment (EDM, Sec.\
\ref{SectEDM}) have also been discussed \citep{Sch93,Wei93EDM}.
Note, however, that a competitive EDM experiment would need a very
intense optically thick atomic beam, such as the source built for
experiments with atomic thallium by \citet{Dem94a}.

\subsubsection{Connection with the Ramsey separated-oscillatory-field method}
\label{subsection:NMOR_Ramsey_Connection}

Faraday-Ramsey spectroscopy is in fact a Ramsey double-resonance
technique \citep{Ram90} without oscillating fields. In
conventional Ramsey spectroscopy, a spin-polarized beam traverses
two identical spatially separated regions in which the particles
are exposed to phase-locked radio-frequency magnetic fields
$B_1\prn{t}$ oscillating at the same frequency. The whole
arrangement is in a homogeneous field $B_0$. In the first rf
region a $\pi/2$-pulse tips the spins to an orientation
perpendicular to $B_0$. After precessing freely the spin is
projected back onto its original direction in the second region.
The rf regions thus play the role of start/stop pulses for the
spin precession. The polarization recovered in the second rf
region depends on the phase accumulated during the precession. In
Faraday-Ramsey spectroscopy, the start pulse is provided by an
optical-pumping light pulse (in the atomic frame of reference)
which orients the spin so that it can precess, and the weak probe
pulse detects the precession angle. In Ramsey spectroscopy the
relative phase $\phi_\text{rf}$ of the rf fields determines the
symmetry of the fringe pattern (absorptive for $\phi_\text{rf}=0$,
dispersive for $\phi_\text{rf}=\pi/2$), while in Faraday-Ramsey
spectroscopy, this role is played by the relative orientations of
the light polarizations $\theta_\text{pp}$.

\subsection{Experiments with buffer-gas cells}
\label{subsection:BufGasCells}

\subsubsection{Warm buffer gas}
\label{subsubsection:WarmBufGas}

Early studies of NMOR by \citet{Davies87,Bai89,Bark89b} on
transitions in Sm showed that nonlinear effects are extremely
sensitive to the presence of buffer gas. The
Bennett-structure-related effects (Sec.\ \ref{subsection:BenStr})
are suppressed when the peaks and holes in the velocity
distribution are washed out by velocity-changing collisions.
Generally, coherence effects (Sec.\ \ref{subsection:CohEf}) are
also easily destroyed by depolarizing buffer-gas collisions.
However, a very important exception is effects with atoms with
total electronic angular momentum in the ground state $J=1/2$,
e.g., the alkali atoms. For such atoms, the cross-sections for
depolarizing buffer-gas collisions are typically 4 to 10 orders of
magnitude smaller than those for velocity-changing collisions
\citep{Wal89,Wal97}. This suppression allows long-lived
ground-state polarization in the presence of buffer gas and also
serves to prevent depolarizing wall collisions \citep{Hap72}.
Recently, coherent dark resonances (Sec.\
\ref{subsection:lambdares}) with widths as narrow as 30 Hz were
observed by \citet{Bra97} in Cs and by \citet{Erh2000,Erh2001} in
Rb using neon-buffer-gas vapor cells at room temperature.

\citet{Zibrams2001,ZibRAms2002,NovAc2001,Nov2002Mag} have recently
studied NMOE in buffer-gas cells. Spectral line shapes of optical
rotation quite different from those seen in buffer-gas-free cells
and atomic beams were observed \citep{NovAc2001}. This effect,
discussed in Sec.\ \ref{subsect:ARCoatedCells:Exp}, is due to
velocity-changing collisions and was previously observed by
\citet{Bud2000Sens} in paraffin-coated cells. \citet{Nov2002Imp}
proposed the use of the dependence of the NMOR spectral lineshape
on buffer gas to detect the presence of small amounts of
impurities in the resonant medium.

\subsubsection{Cryogenic buffer gas}
\label{subsubsection:ColdBufGas}

Cold-buffer-gas systems are currently under investigation by
\citet{Hat2000,Yas2002Drop,Kim2001,Hat2002} as a method of
reducing collisional relaxation. At low temperatures, atomic
collisions approach the S-wave scattering regime, in which spin
relaxation is suppressed. Theoretical estimates of spin relaxation
in spin-rotation interactions by \citet{Sus2001}, based on the
approach of \citet{Wal97}, suggest that for the Cs-He case, a
reduction of the relaxation cross-section from its room
temperature value by a factor of $\sim$20--50 may be expected at
liquid-Nitrogen temperatures. Recently, \citet{Hat2000} showed
experimentally that at temperatures below $\sim$2 K, the spin
relaxation cross-section of Rb atoms in collisions with
He-buffer-gas atoms is orders of magnitude smaller than its value
at room temperature. The results of \citeauthor{Hat2000}\ imply
that relaxation times of minutes or even longer (corresponding to
resonance widths on the order of 10 mHz) can be obtained. Such
narrow resonances may be applied to atomic tests of discrete
symmetries (Sec.\ \ref{SectEDM}) and to very sensitive
measurements of magnetic fields (Sec.\ \ref{subSect:Magnetom}).

The crucial experimental challenge with these systems is creating,
in the cold buffer gas, atomic vapor densities comparable to those
of room temperature experiments, i.e., on the order of
$10^{10}$--$10^{12}\ \mr{cm^{-3}}$. To accomplish this,
\citeauthor{Hat2000}\ use light-induced desorption of alkali atoms
from the surface of the liquid He film inside their cell [this is
presumably similar to light-induced desorption observed with
siloxane \citep[by, for example,][]{Atu99} and paraffin
\citep{AleLIAD,Kim2001} antirelaxation coatings].
\citeauthor{Hat2000}\ can successfully\footnote{It turns out that
the injection efficiency decreases with repetitions of the
injection cycle. It is found that the efficiency recovers when the
cell is heated to room temperature and then cooled again.
\citet{Hat2002} report that they cannot inject Cs atoms with their
method.} inject Rb atoms into the He gas by irradiating the cell
with about 200 mW of (750-nm) Ti:sapphire laser radiation for 10
s. \citet{Yas2002Drop} are exploring a significantly different
method of injecting atoms into cold He buffer gas---laser
evaporation of micron-sized droplets falling through the gas
\citep[see also discussion by][]{Kim2001}.

\subsection{Antirelaxation-coated cells}
\label{subsect:ARCoatedCells}

Antirelaxation-coated cells provide long spin-relaxation times due
to the greatly reduced rate of depolarization in wall collisions.
Three nonlinear effects can be observed ``nested'' in the
magnetic-field dependence of NMOR produced by such a cell (Fig.\
\ref{fig:ThreeEffects}).
%-------------------------------------------------------------------
\begin{figure}
\includegraphics{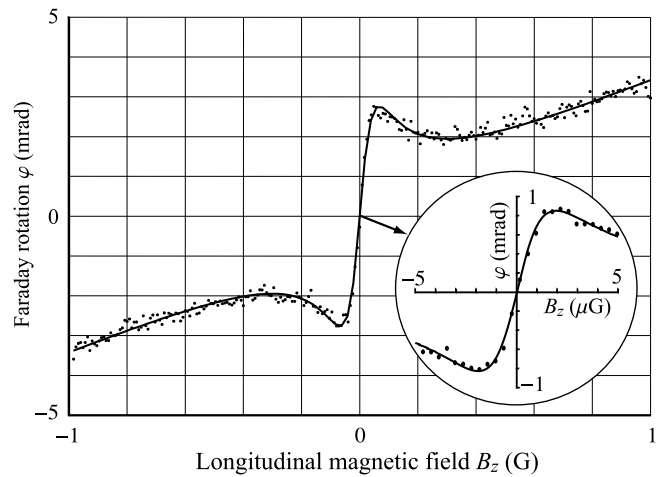}
\caption{Longitudinal magnetic-field dependence of optical
rotation in a paraffin-coated $^{85}$Rb-vapor cell \citep{Bud98}.
The background slope is due to the Bennett-structure effect. The
dispersion-like structure is due to the transit effect. The inset
shows the near-zero $B_z$-field behavior at a $2\times10^5$
magnification of the magnetic-field scale. Light intensity is
$\sim$$100\ \mr{\mu{W}\,cm^2}$. The laser is tuned $\sim$150 MHz
to the high frequency side of the $F=3\rightarrow{F'}$ absorption
peak.}\label{fig:ThreeEffects}
\end{figure}
%-------------------------------------------------------------------
The widest feature is due to Bennett-structure effect (Sec.\
\ref{subsection:BenStr}), followed by the feature due to the
transit effect, which also occurs in uncoated cells (Sec.\
\ref{subsect:buff_gas_free_cells}). The narrowest feature is due
to the wall-induced Ramsey effect, a variant of the
separated-field transit effect in which atoms leave the light beam
after being optically pumped and are later probed after colliding
with the cell walls and returning to the beam.

\subsubsection{Experiments}
\label{subsect:ARCoatedCells:Exp}

 In their early work on optical
pumping, \citet{Rob58} showed that by coating the walls of a vapor
cell with a chemically inert substance such as paraffin (chemical
formula $\mr{C}_n\mr{H}_{2n+2}$), the relaxation of atomic
polarization due to wall collisions could be significantly
reduced.

Recently, working with a paraffin-coated Cs vapor cell,
\citet{Kan95} discovered a narrow feature (of width $\sim$1 mG) in
the magnetic-field dependence of Faraday rotation. \citet{Kan95}
described the feature as a Ramsey resonance induced by multiple
wall collisions.\footnote{It is interesting to note that Ramsey
himself \citep*{Kle58}, in order to decrease the resonance widths
in experiments with separated oscillatory fields, constructed a
``storage box'' with Teflon-coated walls in which atoms would
bounce around for a period of time before emerging to pass through
the second oscillatory field.} In antirelaxation-coated cells, the
precession stage of the three-step coherence-effect process (Sec.\
\ref{subsection:CohEf}) occurs after the atom is optically pumped
and then leaves the light beam. The atom travels about the cell,
undergoing many---up to $10^4$ [\citet{BouBro66}; \citet{Ale92};
\citet{AlexandrovLPh96}], velocity-changing wall
collisions---before it flies through the light beam once more and
the probe interaction occurs. Thus the time between pumping and
probing can be much longer for the wall-induced Ramsey effect than
in the transit effect (Sec.\ \ref{subsect:buff_gas_free_cells}),
leading to much narrower features in the magnetic-field dependence
of NMOR.

\citet{Bud98} performed an investigation of the wall-induced
Ramsey effect in NMOR using Rb atoms contained in paraffin-coated
cells \citep{AlexandrovLPh96}. The apparatus employed in these
investigations is discussed in detail in Sec.\
\ref{SS:TypNMOEset}. \citet{Bud98} observed $\sim$1-$\mr{\mu
G}$-width features in the magnetic-field dependence of NMOR (Fig.
\ref{fig:ThreeEffects}).

\citet{Bud2000Sens} investigated the dependence on atomic density
and light frequency and intensity of NMOR due to the wall-induced
Ramsey effect. The NMOR spectra for the wall-induced Ramsey effect
were found to be quite different from those for the transit effect
(Fig.\ \ref{fig:D1sens}).
%----------------------------------------------------------------
\begin{figure}
\includegraphics{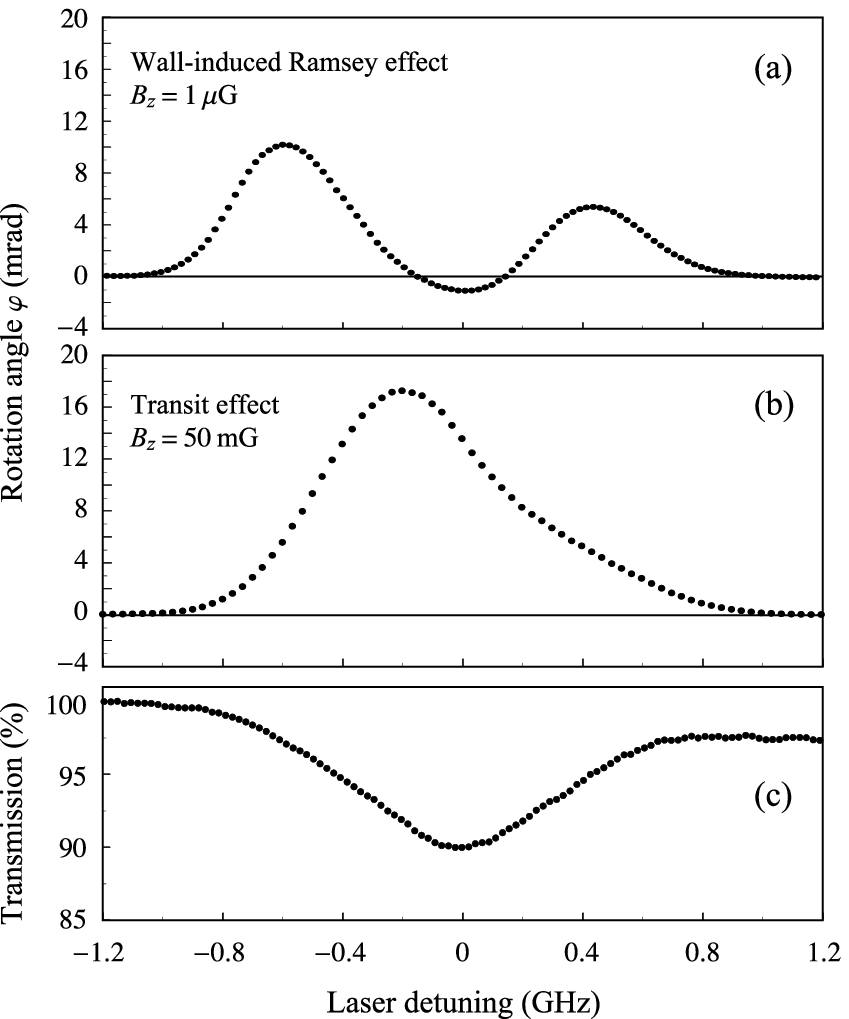}
\caption{(a) Wall-induced Ramsey rotation spectrum for the
$F=3\rightarrow F'$ component of the $D1$ line of $^{85}$Rb
obtained by \citet{Bud2000Sens} for light intensity 1.2
$\mr{mW\,cm^{-2}}$ and beam diameter $\sim$3 mm. (b) Transit
effect rotation spectrum, for light intensity 0.6
$\mr{mW\,cm^{-2}}$. (c) Light transmission spectrum for light
intensity 1.2 $\mr{mW\,cm^2}$. Background slope in light
transmission is due to change in incident laser power during the
frequency scan.}\label{fig:D1sens}
\end{figure}
%----------------------------------------------------------------
In the wall-induced Ramsey effect, atoms undergo many
velocity-changing collisions between pump and probe interactions.
When Doppler-broadened hyperfine transitions overlap, it is
possible for the light to be resonant with one transition during
pumping and another transition during probing. Both the
ground-state polarization produced by optical pumping and the
effect on the light of the atomic polarization that has evolved in
the magnetic field depend on the nature of the transition.  For
the $F=3\rightarrow{F'=2,3}$ component of the $^{85}$Rb $D1$ line,
the contribution to optical rotation from atoms pumped and probed
on different transitions has opposite sign as that from atoms
pumped and probed on the same transition. Thus the wall-induced
Ramsey NMOR spectrum consists of two peaks, since the
contributions to optical rotation nearly cancel at the center of
the Doppler profile.  In the transit effect for buffer-gas-free
cells, in contrast, atoms remain in a particular velocity group
during both optical pumping and probing. The transit-effect
spectrum has a single peak because for each atom light is resonant
with the same transition during both pumping and probing.

\citet{Bud2000Sens} also found that at the light intensity and
frequency at which highest magnetometric sensitivity is achieved,
the sign of optical rotation is opposite of that obtained for the
low-light-power transit effect. \citet{Bud2000AOC} explained this
as the effect of alignment-to-orientation conversion (Sec.\
\ref{subsection:AOC}) due to the combined action of the optical
electric field and the magnetic field.

\citet{SkallaRam97} studied the wall-induced Ramsey effect in
cells with various geometries (cylindrical, spherical, and
toroidal), and used spatially separated pump and probe fields to
measure Berry's topological phase \citep{Ber84}. \citet{Yas99Sep}
investigated the possibilities of applying the separated optical
field method to improve the sensitivity of NMOR-based
magnetometers (Sec.\ \ref{subSect:Magnetom}).

In addition to their application in precision magnetometry, NMOE
in paraffin-coated cells were investigated in relation to tests of
fundamental symmetries (Sec.\ \ref{SectEDM}) and the study of
light propagation dynamics (Sec.\
\ref{subsection:Slow_Fast_Light}).

\subsubsection{Theoretical analysis}

In order to obtain a theoretical description of NMOE in
paraffin-coated cells, one must extend the density-matrix
calculation of Sec.\ \ref{subsection:DMCalc} to describe both the
illuminated and nonilluminated regions of the cell, and the
effects of velocity-mixing and spin exchange. Expressions for the
effect of alkali-alkali spin-exchange on the density matrix have
been obtained by \citet{Oku94,Oku95a,Val94,Val96}, following work
of \citet{Gro65}. When atomic orientation is nonzero, the
expressions become nonlinear; however, even under conditions in
which alignment-to-orientation conversion (Sec.\
\ref{subsection:AOC}) is important for optical rotation, the total
sample orientation can be small enough for linearized
spin-exchange equations to be applied. Taking velocity mixing and
multiple cell regions into account is straightforward but
computationally intensive. A detailed description of such a
calculation and comparison with experiment is in preparation by D.
Budker and coworkers.

\subsection{Gas discharge}
\label{subsect:gas discharge}

Gas discharge allows one to study ionized species, refractory
materials, and transitions originating from metastable states, and
has been extensively used in atomic spectroscopy and polarization
studies in particular [see, for example, \citet{Lom69};
\citet{Ale72}].

In \emph{optogalvanic spectroscopy} one detects light-induced
transitions by measuring the changes in conductivity of a
discharge. The optogalvanic method has been applied to the study
of nonlinear level crossing and other NMOE \citep{Han81,Sta89}.

Other examples of NMOE work employing gas discharge are the study
of \citet{Low87}, who used Zeeman quantum beats in transmission
(i.e., time-dependent NMOR; see Sec.\ \ref{subsection:TimeDomain})
in a pulsed-pump, cw-probe experiment to determine
polarization-relaxation properties of a Sm vapor produced in a
cathode sputtering discharge, and that of \citet{Ali99} who
studied the nonlinear Voigt effect in the
$2\,^3\!P\rightarrow{3\,^3\!D}$ transition in neutral He.

\subsection{Atoms trapped in solid and liquid helium}
\label{subsection:SolLiqHe}

A recent (and experimentally demanding) technique for reducing
spin relaxation of paramagnetic species is the trapping of the
atoms in condensed (superfluid or solid) ${}^4$He.
Solid-rare-gas-matrix isolation spectroscopy \citep{Cou84,Dun98}
has been extensively used by chemists in the past half-century for
the investigation of reactive atoms, ions, molecules, and
radicals. In all heavy-rare-gas matrices, however, the anisotropy
of the local fields at individual atomic-trapping sites causes
strong perturbation of guest atom spin-polarization via spin-orbit
interactions. As a consequence, no high-resolution magneto-optical
spectra have been recorded in such matrices. However, the quantum
nature (large amplitude zero-point motion) of a condensed helium
matrix makes it extremely compressible, and a
single-valence-electron guest atom can impose its symmetry on the
local environment via Pauli repulsion. Alkali atoms thus form
spherical nanometer-sized cavities (atomic bubbles), whose
isotropy, together with the diamagnetism of the surrounding He
atoms, allows long-lived spin polarization to be created in the
guest atoms via optical pumping. With Cs atoms in the cubic phase
of solid He, longitudinal spin relaxation times $T_1$ of 1 s were
reported by \citet{Arn95}, while transverse relaxation times $T_2$
are known to be larger than 100 ms \citep{Kan96}. The $T_1$ times
are presumably limited by quadrupolar zero-point oscillations of
the atomic bubble, while the $T_2$ times are limited by residual
magnetic-field inhomogeneities. In superfluid He, the coherence
life times are limited by the finite observation time due to
convective motion of the paramagnetic atoms in the He matrix
\citep{Kin94}.

Nonlinear magneto-optical level-crossing signals (the longitudinal
and transverse ground-state Hanle effects) were observed by
\citet{Arn95,Wei95} as rather broad lines due to strong
magnetic-field inhomogeneities. In recent years, spin-physics
experiments in solid helium have used
(radio-frequency/microwave--optical) double resonance techniques.
Optical and magneto-optical studies of defects in condensed helium
were reviewed by \citet{Kan98}.

Applications of spin-polarized atoms in condensed helium include
the study of the structure and dynamics of local trapping sites in
quantum liquids/solids, and possible use of such samples as a
medium in which to search for a permanent atomic electric-dipole
moment (\citet{Wei97}; Sec.\ \ref{SectEDM}).

\subsection{Laser-cooled and trapped atoms}
\label{subsect:AtomTraps}

Recently, atomic samples with relatively long spin-relaxation
times were produced using laser-cooling and trapping techniques.
When studying NMOE in trapped atoms, one must take into account
the perturbation of Zeeman sublevels by external fields (usually
magnetic and/or optical) involved in the trapping technique. Atoms
can be released from the trapping potentials before measurements
are made in order to eliminate such perturbations. Another point
to consider is that Doppler broadening due to the residual thermal
velocities of trapped atoms is smaller than the natural line width
of the atomic transition. Therefore, in the low-power limit,
optical pumping does not lead to formation of Bennett structures
(Sec.\ \ref{subsection:BenStr}) narrower than the width of the
transition.

There have been several recent studies of the Faraday and Voigt
effects using magneto-optical traps (MOTs). \citet{Isa99} trapped
$\sim$$10^8$ $^{85}$Rb atoms in a MOT and cooled them to about
$10\ \mr{\mu{K}}$. A pump beam was then used to spin-polarize the
sample of cold atoms orthogonally to an applied magnetic field of
$\sim$2 mG. Larmor precession of the atoms was observed by
monitoring the time-dependent polarization rotation of a probe
beam. The limiting factor in the observation time ($\sim$11 ms)
for this experiment was the loss of atoms from the region of
interest due to free fall in the Earth's gravitational field.
\citet{Lab2001} studied magneto-optical rotation and induced
ellipticity in $^{85}$Rb atoms trapped and cooled in a MOT that
was cycled on and off. The Faraday rotation was measured during
the brief time ($\sim$8 ms) that the trap was off. The time that
the MOT was off was sufficiently short so that most of the atoms
were recaptured when the MOT was turned back on. \citet{Fra2001}
performed measurements of both the Faraday and Voigt effects
(Sec.\ \ref{section:Intro}) in an ensemble of cold $^7$Li atoms.
Optical rotation of a weak far-detuned probe beam due to
interaction with polarized atoms was also used as a technique for
non-destructive imaging of Bose-Einstein condensates \citep[as in
the work of, for example,][]{MatthewsThesis}.

\citet{Narducci} has recently begun to explore the possibility of
measuring NMOE in cold atomic fountains.  The experimental
geometry and technique of such an experiment would be similar to
the setup for Faraday-Ramsey spectroscopy described in Sec.\
\ref{subsection:FRS}, but the time-of-flight between the pump and
probe regions could be made quite long (a few seconds).

With far-off-resonant blue-detuned optical dipole traps, one can
produce alkali vapors with ground-state relaxation rates $\sim$0.1
Hz [for example, \citet{Dav95} observed such relaxation rates for
ground-state hyperfine coherences]. Correspondingly narrow
features in the magnetic-field dependence of NMOE should be
observable.  However, we know of no experiments investigating NMOE
in far-off-resonant optical dipole traps.

\section{Magneto-optical effects in selective reflection}
\label{section:SelRefl}

\subsubsection{Linear effects}

The standard detection techniques in atomic spectroscopy are the
monitoring of the fluorescence light and of the light transmitted
through the sample. A complementary, though less often used
technique is monitoring the light reflected from the sample, or,
more precisely, from the window-sample interface. The reflected
light intensity shows resonant features when the light frequency
is tuned across an atomic absorption line \citep{Coj54}. The
technique was discovered by \citet{Woo09} and is called
\emph{selective reflection spectroscopy}. A peculiar feature of
selective reflection spectroscopy was discovered after the advent
of narrow-band tunable lasers: when the light beam is near normal
to the interface, the spectral profile of the reflection
coefficient shows a narrow absorptive feature of sub-Doppler width
superimposed on a Doppler-broadened dispersive pedestal
\citep{Woe75}. This peculiar line shape is due to the quenching of
the optical dipole by atom-window collisions, which breaks the
symmetry of the atomic velocity distribution.
%and results in a non-local optical response of the medium.

A theoretical model of this effect developed by \citet{Sch76} for
low vapor pressure and low light intensity yields for the resonant
modification of the reflection coefficient:
%-------------------------------------------------------------
\begin{equation}
\delta{R}\prn{y\!=\!\frac{\Delta}{\Gamma_D}}\propto\re\int_0^\infty\frac{e^{-x^2}}{(x-y)-i\gamma_0/\prn{2\Gamma_D}}\,dx,
\label{SRlineshape}
\end{equation}
%-------------------------------------------------------------
where $\gamma_0$ and $\Gamma_D$ are the homogeneous and Doppler
widths of the transition, respectively, and $\Delta$ is the light
detuning. The expression differs in the lower bound of the
integral from the usual Voigt integral of the index of refraction.
In the limit of $\Gamma_D\gg\gamma$, the frequency derivative of
Eq.\ (\ref{SRlineshape}) is a dispersive Lorentzian of width
$\gamma_0$, which illustrates the Doppler-free linear nature of
the effect. As only atoms in the close vicinity ($\lambda/2\pi$)
of the interface contribute significantly to the reflection
signal, selective reflection spectroscopy is well suited for
experiments in optically thick media and for the study of
atom-window van der Waals interactions.

The first theoretical discussion of magnetic-field-induced effects
in selective reflection was given by \citet{Ser67SRS}. However,
that treatment was incorrect, as it did not take into account the
above-mentioned sub-Doppler features. The first magneto-optical
experiment using selective reflection spectroscopy was performed
by \citet{Sta74SRS} on mercury vapor who studied
magnetic-field-induced excited-state level crossings with
broad-band excitation. The theoretical description of that
experiment was given by \citet{Sch76b} who considered correctly
the modifications induced by window collisions. The linear
magneto-optical circular dichroism and rotation in the selective
reflection spectra from Cs vapor in a weak longitudinal magnetic
field were studied by \citet{Wei93sel}, who recorded the spectral
profiles of both effects at a fixed magnetic field and provided a
theoretical analysis of the observed line shapes. In an experiment
of \citet{Pap94}, the narrow spectral lines in selective
reflection spectroscopy allowed resolution of Zeeman components in
the reflection spectra of $\sigma^+$- and $\sigma^-$-polarized
light in fixed (120--280 G) longitudinal magnetic fields. For
understanding the line shapes observed in both experiments, it
proved to be essential to take into account the wave-function
mixing discussed in Sec.\ \ref{subsection:LinMOEmechs}.

\subsubsection{Nonlinear effects}

To our knowledge the only study of NMOE using selective reflection
spectroscopy was the observation by \citet{Wei92} of ground-state
Zeeman coherences in a magneto-optical rotation experiment on Cs
vapor. In this experiment, the laser light was kept on resonance
and the magnetic field was scanned, revealing the familiar narrow
dispersive resonance of NMOR. The dependence of the width of this
resonance on the light intensity and on the atomic number density
was studied under conditions of high optical opacity.

\section{Linear and nonlinear electro-optical effects}
\label{section:NEOE}

Although linear electro-optical effects in resonant media have
been studied since the 1920s \citep[a comprehensive review was
written by][]{Bon67}, there has been relatively little work on
nonlinear electro-optical effects (NEOE). \citet{Dav88}
experimentally and theoretically investigated linear and nonlinear
light polarization rotation in the vicinity of
$F=0\rightarrow{F'=1}$ transitions in atomic samarium, and
\citet{Fom95} performed additional theoretical studies of NEOE for
$0\rightarrow1$ and $1\rightarrow0$ transitions.  There are many
similarities between NEOE and NMOE: the shift of Zeeman sublevels
due to the applied fields is the fundamental cause of both
effects, the nonlinear enhancement in both cases can arise from
the formation of Bennett structures in the atomic velocity
distribution \citep{Bud2002Bennett} or the evolution of
ground-state atomic polarization, and these mechanisms change the
optical properties of the atomic vapor, thereby modifying the
polarization properties of the light field.

Consider a Doppler-broadened atomic vapor subject to a dc electric
field $\mb{E}$ and light near-resonant with an atomic transition.
The light propagates in a direction orthogonal to $\mb{E}$ and is
linearly polarized along an axis at $45^\circ$ to $\mb{E}$. Due to
the Stark shifts $\Delta_s$, the component of the light field
along $\mb{E}$ sees a different refractive index than the
component of the light field perpendicular to $\mb{E}$. The
difference in the real parts of the refractive indices leads to
linear birefringence (causing the light to acquire ellipticity)
and the difference in the imaginary parts of the refractive
indices leads to linear dichroism (causing optical rotation).

If the light intensity is large enough to form Bennett structures
in the atomic velocity distribution, the narrow features (of width
$\sim\,$$\gamma_0$) in the indices of refraction cause the maximum
ellipticity to become proportional to $\Delta_s/\gamma_0$, rather
than $\Delta_s/\,\Gamma_D$ as in the linear case. Optical rotation
is not enhanced by Bennett structures in this case because it is
proportional to the difference of bell-shaped features in the
imaginary part of the indices of refraction, which is zero for the
resonant velocity group (and of opposite signs for velocity groups
to either side of the resonance).

There are also coherence effects in NEOE for electric fields such
that $\Delta_s\ll\gamma_{\text{rel}}$, which result from the
evolution of optically pumped ground-state atomic polarization in
the electric field (Fig.\ \ref{fig:J1EzMovie}). The maximum
ellipticity induced in the light field due to the coherence effect
is proportional to $\Delta_s/\gamma_{\text{rel}}$.

While much of the recent work on coherence NMOE has been done with
alkali atoms, these are not necessarily the best choice for NEOE
because tensor polarizabilities are suppressed for states with
$J=1/2$ as discussed in Sec.\ \ref{subsect:TensPol}.

\citet{Her86} studied a different kind of NEOE in which an
electric field is applied to a sample of atoms, and normally
forbidden two-photon transitions become allowed due to Stark
mixing. The large polarizabilities of Rydberg states make
two-photon Stark spectroscopy of transitions involving Rydberg
levels an extremely sensitive probe for small electric fields.

\section{Experimental techniques}
\label{section:ExpTech}

In this Section, we give details of some of the experimental
techniques that are employed for achieving long ground-state
relaxation times (atomic beams with separated light-interaction
regions, buffer-gas cells, and cells with antirelaxation wall
coating), as well as the techniques used for sensitive detection
of NMOE and NEOE, especially spectropolarimetry (Sec.\
\ref{subsection:Polarim}).

We begin the discussion by describing a representative
experimental setup (Sec.\ \ref{SS:TypNMOEset}). Using this example
we formulate some general requirements for experimental apparatus
of this sort, [for example, laser frequency tunability and
stability (Sec.\ \ref{subsection:LaserStab}), and magnetic
shielding (Sec.\ \ref{subsection:MagShield})], and outline how
these requirements have been met in practice.

\subsection{A typical NMOE experiment}
\label{SS:TypNMOEset}

Here we give an overview of the Berkeley NMOE apparatus, which has
been used to investigate various aspects of the physics and
applications of narrow
\citep[$\sim$$2\pi\!\times\!1$-Hz,][]{Bud98} resonances, such as
alignment-to-orientation conversion \citep{Bud2000AOC}, self
rotation \citep{Roc2001SR}, and reduced group velocity of light
\citep{BudGroupVel99}. The setup has also been used to investigate
NEOE and to conduct exploratory experiments whose ultimate goal is
testing fundamental symmetries \citep{Yas99,Kim2001}. Due to the
large enhancement of small-field optical rotation produced by the
narrow resonance, this setup is a sensitive low-field magnetometer
\citep{Bud2000Sens}. With a few modifications, this apparatus can
also be applied to extremely high-sensitivity magnetometry in the
Earth-field range \citep{Bud2002FM}.

The Berkeley apparatus is shown in Fig.\ \ref{fig:NMOEsetup}.
%----------------------------------------------------------------
\begin{figure}
\includegraphics{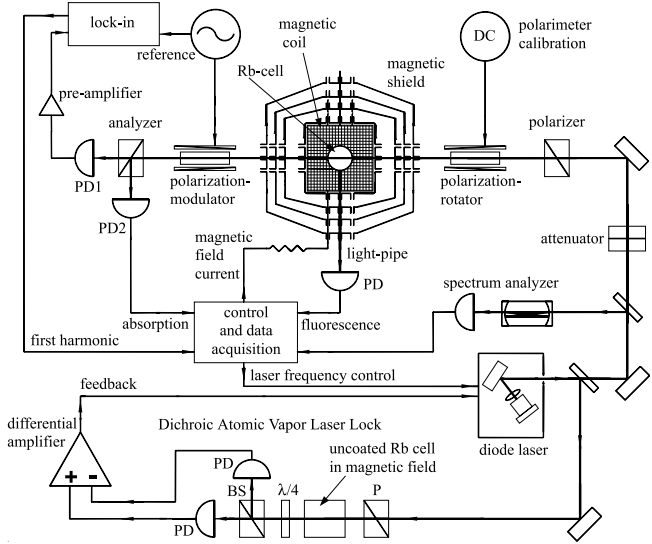}
\caption{Schematic diagram of the modulation polarimetry-based
experimental setup used by \citet{Bud98} to investigate NMOE in
paraffin-coated alkali-metal-vapor cells.}\label{fig:NMOEsetup}
\end{figure}
%----------------------------------------------------------------
Rubidium and/or cesium atoms are contained in a vapor cell with
paraffin coating \citep{AlexandrovLPh96}, used to suppress
relaxation of atomic ground-state polarization in wall collisions
(Sec.\ \ref{subsect:ARCoatedCells}). The experiments are performed
on the $D1$ and $D2$ resonance lines. A tunable extended-cavity
diode laser is used as the light source. The laser frequency is
actively stabilized and can be locked to an arbitrary point on the
resonance line using the dichroic-atomic-vapor laser-lock
technique (Sec.\ \ref{subsection:LaserStab}). The laser line
width, measured with a Fabry-Perot spectrum analyzer, is $\leq$7
MHz. Optical rotation is detected with a conventional
spectropolarimeter using a polarization-modulation technique
(Sec.\ \ref{subsection:Polarim}). The polarimeter incorporates a
crossed Glan prism polarizer and polarizing beam splitter used as
an analyzer. A Faraday glass element modulates the direction of
the linear polarization of the light at a frequency
$\omega_m\simeq2\pi\!\times\!1$ kHz with an amplitude
$\varphi_m\simeq5\!\times\!10^{-3}$ rad. The first harmonic of the
signal from the photodiode in the darker channel of the analyzer,
is detected with a lock-in amplifier. It is proportional to the
angle of the optical rotation caused by the atoms in the cell,
$\varphi_s$ (Sec.\ \ref{subsection:Polarim}). This signal,
normalized to the transmitted light power detected in the brighter
channel of the analyzer, is a measure of the optical rotation in
the vapor cell.

A coated 10-cm-diameter cell containing the alkali vapor is placed
inside a four-layer magnetic shield and surrounded with three
mutually perpendicular magnetic coils \citep{YasShield2002}. The
three outer layers of the shield are nearly spherical in shape,
while the innermost shield is a cube (this facilitates application
of uniform fields to the cell). The shield provides nearly
isotropic shielding of external dc fields by a factor of
$\sim$$10^6$ \citep{YasShield2002}. Although a very primitive
degaussing procedure is used (Sec.\ \ref{subsection:MagShield}),
the residual magnetic fields averaged over the volume of the vapor
cell are typically found to be at a level of a few tens of
microgauss before compensation with the magnetic coils, and a few
tenths of a microgauss after such compensation.

\subsection{Polarimetry}
\label{subsection:Polarim}

Successful application of the NMOE-based experimental methods
(Sec. \ref{Sect:NMORApplications}) depends on one's ability to
perform precision polarimetry or, more specifically, laser
spectropolarimetry.

In a ``balanced polarimeter'' \citep{Huard}, a sample is placed
between a polarizer and a polarizing beam splitter (analyzer)
whose transmission axes are oriented at $45^\circ$ to one another.
The optical rotation $\varphi$ due to the optical activity of the
sample can be found from a simple expression valid for
$\varphi\ll1$
%--------------------------------------------------------------------
\begin{equation}
    \varphi=\frac{I_1-I_2}{2\prn{I_1+I_2}}, \label{eqn:BalPolSignal}
\end{equation}
%--------------------------------------------------------------------
where $I_1$ and $I_1$ are the light intensities detected in the
two output channels of the analyzer. In order to measure the
ellipticity $\epsilon$ of the transmitted light, a
$\lambda/4$-plate is placed in front of the analyzer, oriented at
$45^\circ$ to its axis, to form a circular analyzer \citep[][see
also Sec.\ \ref{subsection:LaserStab}]{Huard}.

Another method of polarimetry employs a crossed polarizer and
analyzer (Fig.\ \ref{fig:FRE} and Sec.\ \ref{subsection:FSandLC})
and polarization modulation. A polarization modulator using a
Faraday-glass element is shown in Fig. \ref{fig:NMOEsetup}. The
signal in the ``dark'' (crossed) channel of the analyzer is given
by
%--------------------------------------------------------------------
\begin{equation}\label{eqn:PolarSignal}
\begin{split}
I_s(t)\simeq{}&\chi{I_0}\prn{r_e+\frac{1}{2}\alpha_m^2\!+\varphi^2}\\
&{}+2\chi{I_0}\alpha_m\varphi\sin\omega_mt-\frac{1}{2}\chi{I_0}\alpha_m^2\!\cos2\omega_mt.
\end{split}
\end{equation}
%--------------------------------------------------------------------
Here $\chi$ is the  coefficient defined by absorption and
scattering of light by the atomic vapor cell, $I_0$ is the total
photon flux of the linearly polarized light (in photons per unit
time) transmitted through the polarizer, $r_e$ is the
polarizer/analyzer extinction ratio, and $\alpha_m$ and $\omega_m$
are polarization modulation amplitude and frequency, respectively.
The amplitude of the first harmonic of the signal Eq.\
(\ref{eqn:PolarSignal}) is a measure of the polarization rotation
caused by the sample. The first-harmonic amplitude is also
proportional to the modulation amplitude $\alpha_m$. Typically,
with a Faraday modulator with a high-Verdet-constant glass (such
as Hoya FR-5), the modulation $\alpha_m$ is $\sim$$10^{-2}$ rad
\citep{Wol90,Wol91} at modulation frequencies $\sim$1 kHz. (High
frequency modulation is often limited by the inductance of the
Faraday modulator solenoid.) Resonant photoelastic modulators can
also be used for polarimetry \citep[as in the work of, for
example,][]{Oak95,Wang99}, allowing for larger polarization
modulation and higher modulation frequencies (several tens of
kilohertz).

From Eq.\ (\ref{eqn:PolarSignal}), one can obtain the sensitivity
of the polarimeter with data accumulation time $T$ for
shot-noise-limited detection of the first-harmonic signal in the
case of an ``ideal polarimeter'' ($\alpha_m^2\gg{r_e}+\varphi^2$):
%--------------------------------------------------------------------
\begin{equation}
    \delta\varphi_s\simeq\frac{1}{2\sqrt{\chi{I_0}T}}.
    \label{eqn:PolarSensitivity}
\end{equation}
%--------------------------------------------------------------------
For a 1-mW visible-light beam and $\chi\simeq0.5$, this
corresponds to $\sim$$2\times10^{-8}\ \mr{rad\,Hz^{-1/2}}$.
Similar shot-noise-limited sensitivity can be achieved with a
balanced polarimeter \citep{Bir94}.

Various modifications of the spectropolarimetry techniques have
been developed to meet specific experimental requirements, with
the common challenge of realizing the shot-noise limit of Eq.\
(\ref{eqn:PolarSensitivity}). A review of magneto-optics and
polarimetry of condensed matter was given by \citet{Zap75}.
\citet{Bir94} considered in detail experimental approaches to
measurement of small optical rotation and their limiting factors.

\subsection{Nonlinear magneto-optical rotation with
frequency-modulated light}
\label{subsection:FM_NMOR}

\citet{Bud2000Sens} showed that NMOR can be used for measurements
of sub-microgauss magnetic fields with sensitivity potentially
exceeding $10^{-11}\ \mr{G\,Hz^{-1/2}}$ (Sec.\
\ref{subsect:ARCoatedCells}). However, for many applications (such
as geophysics, magnetic prospecting, and navigation), it is
necessary to have a magnetometer with dynamic range $\sim$1 G.
\citet{Bud2002FM} recently demonstrated that if
frequency-modulated light is used to induce and detect nonlinear
magneto-optical rotation (FM NMOR), the ultra-narrow features in
the magnetic-field dependence of optical rotation normally
centered at $B=0$ can be translated to much larger magnetic
fields. In this setup, light polarization modulation (Fig.\
\ref{fig:NMOEsetup}) is replaced by frequency modulation of the
laser, and the time-dependent optical rotation is measured at a
harmonic of the light modulation frequency $\Omega_m$. The
frequency modulation affects both optical pumping and probing of
atomic polarization. This technique should enable the dynamic
range of an NMOR-based magnetometer to extend beyond the
Earth-field range.

For sufficiently low light intensities (so that the optical
pumping saturation parameter does not exceed unity), FM NMOR can
be understood as a three-stage (pump, precession, probe) process
(Sec.\ \ref{subsection:CohEf}). Due to the frequency modulation of
the laser light, the optical pumping and probing acquire a
periodic time dependence. When the pumping rate is synchronized
with the precession of atomic polarization, a resonance occurs and
the atomic medium is pumped into an aligned state whose axis
rotates at the Larmor frequency $\Omega_L$. The optical properties
of the medium are modulated at $2\Omega_L$, due to the symmetry of
atomic alignment. This periodic change of the optical properties
of the atomic vapor modulates the angle of the light polarization,
leading to the high-field FM NMOR resonances. If the
time-dependent optical rotation is measured at the first harmonic
of $\Omega_m$, a resonance occurs when $\Omega_m$ coincides with
$2\Omega_L$. Additional resonances can be observed at higher
harmonics.

It should be noted that this technique is closely related to the
work of \citet{Bel61a}, in which the intensity of pump light was
modulated in order to optically pump the atomic medium into a
polarized state which precessed with the Larmor frequency. Also
related is the work on $^4$He optical pumping magnetometers with
light frequency and intensity modulation and transmission
monitoring [\citet{Gil2001} and references therein].

\subsection{Magnetic shielding}
\label{subsection:MagShield}

Nonlinear magneto-optical effects are generally very sensitive to
magnetic-field magnitude, direction, and gradients. For
ground-state spin-relaxation times of $\lesssim1$ s, NMOE are
maximum at sub-microgauss magnetic fields [Eq.\
(\ref{Eqn:phi_vs_B})], necessitating their control at this
level.\footnote{There are specific requirements to magnetic
shielding for the work with atoms in a cryogenic-buffer-gas
environment discussed in Sec. \ref{subsubsection:ColdBufGas}, in
which relaxation times of up to minutes are expected.} This
control can be achieved with ferromagnetic shielding in spite of
the fact that residual fields due to imperfections of the shield
material are usually $\sim$10--50 $\mu$G \citep{KozTar82}. With a
carefully designed four-layer magnetic shield and
three-dimensional coil system inside the shield (see Fig.\
\ref{fig:NMOEsetup}), it is possible to achieve long-term
magnetic-field stability inside the innermost shield at the level
of $0.1\ \mu$G \citep{Bud98,Yas99,YasShield2002}.

Closed ferromagnetic shells shield external fields of different
frequencies via different physical mechanisms. For static and very
low-frequency external fields, the most important mechanism is
flux shunting due to the high permeability of the material. At
high frequencies, the skin effect becomes the most important
mechanism. A review of the early work on the shielding problem, a
theoretical treatment, and a survey of practical realizations was
given by \citet{Sum87}; see also books by \citet{RikBook87} and
\citet{KhripLamor}.

For static fields, the dependence of the shielding ratio on the
shape and size of $n$ shielding layers is given by an approximate
formula:
%--------------------------------------------------------------------
\begin{gather}
S_{tot}\equiv\frac{B_{in}}{B_0}\simeq{S_n}\prod_{i=1}^{n-1}S_i\sbrk{1-\prn{\frac{X_{i+1}}{X_i}}^k}^{-1},\\
\text{where }S_i\simeq\frac{X_i}{\mu_it_i}.\notag
\label{Eqn:ShieldRatio}
\end{gather}
%--------------------------------------------------------------------
Here $B_0$ is the magnetic field applied to the shield, $B_{in}$
is the field inside the shield, $S_i$ is the shielding factor of
the $i$-th layer, $X_i$ is the layer's radius or length (depending
on the relative orientation of the magnetic field and the layer),
and $t_i$ and $\mu_i$ are the thickness and magnetic permeability
of each layer, respectively. We assume $X_i>X_{i+1}$ and
$\mu_i\gg{X_i}/t_i$. The power $k$ depends on the geometry of the
shield: $k\simeq3$ for a spherical shield; $k\simeq2$ for the
transverse and $k\simeq1$ for the axial shielding factor of a
cylindrical shield with flat lids. Thus, for shields of comparable
dimensions, spherical shells provide the best shielding. In the
design of the three outer layers of the four-layer shield shown in
Fig.\ \ref{fig:NMOEsetup}, an approximation to a spherical shape,
simpler to manufacture than a true sphere, is used. The overall
shielding ratio is measured to be $\sim$$10^{-6}$, roughly the
same in all directions. The innermost shield is in the shape of a
cube with rounded edges. This allows compensation of the residual
magnetic field and its gradients as well as application of
relatively homogeneous fields with a simple system of nested 3-D
coils of cubic shape \citep{YasShield2002}. The field homogeneity
is increased by image currents, due to the boundary conditions at
the interface of the high-permeability material, which effectively
make the short coils into essentially infinite solenoidal
windings.

An important characteristic of a magnetic shield, in addition to
the shielding factor and the residual fields and gradients within
the shielded volume, is the residual magnetic noise. One source of
such noise is the Johnson thermal currents
\citep{Nen96,Lam99,All2002} in the shield itself.
% \citet{All2002}
%found that the noise related to this effect was $\sim$$10^{-10}\
%\mr{G\,Hz^{-1/2}}$ in the center of a multilayered cylindrical
%shield (of inner diameter $\sim$40 cm).

Superconducting shielding [\citet{Cabrera73}; \citet{Cabr88} and
references therein] yields the highest field stability. The
shielding properties of superconductors originate in the Meissner
effect, i.e., the exclusion, due to persistent currents, of
magnetic flux from the bulk of a superconductor.
\citet{Ham70,Tab93} developed a technique for reducing the
magnetic field inside a superconducting shield by expanding the
shield. When the dimensions of the superconducting enclosure
increase, the enclosed magnetic field decreases due to
conservation of magnetic flux ``frozen'' into the superconductor.
In one experiment, expansion of a folded lead-foil (100
$\mu$m-thick) balloon was used; an ultra-low residual magnetic
field ($\sim$ 0.06 $\mu$G) was achieved over a liter-sized volume
\citep{Cabrera73}. With expanded lead bags as superconducting
shields and a surrounding conventional ferromagnetic shield, a
magnetic field of $<0.1\ \mu$G in the flight dewar of the Gravity
Probe B Relativity Mission was maintained in a 1.3 m-by-0.25
m-diameter volume enclosing a gyroscope \citep{Mes2000}, giving a
shielding ratio of about $5\times 10^{-9}$. Some limitations of
superconducting shields were discussed by \citet{Die90}.

An essential difference between superconducting and ferromagnetic
shields is their opposite boundary conditions. Since a
superconductor is an ideal diamagnetic material, magnetic field
lines cannot penetrate into it, and so image currents are of
opposite sign as those in ferromagnetic material. This is
important for determining the shield geometry that provides the
best field homogeneity.

\subsection{Laser-frequency stabilization using magneto-optical effects}
\label{subsection:LaserStab}

The development of frequency-stabilized diode laser systems has
led to recent progress in experimental atomic and molecular
physics in general, and the study of NMOE in particular.
Conversely, some of the simplest and most effective methods of
laser-frequency stabilization are based on linear and nonlinear
magneto-optics.

A simple laser-locking system that uses no electronics is based on
the linear Macaluso-Corbino effect, employing a cell in a
longitudinal magnetic field and a polarizer placed inside the
laser cavity and forming a frequency-selective element
\citep[][and references therein]{Luk86,Wan92IEEE}. A similar
method based on narrow (Doppler-free) peaks in the spectrum of the
nonlinear magneto-optical activity for laser-frequency
stabilization was suggested by \citet{Cyr91} and adopted for laser
stabilization by, for example,
\citet{LeeWD92,LeeWD93,Kit95,Kit94,Was2002}.

A method based on magnetic-field-induced circular dichroism of an
atomic vapor was developed by \citet{Che94} for frequency
stabilization of single-frequency solid-state
$\mr{La}_x\mr{Nd}_{1-x}\mr{MgAl}_{11}\mr{O}_{19}$ (LNA) lasers.
This technique was subsequently adapted to diode lasers by
\citet{Cor98}, who coined the term dichroic-atomic-vapor laser
lock (DAVLL).

Figure \ref{fig:NMOEsetup} shows an example of DAVLL application
\citep{Bud98}. The optical scheme incorporates a polarizer (P), an
uncoated Rb-vapor cell placed in a static longitudinal magnetic
field, a $\lambda/4$-plate, and a polarizing beam splitter (BS).
With the fast axis of the $\lambda/4$-plate oriented at
$45^{\circ}$ to the axis of the polarizing beam splitter, as was
originally suggested by \citet{Che94}, the scheme realizes a
circular analyzer sensitive to the outgoing light ellipticity
induced by the circular dichroism of the atomic vapor. The output
signal from the differential amplifier has dispersion-like
frequency dependence corresponding to the difference between
absorption spectra for left and right circular components of the
light. The differential signal is used to frequency lock an
external-cavity diode laser to the center of the absorption line
by feeding back voltage to a piezoelectric transducer which
adjusts a diffraction grating in the laser cavity [\citet{Wie91};
\citet{Pat91}; \citet{Fox97}]. This setup is shown schematically
in Fig.\ \ref{fig:NMOEsetup}. Frequency tuning in the vicinity of
the line center can be achieved by adjusting the angle of the
$\lambda/4$-plate, or by applying an appropriate bias in the
electronic feedback circuit.

Various modifications of the DAVLL technique were devised to meet
specific experimental requirements. \citet{YasRSI2000} showed that
a simple readjustment of the respective angles of optical elements
allows one to extend the DAVLL frequency tuning range to the wings
of a resonance line. The DAVLL system of \citeauthor{YasRSI2000}\
was developed for use in the vicinity of equipment sensitive to
magnetic fields. The cell-magnet system, the core component of the
device, was designed to suppress the magnetic-field spillage from
the $\sim$200 G magnetic field that is applied over the Rb-cell
volume. \citet{Bev2001} developed a DAVLL device with fast
response in which the laser cavity optical length is adjusted with
an intra-cavity electro-optical modulator instead of much slower
mechanical motion of the grating. \citet{Cli2000} showed that the
zero crossing of the DAVLL signal (and thus the laser locking
frequency) is dependent on the magnitude of the magnetic field
applied to the Rb- or Cs-vapor cell and used this for tuning the
locked laser by varying the applied magnetic field.

Zeeman shifts of resonances in linear \citep[discussed by, for
example,][]{Rik91} and nonlinear [\citet{Wei88}; \citet{Ike89};
\citet{Din92}; \citet{Lec2000}] spectroscopy can be used for
laser-frequency locking and tuning. Using nonlinear saturation
spectroscopy, laser output is locked to a saturated absorption
line whose frequency is modulated and shifted by a combination of
ac and dc magnetic fields.  With this method, a line width of 15
kHz and long-term stability of 10 kHz can be achieved. By changing
the dc magnetic field applied to the atomic cell, the frequency of
the laser can be tuned almost linearly in the range of hundreds of
megahertz without modulation of laser intensity or frequency.

\section{Applications}
\label{Sect:NMORApplications}

\subsection{Magnetometry}
\label{subSect:Magnetom}

In Sec. \ref{subsection:OpticalPumping}, we mentioned that
\citet{Dup69,Coh69b,Dup70} performed ultra-sensitive
($\sim$$10^{-9}\ \mr{G\,Hz^{-1/2}}$) magnetometry using the
ground-state Hanle effect. Since then, the technology of
optical-pumping magnetometers has been further refined \citep[by,
for example,][]{Ale87}, and such magnetometers, typically
employing the rf--optical double-resonance method, are now used in
a variety of applications \citep[by, for example,][]{Ale92},
particularly for measuring geomagnetic fields.

The sensitivity of NMOR to small magnetic fields naturally
suggests it as a magnetometry technique \citep{Bark89b}. In this
Section, we discuss the limits of the sensitivity and recent
experimental work on NMOR magnetometry.

\subsubsection{Quantum noise limits}
\label{subsubsection:UltimateSensLimits}

The shot-noise-limited sensitivity of a magnetic field measurement
performed for a time $T$ with an ensemble of $N$ particles with
spin-coherence time $\tau$ is
%--------------------------------------------------------------------
\begin{equation}
\delta{B}\simeq\frac{1}{g\mu}\frac{\hbar}{\sqrt{N\tau{T}}}\,.
\label{Eqn:SQL}
\end{equation}
%--------------------------------------------------------------------
In this expression, we have neglected factors of order unity that
depend on particulars of the system (for example, the value of
$F$, and the relative contributions of different Zeeman
sublevels). When light is used to interrogate the state of the
spins, as in NMOR, one also needs to consider the photon
shot-noise [Eq.\ (\ref{eqn:PolarSensitivity})].\footnote{We do not
consider the use of the so-called spin-squeezed quantum states
\citep[discussed by, for example,][and references
therein]{Ula2001} or squeezed states of light \citep{Gra87}, which
can allow sensitivity beyond the standard quantum limit.}
Depending on the details of a particular measurement, either the
spin noise (\ref{Eqn:SQL}) or the photon noise
(\ref{eqn:PolarSensitivity}) may dominate. If a measurement is
optimized for statistical sensitivity, the two contributions to
the noise are found to be comparable \citep{Bud2000Sens}.

\citet{Fle2000} considered an additional source of noise in
polarimetric spin measurements. When the input light is
off-resonant, independent quantum fluctuations in intensity of the
two oppositely circularly polarized components of the light couple
to the atomic Zeeman sublevels via ac Stark shifts and cause
excess noise in the direction of the atomic spins. \citet{Fle2000}
concluded that this source of noise, while negligible at low light
power, actually dominates over the photon shot noise above a
certain critical power. Estimates based on the formulae derived by
\citeauthor{Fle2000}\ show that for the optimal conditions found
by \citet{Bud2000Sens} this effect could contribute to the overall
noise at a level comparable to the photon shot noise. However, the
effect of the ac Stark shifts can, in certain cases, be minimized
by tuning the light frequency so that the shifts due to different
off-resonance levels compensate each other \citep{NovAc2001}. Some
other possibilities for minimizing the additional noise may
include compensation of the effect by different atomic isotopes
\citep{NovAc2001}, or the use, instead of a polarization rotation
measurement, of another combination of the Stokes parameters
(Appendix \ref{ApStokes}) chosen to maximize the signal-to-noise
ratio.

\subsubsection{Experiments}
\label{subsubsection:Magnetom_Exper}

One approach to using NMOR for precision magnetometry is to take
advantage of the ultra-narrow (width $\sim$1 $\mu$G) resonance
widths obtainable in paraffin-coated cells (Sec.\
\ref{subsect:ARCoatedCells}). Using the experimental setup
described in Sec.\ \ref{SS:TypNMOEset}, \citet{Bud2000Sens}
optimized the sensitivity of an NMOR-based magnetometer to
sub-microgauss magnetic fields with respect to atomic density,
light intensity, and light frequency near the $D1$ and $D2$ lines
of $^{85}$Rb. They found that a shot-noise-limited magnetometric
sensitivity of $\sim$$3\times10^{-12}\ \mr{G\,Hz^{-1/2}}$ was
achievable in this system. This sensitivity was close to the
shot-noise limit for an ideal measurement [Eq.\ (\ref{Eqn:SQL})]
with the given number of atoms in the vapor cell ($\sim$$10^{12}$
at $20^\circ$ C and rate of ground-state relaxation ($\sim$1 Hz),
indicating that, in principle, NMOR is a nearly optimal technique
for measuring the precession of polarized atoms in external
fields. If limitations due to technical sources of noise can be
overcome, the sensitivity of an NMOR-based magnetometer may
surpass that of current optical pumping \citep{AlexandrovLPh96}
and SQUID (superconducting quantum interference device)
magnetometers \citep{Cla96}, both of which operate near their
shot-noise-limit, by an order of magnitude. Even higher
sensitivities, up to $2 \times 10^{-14}\ \mr{G\,Hz^{-1/2}}$, may
be achievable with an ingenious magnetometric setup of
\citet{All2002} in which K vapor at a density of $10^{14}\
\mr{cm^{-3}}$ is used and the effect of spin-exchange relaxation
is reduced by ``locking'' the precession of the two ground-state
hyperfine components together via the spin-exchange collisions
themselves \citep{Hap73}.

The magnetic-field dependence of NMOR is strongly affected by the
magnitude and direction of transverse magnetic fields
\citep{Bud98,Bud98ICAP}.  At low light powers, the
transverse-field dependence can be quantitatively understood using
a straightforward extension of the Kanorsky-Weis model discussed
in Sec.\ \ref{subsection:KanWeiTheory}, which opens the
possibility of sensitive three-dimensional magnetic-field
measurements.

As discussed in detail in Sec.\ \ref{subsection:FM_NMOR}, if
frequency-modulated light is used to induce and detect NMOR, the
dynamic range of an NMOR-based magnetometer may be increased
beyond the microgauss range to $>1$ G without appreciable loss of
sensitivity.

\citet{NovAc2001,Nov2002Mag}, using buffer-gas-free uncoated
cells, investigated the application of NMOR in optically thick
media to magnetometry (Sec. \ref{subsubsection:OptThick}).
Although the ultimate sensitivity that may be obtained with this
method appears to be some two orders of magnitude inferior to that
obtained with paraffin-coated cells,\footnote{The limitation in
practical realizations of thick-medium magnetometry may come from
the effects of radiation trapping
\citep{Nov2002Mag,MatRadTrap2001,Mat2002Rad}.} this method does
provide a broad dynamic range, so that Earth-field ($\sim$0.4 G)
values could be measured with a standard polarimeter.

\citet{Roc2002} theoretically analyzed magnetometric sensitivity
of thick medium NMOR measurements optimized with respect to light
intensity in the case of negligible Doppler broadening, and in the
case of large Doppler broadening. In the former case, the
sensitivity improves as the square root of optical density, while
in the latter, it improves linearly---a result which can be
obtained from standard-quantum-limit considerations (Sec.\
\ref{subsubsection:UltimateSensLimits}).

Cold atoms prepared by laser trapping and cooling were also
recently used for NMOR-based magnetometry. \citet{Isa99} employed
a pump/probe geometry with $\sim$$10^8$ cold $^{85}$Rb atoms that
were trapped in a MOT and then released (Sec.\
\ref{subsect:AtomTraps}) for measurement of an applied magnetic
field of $\sim$2 mG. The observation time for a single measurement
was limited to about 10 ms due to the free fall of the atoms in
the Earth's gravitational field; after 30 such measurements
\citet{Isa99} obtained a precision of $0.18\ \mr{\mu G}$ in the
determination of the applied magnetic field. It is expected that
by using atomic fountains \citep{Narducci} or far-off-resonant
optical dipole traps \citep{Dav95}, the observation time and
overall sensitivity to magnetic fields can be considerably
improved.

\subsection{Electric-dipole moment searches}
\label{SectEDM}

The existence of a permanent electric-dipole moment \citep[EDM,
reviewed by, for example,][]{KhripLamor} of an elementary or
composite particle such as an atom would violate parity- ($P$) and
time-reversal ($T$) invariance. The nonrelativistic Hamiltonian
$H_\text{EDM}$ describing the interaction of an EDM $\mb{d}$ with
an electric field $\mb{E}$ is given by
%--------------------------------------------------------------------
\begin{equation}
  H_\text{EDM}=-\mb{d}\cdot\mb{E}\propto\mb{F}\cdot\mb{E}.
\label{EDM_Ham}
\end{equation}
%--------------------------------------------------------------------
Under the parity operator $\hat{P}$ (the space-inversion operator
that transforms $\mb{r}\rightarrow-\mb{r}$), the axial vector
$\mb{F}$ does not change sign, whereas the polar vector $\mb{E}$
does. Therefore, $H_\text{EDM}$ is $P$-odd, i.e., violates parity.
Under the time-reversal operator $\hat{T}$, $\mb{E}$ is invariant
and $\mb{F}$ changes sign, so $H_\text{EDM}$ is also $T$-odd.

Ever since the discovery in 1964 of $CP$-violation in the neutral
kaon system ($C$ is charge-conjugation), there has been
considerable interest in the search for EDMs of elementary
particles. ($CP$-violation implies $T$-violation if the combined
$CPT$ symmetry is valid, as is generally believed). Many
theoretical attempts to explain $CP$-violation in the neutral kaon
system, such as supersymmetry, predict EDMs of the electron and
neutron that are near the current experimental sensitivities
\citep[see discussion by][]{KhripLamor}.  Measurements of the EDM
of paramagnetic atoms and molecules (that have unpaired electrons)
are primarily sensitive to the electron EDM $d_e$.  It is
important to note that due to relativistic effects in heavy atoms,
the atomic EDM can be several orders of magnitude larger than
$d_e$.

If an atom has an EDM, its Zeeman sublevels will experience linear
electric-field-induced splitting much like the usual Zeeman
effect. Most methods of searching for an EDM have thus been based
on detection (using magnetic-resonance methods or light
absorption) of changes in the Larmor precession frequency of
polarized atoms when a strong electric field is applied, for
example, parallel or anti-parallel to a magnetic field.

Another (related) way of searching for EDMs is to use induced
optical activity. As was pointed out by \citet{Bar77} and
\citet{Sus78}, a vapor of EDM-possessing atoms subject to an
electric field will cause rotation of the polarization plane of
light propagating along the direction of the electric field---the
analog of Faraday rotation. \citet{Bark88b} demonstrated that,
just as in magneto-optics, nonlinear optical rotation is
significantly more sensitive than linear optical rotation to the
EDM of an atom or molecule \citep[see also discussion
by][]{Hun91,Sch93,Wei93EDM}.

\citet{Yas99}; \citet{Kim2001} recently investigated the
possibility of performing a search for the electron EDM $d_e$
using nonlinear optical rotation in a paraffin-coated Cs-vapor
cell subjected to a longitudinal electric field. In addition to
Cs, which has an enhancement factor for the electron EDM of
$\sim$120, the cell would also contain Rb, which has much smaller
enhancement factor and would be used as a ``co-magnetometer.'' The
estimated shot-noise-limited sensitivity to $d_e$ for such an
experiment is $\sim$$10^{-26} e\,\mr{cm\,Hz^{-1/2}}$ (for a 10
$\mr{kV\,cm^{-1}}$ electric field). This statistical sensitivity
should enable a NMOR-based EDM search to compete with the best
present limits on $d_e$ from measurements in Tl \citep{Reg2002},
which has an enhancement factor of $\sim$600, and Cs
\citep{Mur89}. However, in order to reach this projected
sensitivity, there are several problems that must be overcome. The
first is a significant change in the atomic density when electric
fields are applied to the cell. The second is a coupling of the
atomic polarizations of Cs and Rb via spin-exchange collisions,
which would prevent Rb from functioning as an independent
co-magnetometer. These issues are discussed in more detail by
\citet{Kim2001}.

There are also molecular (YbF) beam experiments \citep{Sau2001}
searching for an electron EDM using a separated light pump and
probe technique that is a variant of the Faraday-Ramsey
spectroscopic method (Sec.\ \ref{subsection:FRS}).

In another class of experiments, one searches for EDMs of
diamagnetic atoms and molecules. Such experiments are less
sensitive to the EDM of the electron than those employing
paramagnetic atoms and molecules. However, they probe
$CP$-violating interactions within nuclei, and are sensitive to
EDMs of the nucleons. \citet{Rom2001PRL,Rom2001} have conducted
the most sensitive experiment of this kind in $^{199}$Hg. A
cylindrical quartz vapor cell at room temperature (vapor
concentration $\sim$$5\times10^{13}\ \mr{cm^{-3}}$) is subjected
to an electric field of up to $10\ \mr{kV\,cm^{-1}}$, applied via
conductive $\mr{SnO_2}$ electrodes deposited on the inner surfaces
of the flat top and bottom quartz plates. The entire inner surface
of the cell including the electrodes is coated with paraffin. The
cell is filled with a N$_2$/CO buffer-gas mixture at a total
pressure of several hundred Torr. The buffer gas is used to
maintain high breakdown voltage, and to quench metastable states
of mercury that are populated by the UV light employed. The
nuclear-spin-relaxation time ($I=1/2$ for $^{199}$Hg) is
$\sim$100--200 s. The experiment utilizes the
$6\,^1\!S_0\rightarrow 6\,^3\!P_1$ transition at 253.7 nm, excited
with light from a home-made cw laser system. A magnetic field of
17 mG and the electric field are applied to the cell, and
circularly polarized resonant light, chopped at the Larmor
frequency, illuminates the cell in a direction perpendicular to
that of the fields. After the atoms are pumped for $\sim$30 s with
70 $\mu$W of light power to establish transverse circular
polarization, optical rotation of linearly polarized probe light,
detuned from resonance by 20 GHz, is detected (Fig.\
\ref{fig:HgSignal}).\footnote{A previous (several times less
sensitive) version of this experiment \citep{Jac95} used the same
light beam for both pumping and probing.}
%-------------------------------------------------------------
\begin{figure}
\includegraphics{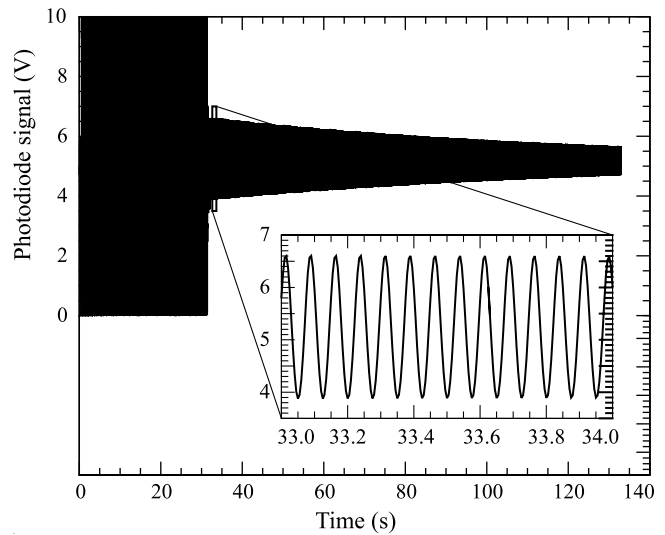}
\caption{Experimental optical rotation signal obtained by
\citet{Rom2001} produced by Hg atomic polarization in a magnetic
field of 17 mG. The inset shows a one-second segment of the
data.}\label{fig:HgSignal}
\end{figure}
%-------------------------------------------------------------
The spin-precession frequency is measured in two identical cells
with opposite directions of the electric field with respect to the
magnetic field. This device has a sensitivity to energy level
shifts of $\sim$0.3 $\mr{{\mu}Hz/Hz^{1/2}}$ over a measurement
time of a hundred seconds, better than any other existing device
(see Sec.\ \ref{subSect:Magnetom}). With several months of data
taking, it has set a limit on the atomic EDM of
$\abs{d\prn{^{199}\mr{Hg}}}< 2.1\times10^{-28}\ e\,\text{cm}$
(95\% confidence).

\subsection{The Aharonov-Casher phase shift} \label{ExV}

An atom moving in an electric field $\mb{E}$ with velocity
$\mb{v}$ experiences a magnetic field
$\mb{B}=\mb{E}\times\mb{v}/c$. This ``motional'' magnetic field
induces phase shifts of the atomic magnetic sublevels and thus
affects nonlinear magneto-optical signals. The
$\mb{E}\times\mb{v}$ effect is a linear Stark effect that
represents a severe systematic problem in many EDM experiments.
The induced phase shift is given by the Aharonov-Casher
(\citeyear{Aha84}) phase
%--------------------------------------------------------------------
\begin{equation}
\phi_\text{AC}=\frac{1}{\hbar{c}}\int_{(c)}\bs{\mu}\times\mb{E(s)}\cdot
d\mb{s} \label{eqn:AC_phase}
\end{equation}
%--------------------------------------------------------------------
acquired by a magnetic moment $\bs{\mu}$ carried on a path
$(c)$.\footnote{Equation (\ref{eqn:AC_phase}) can be easily
derived by considering a particle with magnetic moment $\bs{\mu}$
moving along a trajectory whose element is given by
$d\mb{s}=\mb{v}dt$. The particle acquires a differential phase
shift
%--------------------------------------------------------------------
\begin{align}\label{ExV1}
    d\phi
        =\frac{\bs{\mu}\cdot\mb{B}}{\hbar}dt
        =\frac{\bs{\mu}\cdot\mb{E}\times\mb{v}}{\hbar c}dt
        =\frac{\bs{\mu}\cdot\mb{E}\times d\mb{s}}{\hbar c},
\end{align}
%--------------------------------------------------------------------
which gives (\ref{eqn:AC_phase}) upon integration.} The phase
(\ref{eqn:AC_phase}) is independent of the shape of the trajectory
$(c)$ and is thus referred to as a ``topological phase''.

The Aharonov-Casher effect was studied interferometrically with
neutrons \citep{Cim89} and with the fluorine nucleus in the TlF
molecule \citep{San93,San95} using conventional Ramsey molecular
beam spectroscopy. Faraday-Ramsey spectroscopy (Sec.\
\ref{subsection:FRS}) is a convenient method for measuring the
Aharonov-Casher effect with atoms. \citet{Gor95} performed such an
experiment with Rb atoms and found agreement with the theoretical
prediction at the level of 1.4\%. They verified the linear
dependence of the phase shift on the strength of the electric
field as well as its nondispersive nature (independence of
velocity). An atomic Aharonov-Casher effect was also measured by
\citet{Zei95} using an atomic interferometer.

In Faraday-Ramsey geometry, the phase shift accumulated in a
homogeneous static electric field between $\Delta M=2$ sublevels
of an aligned atom is given by
%--------------------------------------------------------------------
\begin{equation}
\phi_\text{AC}=\frac{2g\mu}{\hbar{c}}\int{E}dL,
\end{equation}
%--------------------------------------------------------------------
where the field integral extends over the flight region between
the two optical interaction regions. This phase shift enters
directly in the Faraday-Ramsey line-shape function given by Eq.\
(\ref{eq:FRSshape}). When using atoms with a well known
$g$-factor, the measurement of $\phi_\text{AC}$ can thus be used
to experimentally determine the electric field integral
\cite{Ras2001}.

\subsection{Measurement of tensor electric
polarizabilities} \label{subsect:TensPol}

The quadratic Stark effect in a Zeeman manifold $\ket{F,M}$ in a
homogeneous static electric field $E$ can be parameterized by
%--------------------------------------------------------------
\begin{equation}
\Delta_s(F,M)=-\frac{1}{2}\alpha{E^2},
\end{equation}
%--------------------------------------------------------------
where the static electric polarizability $\alpha$ can be
decomposed into scalar ($\alpha_0$) and tensor ($\alpha_2$) parts
according to \citep{Ang68}:
%--------------------------------------------------------------
\begin{equation}
\alpha=\alpha_0+\frac{3M^2-F\prn{F+1}}{F\prn{2F-1}}\,\alpha_2.
\label{eq:el_pol_decomp}
\end{equation}
%--------------------------------------------------------------

States with $J=1/2$, such as the alkali ground states, can only
have tensor properties due to nonzero nuclear spin. Tensor
polarizabilities of such states are suppressed compared to the
scalar polarizabilities by the ratio of the hyperfine splitting to
the energy separation between the state of interest and
electric-dipole-coupled states of opposite parity. For the alkali
ground states, the suppression is $\sim$$10^6$--$10^7$.

When combined with the magnetic shifts of Zeeman sublevels, the
tensor electric shifts lead to complex signals in level-crossing
experiments (Sec. \ref{subsubsection:Hanle_LC}) when magnetic
field is scanned. This technique has been used for several decades
to measure excited-state tensor polarizabilities. The combination
of electric and magnetic shifts is also responsible for the
alignment-to-orientation conversion processes (Sec.\
\ref{subsection:AOC}).

Measurements of ground-state tensor polarizabilities in the
alkalis were performed in the 1960s using conventional atomic beam
Ramsey resonance spectroscopy [\citet{San64};
\citet{Car68,Gou69}]. A. Weis and co-workers are currently using
the effect of the electric-field-dependent energy shifts on the
nonlinear magneto-optical properties of the atomic medium to
remeasure these polarizabilities \citep{Ras2001,Wei2001,Ras2001b}.
The technique is an extension of Faraday-Ramsey spectroscopy
(Sec.\ \ref{subsection:FRS}) to electric interactions in the
precession region. The renewed interest in the tensor
polarizabilities is related to a discrepancy between previous
experimental and theoretical values and it is believed that
increased precision in both (at the sub-1\% level) will provide a
valuable test of atomic structure calculations.

\subsection{Electromagnetic field tomography}
\label{subSectEMTomogr}

Various techniques involving NMOE and NEOE can be used to perform
spatially-resolved measurements of magnetic and electric fields
(electromagnetic field tomography). \citet{SkallaPines97} used
NMOR to measure the precession of spin-polarized $^{85}$Rb atoms
contained in a cell filled with dense $\mr{N_2}$ buffer gas to
ensure sufficiently long diffusion times for Rb atoms. The Rb
atoms were spin-polarized with a set of spatially separated,
pulsed pump beams, and subjected to a magnetic field with a
gradient. Since the precession frequency of the Rb atoms depended
on the position of the atoms within the cell, the spatial
distribution of the magnetic field within the cell could be
determined from the detected Larmor precession frequencies.
\citet{SkallaDiff97} and \citet{Gie2000} used similar techniques
to measure the diffusion of spin-polarized alkali atoms in
buffer-gas-filled cells. \citet{All2002} have investigated the
possibility of biomagnetic imaging applications of NMOE.

It is possible to localize regions of interest by intersecting
pump and probe light beams at an angle. Analysis of the
polarization properties of the probe beam could, in principle,
allow the reconstruction of the electric and magnetic fields
inside the volume where the pump and probe beams overlap. This
reconstruction would be aided by the fact that NMOE enhance
optical rotation while NEOE enhance induced ellipticity. Scanning
the region of intersection would allow one to create a
three-dimensional map of the fields.

\subsection{Parity violation in atoms}
\label{subSect:NL_PNC_Appl}

We have discussed applications of linear magneto-optics to the
study of parity violation in Sec. \ref{subsubsection:PNC}.

In work related to a study of parity-violation, \citet{Bou95}
studied both theoretically and experimentally
magnetic-field-induced modification of polarization of light
propagating through Cs vapor whose frequency was tuned near a
resonance between two excited states. The population of the upper
state of the transition was created by subjecting atoms to a pump
laser pulse connecting this state to the ground state (in the work
of \citet{Bou95}, this is a nominally forbidden M1-transition
which has a parity-violating E1-contribution, and also an
E1-contribution induced by an applied electric field). A
particular feature incorporated in the design of this experiment
is that there is probe beam amplification rather than absorption.

\citet{Koz90} considered theoretically the possibility of applying
an effect directly analogous to BSR NMOR (Sec.\
\ref{subsection:BenStr}) to the measurement of parity violation
(of the $P$-odd, $T$-even rotational invariant
$\mb{k}\cdot\mb{B}$) in the vicinity of atomic transitions with
unsuppressed M1 amplitude. They found that, while the maximum
effect is not enhanced compared to the linear case, the magnitude
of the field $B$ at which the maximum occurs is reduced in the
nonlinear case by the ratio $\gamma_0/\Gamma_D$ of the homogenous
width to the Doppler width of the transition.

\citet{Cro98} attempted to use electromagnetically induced
transparency (EIT; Sec.\ \ref{subsection:lambdares}) to address
the problems of the traditional optical-rotation parity-violation
experiments, namely, the problem of detailed understanding of a
complicated spectral lineshape, and the absence of reversals that
would help distinguish a true $P$-violating rotation from
systematics. In addition to a probe beam tuned to an
M1-transition, there is a second, counter-propagating pump laser
beam present whose frequency is tuned to an adjacent fully allowed
transition. The presence of this pump beam modifies the effective
refractive index ``seen" by the probe beam in a drastic way. The
main advantages of using this scheme for a parity violation
measurement are that the optical rotation lineshapes are now
Doppler-free and can be turned on and off by modulating the pump
beam intensity.

Other examples of application of nonlinear spectroscopic methods
to the study of $P$-violation were reviewed by \citet{BudPNCRev}.

\section{Closely-related phenomena and techniques}

\subsection{Dark and bright resonances} \label{subsection:lambdares}

Zero-field level-crossing phenomena involving linearly polarized
light such as those occurring in the nonlinear Faraday effect or
the ground-state Hanle effect (Sec.\ \ref{subsection:FSandLC}) can
be interpreted in terms of \emph{dark resonances}, or \emph{lambda
resonances}. Such resonances occur when a light field consisting
of two phase-coherent components of frequency $\omega_1$ and
$\omega_2$ is near resonance with a atomic $\Lambda$-system [a
three-level system having two close-lying lower levels\footnote{In
many dark-resonance experiments, the lower-state splitting is the
ground-state hyperfine splitting (for example, as with the
$F=I\pm1/2$ states in alkali atoms). As the coherent superposition
of hyperfine levels can be very long lived, the lambda resonances
can have correspondingly small line widths. In Cs and Rb, line
widths below 50 Hz have been observed
\citep{Bra97,Erh2000,Erh2001}. In a magnetic field, the dark
resonance of an alkali atom splits into $4I+1$ components; the
outermost components correspond to $\Delta{M}=\pm2I$ coherences
which have a Zeeman shift of $\sim$$\pm 4\mu{BI}/\prn{2I+1}$
\citep{Wyn98}. For cesium ($I=7/2$) this yields a seven-fold
enhanced sensitivity to magnetic fields relative to experiments
detecting $\Delta M=\pm 1$ coherences. A magnetometer based on
dark resonances with a sensitivity of 120 $\mr{nG\,Hz^{-1/2}}$ has
recently been demonstrated by \citet{Sta2001}.} and one excited
level, see Fig.\ \ref{fig:LambdaRes}(a)].
%---------------------------------------------------------------
\begin{figure}
\includegraphics{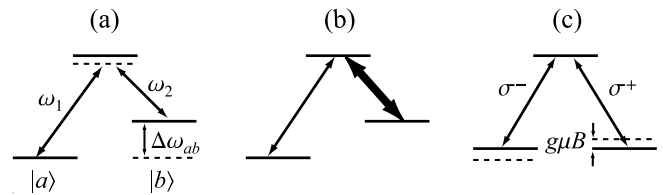}
\caption{Illustration of the connection between lambda resonances,
electromagnetically-induced transparency (EIT), and NMOE. (a)
Lambda resonance in a three-level system. Two phase-coherent
optical fields at frequencies $\omega_1$ and $\omega_2$ pump the
system into a ``dark'' state in which absorption and fluorescence
are suppressed. (b) EIT in the $\Lambda$-configuration. In this
case, one component of the light field is much stronger than the
other. In the absence of the coupling field absorption on the
probe transition is large. When the coupling field is applied, a
dip, with width given by the relaxation rate of the coherence
between the two lower states, appears in the absorption profile.
(c) NMOE in the $\Lambda$-configuration. In this case, the two
light fields have the same intensity and frequency and the two
transitions have the same strength. Here the frequency splitting
of the transitions---rather than that of the light components---is
tuned, using a magnetic field.}\label{fig:LambdaRes}
\end{figure}
%---------------------------------------------------------------
When the difference frequency
$\Delta\omega_\text{light}=\omega_1-\omega_2$ is equal to the
frequency splitting $\Delta\omega_{ab}$ of the two lower levels
(the Raman resonance condition), the system is pumped into a
coherent superposition of the two lower states which no longer
absorbs the bichromatic field. Due to the decrease of fluorescence
at the Raman resonance, this situation is called a ``dark
resonance'' \citep[see, for example, the review
by][]{Ari96}.\footnote{As discussed in Sec.
\ref{subsect:buff_gas_free_cells}, optical pumping in the case of
a closed $F\rightarrow F+1$ transition leads to an increased
absorption of the medium, rather than transparency [\citet{Kaz84};
\citet{Lez99,Ren2001}], which leads to \emph{bright resonances}.}
The creation of a nonabsorbing coherent superposition of the lower
states is also referred to as ``coherent population trapping''.
When one frequency component of the light field is much stronger
than the other [Fig.\ \ref{fig:LambdaRes}(b)], it is referred to
as a ``coupling'' or ``drive'' field, and the reduction of the
absorption of the weaker probe field is called
``electromagnetically induced transparency'' \citep[see, for
example, the review\footnote{A list of earlier references dating
back to the 1950s can be found in a review by \citet{Mat2001SL}.}
by][]{Har97}.

An analogous situation occurs in nonlinear magneto-optical
experiments with linearly polarized light resonant with, for
example, a $F=1\rightarrow{F'=0}$ transition [Fig.\
\ref{fig:LambdaRes}(c)]. The $\sigma^+$ and $\sigma^-$ components
of the linear polarization can be viewed as two phase-coherent
fields which happen to have the same frequency and intensity. A
resonance occurs when the atomic levels, tuned via the Zeeman
effect, become degenerate. Even though here it is the level
splitting rather than the light frequency that is tuned, the
origin of the resonance is the same.\footnote{Although the
connection between various nonlinear effects discussed here and
NMOE is straightforward, it has not been widely recognized until
very recently.}

Other closely-related topics discussed in the literature include
``lasing without inversion'' \citep[see, for example, discussion
by][]{Koc92} and phase-coherent atomic ensembles, or
``phaseonium'' \citep{Scu92Phaseonium}.

\subsection{``Slow'' and ``fast'' light}
\label{subsection:Slow_Fast_Light}

When a pulse of weak probe light propagates through a nonlinear
medium in the presence of a strong field, various peculiar
phenomena can be observed in the propagation dynamics. Under
certain conditions, the light pulse is transmitted without much
distortion, and the apparent group velocity can be very slow
($\sim$1 $\mr{m\,s^{-1}}$ in some experiments), faster than $c$,
or negative \citep[see recent comprehensive reviews
by][]{Mat2001SL,Boy2002}. The system, from the point of view of an
observer detecting the probe pulse shape and timing at the input
and the output, appears to be an effectively \emph{linear} optical
medium but with unusually large magnitude of dispersion. The
``slow'' and ``fast'' light effects, as well as similar effects in
linear optics, are due to different amplification or absorption of
the leading and trailing edges of the pulse (``pulse reshaping'')
\citep[see, for example, discussion by][]{Chi96}.

The connection between NMOE and ``fast'' and ``slow'' light was
established by \citet{BudGroupVel99} in an investigation of the
dynamics of resonant light propagation in Rb vapor in a cell with
antirelaxation wall coating. They modulated the direction of the
input light polarization and measured the time dependence of the
polarization after the cell. The light propagation dynamics that
they observed, including negative ``group delays'' associated with
electromagnetically induced opacity \citep{Aku98,Lez99}, were
analogous to those in EIT experiments. The spectral dependence of
light pulse delays was measured to be similar to that of NMOR,
confirming a theoretical prediction. In addition, magnetic fields
of a few microgauss were used to control the apparent group
velocity. Further studies of magnetic-field control of light
propagation dynamics were carried out by \citet{Nov99SL,Mai2002}.

Recently, \citet{Shv2002} pointed out that EIT and ``slow'' light
phenomena can also occur in magnetized plasma.\footnote{Earlier
studies of EIT phenomena in plasma were conducted by
\citet{Har96,Mat98,Gor2000a,Gor2000b}.} They observe that in such
a medium, EIT, commonly described as a quantum interference effect
(Sec.\ \ref{subsection:lambdares}), can be completely described by
classical physics.

\subsection{Self-rotation}
\label{subsection:self-rotation}

If the light used to study NMOR is elliptically polarized,
additional optical rotation (present in the absence of a magnetic
field) can occur due to nonlinear self-rotation (SR).
Self-rotation arises when the elliptically polarized light field
causes the atomic medium to acquire circular birefringence and
linear dichroism, causing optical rotation. There are several
physical mechanisms that can lead to SR in atomic
media,\footnote{Self-rotation was first observed in molecular
liquids \citep{Mak64}. The alignment of anisotropic molecules in
the light field can lead to SR \citep{Chi69}.} discussed in detail
by \citet{Roc2001SR}. Optical rotation can be caused by circular
birefringence, created by either a difference in the populations
\citep[due to optical pumping,][]{Dav92} or the energies (due to
ac Stark shifts) of the $\pm{M}$ Zeeman sublevels. At high power,
orientation-to-alignment conversion can generate atomic alignment
not along the axes of light polarization, leading to optical
rotation due to linear dichroism.  In general, the spectra of SR
and NMOR are different, and thus both the magnetic field and light
frequency dependences can be used to distinguish NMOR and SR.

Self-rotation can play an important role in the output
polarization of gas lasers \citep{Ale69} and in high-resolution
polarization spectroscopy [\citet{Sai78}; \citet{Aga84};
\citet{Ado86,Ale88}]. Self-rotation was theoretically considered
in relation to NMOR by \citet{Gir85a}; \citet{Fom95} and as a
systematic effect in the study of atomic parity nonconservation by
\citet{Kos86}. Self rotation in alkali vapors has been studied
experimentally by \citet{Bon73}; \citet{Bak89};
\citet{Dav92,Roc2001SR}.

Recently, it was found that if linearly polarized light propagates
through a medium in which elliptically polarized light would
undergo SR, squeezed vacuum is produced in the orthogonal
polarization [\citet{Tan83}; \citet{Boi96};
\citet{Mar98,Mat2002}]. It may be possible to use this effect to
perform sub-shot-noise polarimetry \citep{Gra87} in the detection
of NMOR.

\section{Conclusion}
\label{Section:Conclusion}

In this Review, we have described the history and recent
developments in the study and application of resonant nonlinear
magneto-optical effects. We have discussed the connections and
parallels between this and other subfields of modern spectroscopy,
and pointed out open questions and directions for future work.
Numerous and diverse applications of NMOE include precision
magnetometry, very high resolution measurements of atomic
parameters, and investigations of the fundamental symmetries of
nature. We hope that this article has succeeded in conveying the
authors' excitement about working in this field.

\section*{Acknowledgments}

We are grateful to D.~English and K.~R.~Kerner for help with
bibliography. The authors have greatly benefited from invaluable
discussions with E.~B.~Alexandrov, M.~G.~Kozlov, S.~K.~Lamoreaux,
A.~B.~Matsko, A.~I.~Okunevich, M.~Romalis, J.~E.~Stalnaker,
A.~O.~Sushkov, and M.~Zolotorev. This work has been supported by
the Office of Naval Research (grant N00014-97-1-0214), by the U.S.
Department of Energy through the LBNL Nuclear Science Division
(Contract No. DE-AC03-76SF00098), by NSF CAREER grant PHY-9733479,
Schweizerischer Nationalfonds grant 21-59451.99, NRC US-Poland
Twinning grant 015369, and by the Polish Committee for Scientific
Research KBN (grants 2P03B06418 and PBZ/KBN/043/PO3/2001) and the
National Laboratory of AMO Physics in Torun, Poland.

\appendix
\addcontentsline{toc}{section}{Appendices}

\section{Description of light polarization in terms of the Stokes
parameters} \label{ApStokes}

In the formula (\ref{lightfield}) for the electric field of light
propagating in the $\mb{\hat{z}}$ direction, the polarization
state of the light is given in terms of the parameters $\mc{E}_0$,
$\varphi$, $\epsilon$, $\phi$---the light field amplitude,
polarization angle, ellipticity, and phase, respectively. Another
parameterization of the polarization state is that given by the
\emph{Stokes parameters} \citep[see, for example, book
by][]{Huard}, which are useful because they are defined in terms
of directly measurable intensities:
%--------------------------------------------------------------------\
\begin{equation}
\begin{aligned}
S_0&=I_x+I_y=I_0,\\
S_1&=I_x-I_y,\\
S_2&=I_{+\pi/4}-I_{-\pi/4},\\
S_3&=I_+-I_-,
\end{aligned}
\end{equation}
%--------------------------------------------------------------------
where $I_x$ and $I_y$ are the intensities of the components along
the $x$- and $y$-axes, $I_{\pm\pi/4}$ are the intensities of the
components at $\pm\pi/4$ to the $x$- and $y$-axes, and $I_+$ and
$I_-$ are the intensities of the left- and right-circularly
polarized components, respectively.

The Stokes parameters for fully polarized light can also be
written in a normalized form that is easily related to the
polarization angle and ellipticity:
%--------------------------------------------------------------------
\begin{equation}
\begin{aligned}
S_1'&=S_1/S_0=\cos2\epsilon\cos2\varphi,\\
S_2'&=S_2/S_0=\cos2\epsilon\sin2\varphi,\\
S_3'&=S_3/S_0=\sin2\epsilon.
\end{aligned}
\end{equation}
%--------------------------------------------------------------------

\section{Description of atomic polarization}

\subsection{State multipoles}\label{Appendix:Multipoles}

The density matrix of an ensemble of atoms in a state with angular
momentum $F$ has $(2F+1)\times(2F+1)$ components $\rho_{M,M'}$.
Since magneto-optical effects involve spin rotation (Larmor
precession) and other more complex forms of atomic polarization
evolution, it is often useful to work with the irreducible
components of $\rho$, i.e, the components $\rho_q^{(\kappa)}$ with
$q=-\kappa,\dots,\kappa$ and $\kappa=0,\dots,2F$, which transform
among themselves under rotations \citep[see, for example, books
by][]{VAr88,Omo77}. The $\rho_q^{(\kappa)}$ are related to the
$\rho_{M,M'}$ by
%--------------------------------------------------------------------
\begin{equation}
\rho_q^{(\kappa)}=\sum_{M,M'=-F}^F(-1)^{F-M'}\cg{F,M,F,-M'}{\kappa,q}\rho_{M,M'},
\end{equation}
%--------------------------------------------------------------------
where $\cg{\dots}{\dots}$ indicate the Clebsch-Gordan
coefficients. The density matrix for atoms in a state with angular
momentum $F$ can be decomposed into irreducible multipole
components according to
%--------------------------------------------------------------------
\begin{equation}
\rho=\sum_{\kappa=0}^{2F}\sum_{q=-\kappa}^{\kappa}\rho_q^{(\kappa)}T_q^{(\kappa)}
\label{rhomultipole},
\end{equation}
%--------------------------------------------------------------------
where the $T_q^{(k)}$ are components of the irreducible tensors
$T^{(\kappa)}$ obtained from coupling $\mb{F}$ with $\mb{F}:$
%--------------------------------------------------------------------
\begin{equation}
F\otimes F=T^{(0)}\oplus T^{(1)}\oplus\dots\oplus T^{(2F)}.
\end{equation}
%--------------------------------------------------------------------
The components $\rho_q^{(\kappa)}$ are called \emph{state
multipoles}. The following terminology is used for the different
multipoles: $\rho ^{(0)}$--monopole moment or \emph{population},
$\rho^{(1)}$--vector moment or \emph{orientation},
$\rho^{(2)}$--quadrupole moment or \emph{alignment},
$\rho^{(3)}$--octupole moment, and $\rho^{(4)}$--hexadecapole
moment.\footnote{There are other definitions of the terms
``orientation'' and ``alignment'' in the literature. For example,
in \citet{Zar88}, alignment designates even moments in atomic
polarization (quadrupole, hexadecapole, etc.), while orientation
designates the odd moments (dipole, octupole, etc.).} Each of the
moments $\rho^{(\kappa)}$ has $2\kappa+1$ components.

The term \emph{polarization} is used for the general case of an
ensemble that has any moment higher than population. When the
Zeeman sublevels are not equally populated,
$\rho_0^{(\kappa)}\neq0$ for some $\kappa>0$, and the medium is
said to have \emph{longitudinal polarization}. When there are
coherences between the sublevels, $\rho_{q}^{(\kappa)}\neq0$ for
some $q\neq0$, and the medium is said to have \emph{transverse
polarization}. If $\rho$ is represented in the basis $\ket{F,M}$,
longitudinal orientation and longitudinal alignment are given by
%--------------------------------------------------------------------
\begin{equation}
\begin{aligned}
P_{z}&\propto\rho_0^{(1)}\!\propto\abrk{F_z},\\
A_{zz}&\propto\rho_0^{(2)}\!\propto\abrk{3F_z^2-\mb{F}^2},
\end{aligned}
\end{equation}
%--------------------------------------------------------------------
respectively.

Note also that optical pumping with circularly polarized light (in
the absence of other external fields) creates multipoles of all
orders ($\kappa\leq 2F$), while pumping with linearly polarized
light creates only even-ordered multipoles. This latter fact is a
consequence of a symmetry that is most clearly seen when the
quantization axis is along the light polarization direction.

\subsection{Visualization of atomic polarization}
\label{AppendixSubsect:AtPolVis}

In this section, we outline a technique for visualizing atomic
polarization by drawing a surface in three dimensions representing
the probability distribution of the angular momentum, as presented
in more detail by \citet{Roc2001}. A similar approach has been
used to describe molecular polarization and its evolution
\citep{Auz95,Auz97}, and more recently to analyze anisotropy
induced in atoms and molecules by elliptically polarized light
\citep{MilPr99,Mil99}.

In order to visualize the angular momentum state of atoms with
total angular momentum $F$, we draw a surface whose distance $r$
from the origin is equal to the probability of finding the
projection $M=F$ along the radial direction. To find the radius in
a direction given by polar angles $\theta$ and $\varphi$, we
rotate the density matrix $\rho$ so that the quantization axis is
along this direction and then take the $\rho_{F,F}$ element:
%--------------------------------------------------------------------
\begin{equation}
r\prn{\theta,\varphi}=\bra{M\!\!=\!F}\mb{R}_{\varphi,\theta,0}^{-1}\,\rho\,\mb{R}_{\varphi,\theta,0}\ket{M\!\!=\!F}.
\end{equation}
%--------------------------------------------------------------------
Here $\mb{R}_{\alpha,\beta,\gamma}$ is the quantum mechanical
rotation matrix \citep[see, for example, discussion
by][]{Edmonds}.

Consider, for example, atoms prepared at $t=0$ in the
$\ket{F=1,M=1}$ (``stretched'') state with the quantization axis
chosen along $\mb{\hat{y}}$. An electric field $\mb{E}$ is applied
along $\mb{\hat{z}}$, causing evolution depicted in Fig.\
\ref{fig:J1EzMovie}.
%--------------------------------------------------------------------
\begin{figure}
\includegraphics{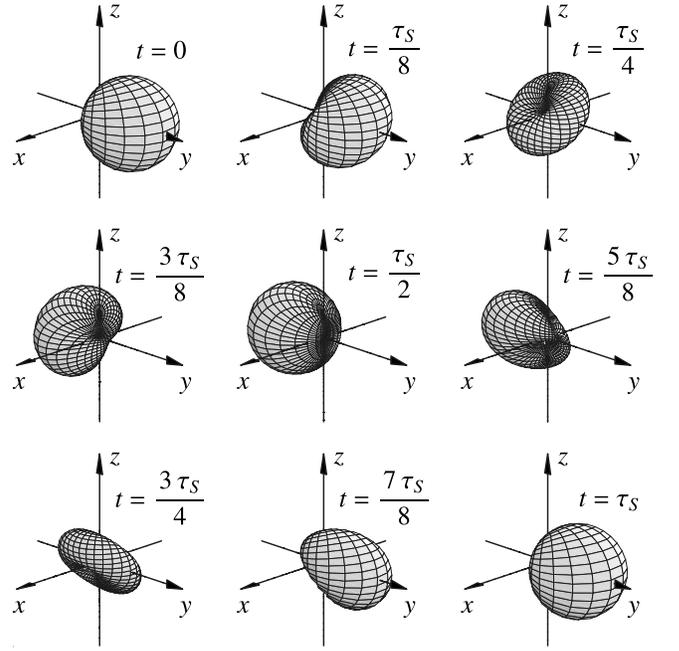}
\caption{A sequence of probability surfaces representing evolution
of a state with $F=1$. The state is initially stretched along
$\mb{\hat{y}}$ at $t=0$ and an electric field is applied along
$\mb{\hat{z}}$, causing Stark beats with period
$\tau_S$.}\label{fig:J1EzMovie}
\end{figure}
%--------------------------------------------------------------------
We see that the state originally stretched along $\mb{\hat{y}}$
oscillates between this state and the one stretched along
$-\mb{\hat{y}}$. In between, the system evolves through states
with no \emph{orientation}, but which are \emph{aligned} along the
$\mb{\hat{z}}\pm\mb{\hat{x}}$ directions. Since the stretched
states have orientation, this evolution is an example of
orientation-to-alignment and alignment-to-orientation conversion
\citep[see, for example, discussion by][]{Blum}.

\section{Abbreviations}

\begin{tabular}{ll}

AOC & Alignment-to-orientation conversion\\
BSR & Bennett-structure-related\\
DAVLL & Dichroic atomic vapor laser lock\\
EDM & A permanent electric-dipole moment\\
EIT & Electromagnetically induced transparency\\
FM NMOR & Nonlinear magneto-optical rotation with\\
& frequency-modulated light\\
FS & Forward scattering\\
FWHM & Full width at half maximum\\
MOT & Magneto-optical trap\\
NEOE & Nonlinear electro-optical effects\\
NMOE & Nonlinear magneto-optical effects\\
NMOR & Nonlinear magneto-optical rotation\\
SQUID & Superconducting quantum interference\\
& device\\
SR & Self-rotation of light polarization\\

\end{tabular}

\bibliography{NMObibl}

\end{document}